\documentclass[useAMS,usenatbib]{mn2e}
\usepackage[dvips]{graphicx}
\usepackage{amsmath}
\usepackage{siunitx}

\title[SNe~Ia SDSS-II host galaxy properties]{SNe~Ia host galaxy properties from Sloan Digital Sky Survey-II spectroscopy}
\author[J. Johansson et al.]{Jonas Johansson$^{1,2}$\thanks{E-mail: jjohansson@mpa-garching.mpg.de}, Daniel Thomas$^{1,3}$, Janine Pforr$^{1,4}$, Claudia Maraston$^{1,3}$, \newauthor
 Robert C. Nichol$^{1,3}$, Mathew Smith$^{5}$, Hubert Lampeitl$^{1}$, Alessandra Beifiori$^{1,6}$,  
 \newauthor
Ravi R. Gupta$^7$, Donald P. Schneider$^{8,9}$\\
$^1$Institute of Cosmology and Gravitation, Dennis Sciama Building, Burnaby Road, Portsmouth PO1 3FX\\
$^2$Max-Planck Institut f{\"u}r Astrophysik, Karl-Schwarzschild-Str. 1, D-85741 Garching, Germany\\
$^3$SEPnet, South East Physics Network\\
$^4$National Optical Astronomy Observatory, 950 N. Cherry Ave., Tucson, AZ, 85719, USA\\
$^5$Department of Physics, University of Western Cape, Bellville 7535, Cape Town, South Africa\\
$^6$Max-Planck-Institut f{\"u}r Extraterrestrische Physik, Giessenbachstrasse, 85748 Garching, Germany\\
$^7$Department of Physics and Astronomy, University of Pennsylvania, 209 South 33rd Street, Philadelphia, PA 19104, USA\\
$^8$Department of Astronomy and Astrophysics, The Pennsylvania State University, 525 Davey Laboratory, University Park, \\~~PA 16802, USA \\
$^9$Institute for Gravitation and the Cosmos, The Pennsylvania State University, 525 Davey Laboratory, University Park, \\~~PA 16802, USA 
}

\begin{document}

\pagerange{\pageref{firstpage}--\pageref{lastpage}} \pubyear{2011}
\maketitle
\label{firstpage}
\begin{abstract} 
We study the stellar populations of Type Ia Supernovae (SNe~Ia) host galaxies using Sloan Digital Sky Survey (SDSS)-II spectroscopy. We focus on the relationships of SNe~Ia properties with stellar velocity dispersion and the stellar population parameters age, metallicity and element abundance ratios. We utilize stellar population models of absorption line indices for deriving the stellar population parameters. Furthermore, we revisit the correlation between Hubble residual and photometric galaxy stellar mass, which has received much attention in the recent literature. We concentrate on a sub-sample of 84 SNe~Ia from the SDSS-II Supernova Survey with available host galaxy spectroscopy. This sub-sample has been selected based upon the quality of the spectroscopy and accuracy of the derived SNe~Ia parameters. In agreement with previous findings, we find that SALT2 stretch factor values show the strongest dependence on stellar population age in terms of a clear anti-correlation. Hence, SNe~Ia peak-luminosity is closely related to the age of the stellar progenitor systems, where more luminous SNe~Ia appear in younger stellar populations. We find no statistically significant trends in the Hubble residual with any of the stellar population parameters studied, including age and metallicity contrary to the literature, as well as with stellar velocity dispersion. We extend the sample to also include SNe~Ia with available SDSS host galaxy photometry only. We find that the method of stellar mass derivation is affecting the Hubble residual-mass relationship when lower number statistics are used, while this effect is weaker for the extended sample. For this larger sample (247 objects) the reported Hubble residual-mass relation is strongly dependent on the stellar mass range studied and behaves as a step function. In the high mass regime, probed by our host spectroscopy sample, the relation between Hubble residual and stellar mass is flat. Below a stellar mass of $\sim2\times10^{10} M_{\odot}$, i.e. close to the evolutionary transition mass of low-redshift galaxies reported in the literature, the trend changes dramatically such that lower mass galaxies possess lower luminosity SNe~Ia after light-curve corrections. This non-linear behaviour of the Hubble residual-mass relationship should be accounted for when using stellar mass as a further parameter for minimising the Hubble residuals.
\end{abstract}

\begin{keywords}
supernovae: general Ð cosmology: observations Ð distance scale Ð cosmological parameters Ð large-scale structure of Universe -- galaxies: abundances
\end{keywords}

\section{introduction}
\label{intro}

Type Ia supernovae (SNe~Ia) are useful for constraining cosmological parameters. Their large peak luminosities can probe vast cosmological distances and connect redshift space to luminosity distance. 
This property led to the discovery of an accelerating expansion of the universe \citep{riess98,perlmutter99}.

To be ideal cosmology indicators, all SNe~Ia explosions should have the same peak luminosity, but this is not the case. Observed SNe~Ia span a range in peak luminosity accompanied by varying decline rates \citep{phillips93}, i.e. the peak luminosity decreases with increasing decline rate. 
This variation can be corrected for to find a standardised peak luminosity. Several light-curve fitting tools are available \citep{jha07,salt2,sifto,kessler09}, where the shape of the light-curve and the colours are corrected to match a standardised peak luminosity. The light-curve shape correction is known as stretch-factor.

Correcting the light-curves increases the precision of the derived luminosity distances and consequently reduces the scatter in the redshift-distance relation, 
thus increasing 
the precision of the derived cosmological parameters. However, even after light-curve corrections this scatter is non-negligible. Understanding systematic uncertainties in the derived SNe~Ia light-curve parameters is therefore key to improving supernovae cosmology. 

The accepted model for SNe~Ia is thermonuclear explosion of a carbon-oxygen white dwarf (WD) that reaches the Chandrasekhar limit \citep{whelan73,hillbrandt00}. 
Two different channels have been proposed, either the single-degenerate (SD) scenario where mass is accreted from an evolved main-sequence binary companion, or the double-degenerate (DD) case of merging of two WDs \citep*[e.g][]{woosley86,branch95,branch01,hoflich95,greggio05,yungelson00}. This raises the possibility that different SNe~Ia populations may be present. On the other hand, it has been shown that a DD system is likely to lead to an accretion-induced collapse rather than a thermonuclear explosion \citep{saio98}. 

The delay time, i.e., the time between progenitor formation and explosion, in the DD scenario is determined by the life-time of the WD progenitors and the orbit of the two binary stars. The delay-time of a SD system partly depends upon the main-sequence lifetime of the companion star. 
A wide variety of delay times have been observationally suggested, ranging from $<$1 Gyr \citep{barris06,aubourg08} to $>2$ Gyr \citep{galyam04,strolger04,strolger05}. Considering several different progenitor systems, theoretical models find the rate of SNe~Ia explosions (SNR) to peak at delay times below or close to 1 Gyr \citep{yungelson00,greggio05,ruiter10}. The SNR of most progenitor systems then smoothly declines and becomes 10-100 times lower at delay times of $\sim$10 Gyr. Comparing observed delay times to theoretical predictions can constrain possible progenitor systems. 

It is well established that the SNR is higher in star-forming late-type than in passively evolving early-type galaxies \citep[e.g.][]{oemler79,vdbergh90,mannucci05,sullivan06,smith12}. 
Moreover, several authors found a dependence on host galaxy mass for the decline rate of SNe~Ia \citep{kelly10,lampeitl10,sullivan10}. Since galaxy mass correlates with stellar population parameters and properties of the inter-stellar medium \citep{tremonti04,gallazzi05,thomas10}, 
more fundamental 
correlations such as with stellar population age, metallicity and element abundance ratios may be expected. The tight relation between SNe~Ia decline rate and peak brightness indicates a primary dependence of luminosity on one of these parameters, thus holding important information about the properties of SNe~Ia progenitor systems. 

The luminosity of SNe Ia arises from the radioactive decay of $^{56}$Ni to $^{56}$Co, that then decays to $^{56}$Fe \citep{colgate69,arnett82}, such that the peak brightness depends on the $^{56}$Ni mass. \citet{timmes03} show theoretically that metallicity effects, in the range 1/3-3 Z$_{\odot}$, can induce a 25\% variation in the $^{56}$Ni mass. Theoretical models have also predicted that the carbon mass fraction, 
determined by the metallicity and mass of the WD progenitor, is responsible for SNe~Ia luminosity variations \citep{umeda99}. Dependencies of SNe~Ia properties on element ratios, such as C/Fe, may thus be expected. 
The ratio of light elements to Fe is also interesting
because Fe dominates the light-curve of SNe~Ia and could well influence some of the SN~Ia properties.

Observationally, it has been found that SNe~Ia in star-forming galaxies show slower decline rates as compared to SNe~Ia in passively evolving galaxies \citep{sullivan06,howell09,neill09,lampeitl10,smith12}. 
Moreover, several authors have recently found dependencies on stellar population age and/or metallicity for SNe~Ia decline rate \citep{hamuy00,gallagher08,howell09,neill09,gupta11}. 
However, either fairly small samples have been used \citep[$<$~30 objects,][]{hamuy00,gallagher08} or metallicity has been measured indirectly \citep{howell09,neill09}.

It is therefore desired to study the full range of stellar population parameters for a statistically significant sample of SNe~Ia host galaxies, which was the main aim of this study.

Correlations between SNe~Ia peak luminosity and stellar population parameters should ideally be eliminated through light-curve fitting, due to the tight correlation between peak luminosity and decline rate. However, if such correlations remain even after light-curve corrections, the previously mentioned non-negligible scatter in the redshift-distance relation could be reduced. 
These relationships are important to identify, especially for large redshift surveys, as the stellar population parameters age and metallicity change over cosmic time. 
Since the scatter in the redshift-distance relation is small, high-quality data are required to detect dependencies on host galaxies. 
To date galaxy mass and metallicity have shown to be the best candidates for such dependencies \citep{gallagher08,howell09,neill09,kelly10,lampeitl10,sullivan10,gupta11,dandrea}. This led to the introduction of stellar mass as a further parameter, besides light-curve stretch-factor and colour, to minimise the scatter in the redshift-distance relation \citep[e.g.][]{lampeitl10}. However, it is desirable to find the fundamental parameter of this scatter and, again, to do this the full range of stellar population parameters is needed for a statistically significant sample.  

In this paper, we study host galaxies of SNe~Ia from the Sloan Digital Sky Survey \citep[SDSS,][]{york00}-II  Supernova Survey \citep{frieman08} for which SDSS-II spectroscopy is available.  
This allows us to investigate SNe~Ia host galaxy stellar populations through Lick absorption line indices \citep[e.g.][]{worthey94,trager98}. These are defined for 25 prominent absorption features in the optical and are useful for breaking the age-metallicity degeneracy, deriving element ratios and they are insensitive to dust reddening \citep{macarthur05}. 
\citet{gallagher08} is to date the only study in the literature that relies on Lick indices for studying SNe~Ia host galaxies, based on a sample of 29 early-type galaxies \citep[][study a sample of five SN~Ia host galaxies through Lick indices]{gallagher08}. 
The SDSS-II Supernova Survey in combination with SDSS-II spectroscopy allow us to significantly improve upon this sample size. 
We use the method presented in \citet{johansson11}, which is based on up-to-date stellar population models of absorption line indices from \citet*[][TMJ]{TMJ11}, to derive age and metallicity together with a range of element abundance ratios including C/Fe as well as O/Fe, Mg/Fe, N/Fe, Ca/Fe and Ti/Fe. This allows us to explore the full range of stellar population parameters to search for the fundamental parameter of the variations of SNe~Ia properties. 

In addition, we determine stellar masses from SDSS-II photometry using different methodologies of photometric SED-fitting to study systematic uncertainties in the proposed dependency of the scatter in the redshift-distance relation on stellar mass. 

The paper is organised as follows. The data sample used is described in Section~\ref{data} along with the description of our derivation of spectroscpic stellar population parameters and stellar masses. The relationships between SN~Ia properties and host galaxy parameters are presented in Section~\ref{results}. We discuss the results in Section~\ref{disc} and concluding remarks are given in Section~\ref{conc}.

\section{data sample}
\label{data}

The sample used in this study is drawn from the SDSS-II Supernova Survey \citep{frieman08}. 
The SDSS 2.5 m telescope \citep{york00,gunn06}, located at the Apache Point Observatory (APO), is equipped with a multi-object spectrograph \citep{smee12} and wide field CCD camera \citep{gunn98}. 
During the eight year period of 2000-2008, the SDSS-I and SDSS-II surveys obtained deep images in the SDSS \textit{ugriz} filters \citep{fukugita96} and spectroscopy for more than 930,000 galaxies.

The SDSS-II SN survey was performed over a three year period (2005-2007) to repeatedly image transient objects in the SDSS  ``Stripe 82'' region. 
Thousands of potential SN candidates were observed. Out of these, 890 were identified as SN~Ia candidates including 551 with spectroscopic confirmation \citep{sako08,holtzman08,sako13}. The SDSS-II SN sample has been used for cosmological analyses \citep{kessler09a,sollerman09,lampeitl10a} as well as photometric and spectroscopic studies of SNe~Ia host galaxy properties \citep{lampeitl10,gupta11,dandrea,smith12}.

Following \citet{lampeitl10}, we include photometrically classified SNe~Ia, to ensure a more complete sample. Based on the Bayesian light-curve fitting of \citet{sako08}, these supernovae have a light-curve consistent with being a Type Ia. 
The contamination of non-SNe~Ia objects using the photometric classification scheme is only  $\sim$3\% \citep{dilday10a}.

In this work we focus on a sub-sample of SNe~Ia with available SDSS-II host galaxy spectroscopy (from now on referred to as the host spectroscopy sample). 
The SN~Ia coordinates and redshifts have been cross-matched with the corresponding parameters in the SDSS-II DR7 catalogue to identify the nearest object with a spectrum within a 0.25" radius, resulting in a host spectroscopy sample of 292 objects. A similar cross-match has been made to acquire co-add photometry from the SDSS  ``Stripe 82'' region \citep{annis11}. 
The host galaxy coordinates from both runs have been further cross-matched to ensure that the spectroscopy and photometry belong to the same object. 
Out of the 292 supernovae in the host spectroscopy sample, 138 (47\%) are spectroscopically confirmed SNe~Ia. For redshifts below 0.21, the fraction of spectroscopically confirmed SNe Ia increases to 
60\%. 
This is also the redshift range covered by the final sample selection (see Section~\ref{cuts}) used in the main analysis.

In addition to the host spectroscopy sample we include in Section~\ref{HR_ext} SNe~Ia host galaxies with available SDSS photometry only, to extend the stellar mass analysis. This extended photometric sample adds another 102 objects from the study of \citet{lampeitl10} and is described in detail in Section~\ref{HR_ext}.

The derived SNe~Ia properties, host galaxy parameters and final selection of the host spectroscopy sample are described in the following sections.

\begin{figure*}
\centering
\includegraphics[clip=true,trim=0cm 1.6cm 1.5cm 0cm, angle=90, scale=0.39]{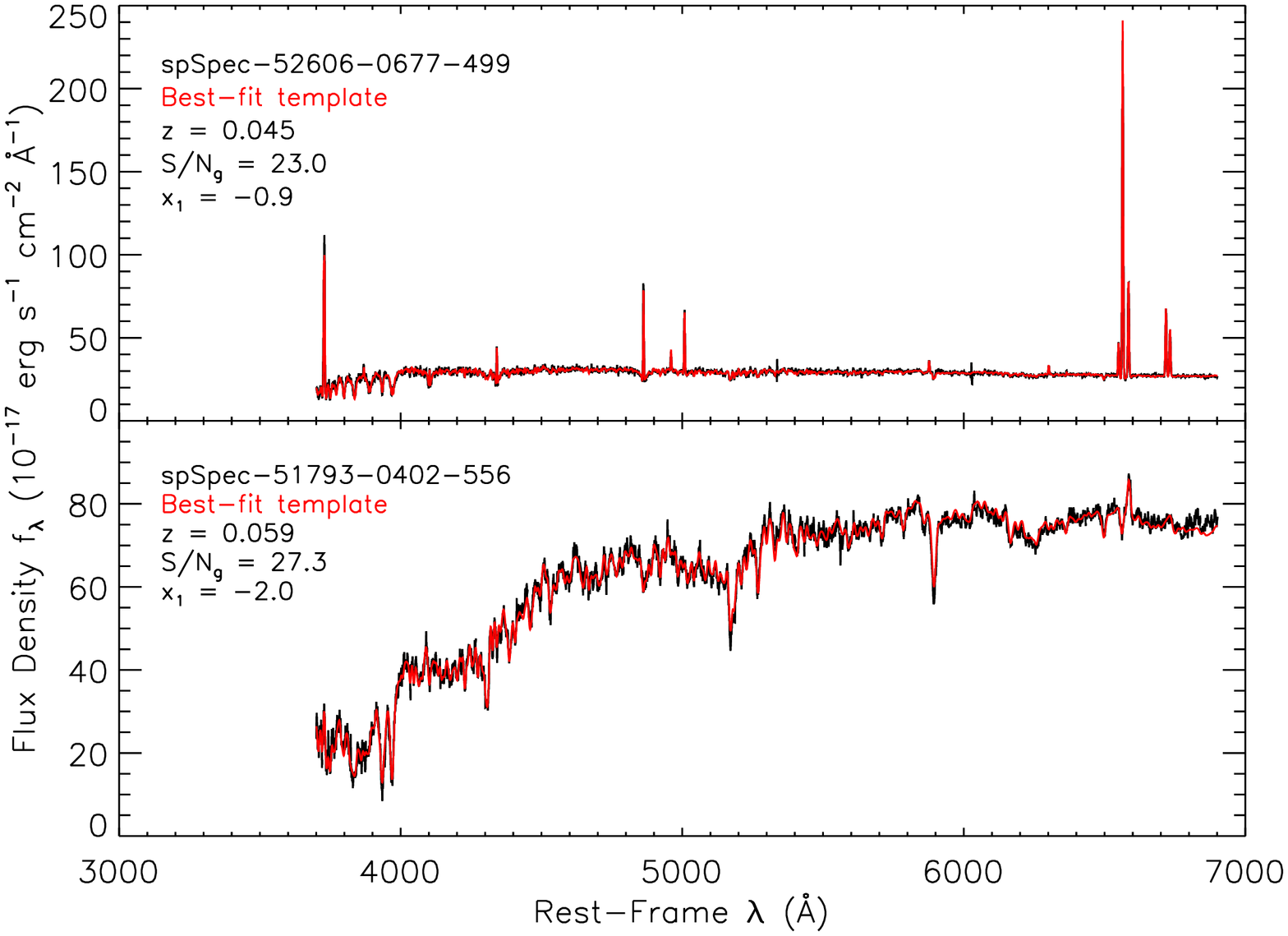}\includegraphics[clip=true,trim=0cm 0cm 1.5cm 0cm, angle=90, scale=0.39]{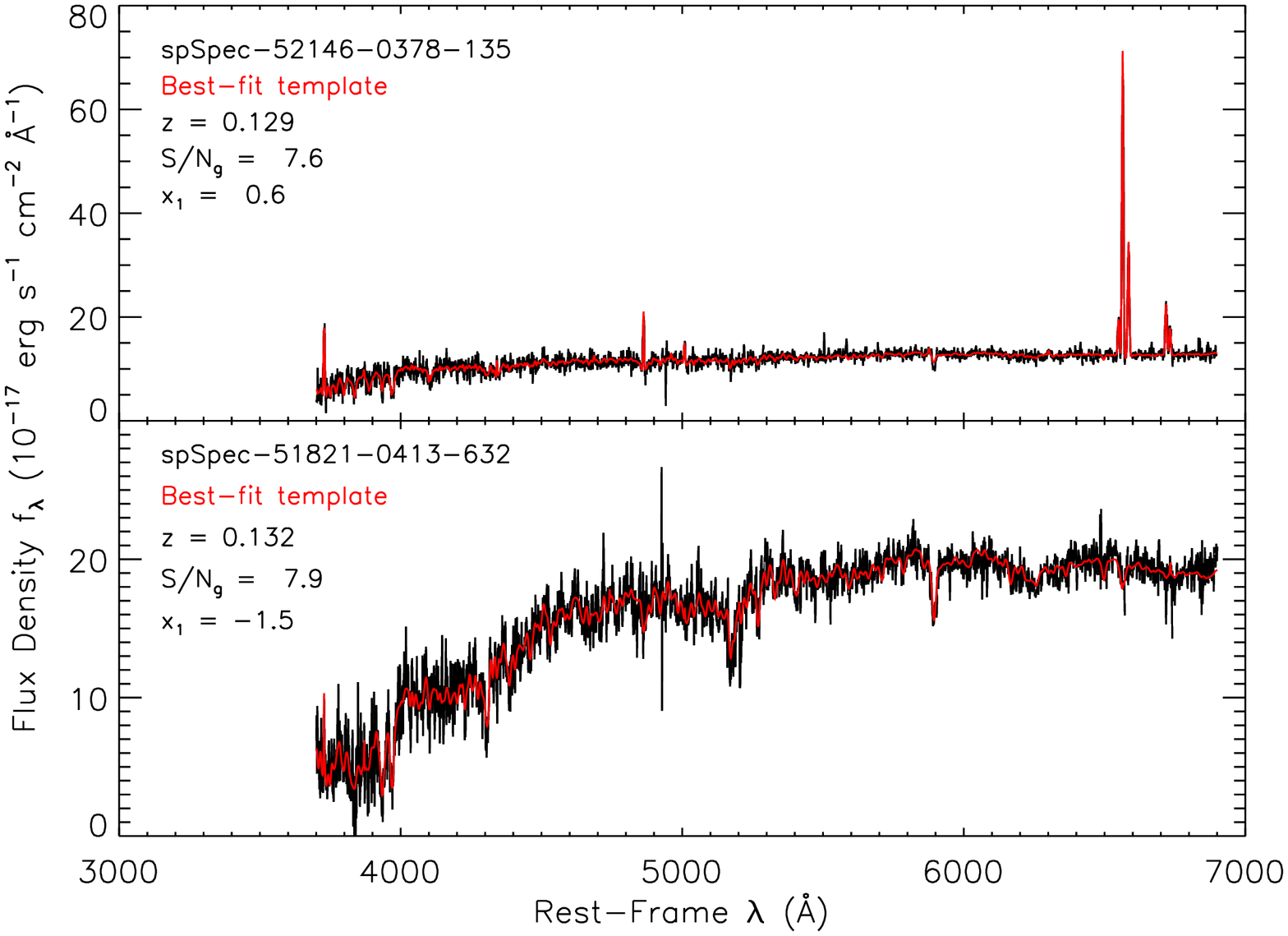}
\caption{Four examples of host galaxy spectra (black SEDs) at rest-frame wavelength and with the best-fit model templates from \texttt{GANDALF/PPXF} (red SEDs) overlaid. The top and lower panels are galaxies hosting SNe Ia with large and small stretch factor values, respectively, as given by the labels. Left and right panels are high and low S/N spectra, with values given by the labels. The redshift for each object is also indicated by the labels.}
\label{SampleSpec}
\end{figure*}

\subsection{SNe~Ia properties}
\label{SNIaprop}

The light-curve fitting technique follows \citet{lampeitl10} to which the reader is referred for more details. A summary of the main features is presented here.

A number of light-curve fitting methods exist in the literature and several authors have confirmed that the results are robust against different analysis algorithms \citep{kelly10,lampeitl10,sullivan10}. SALT2 \citep{salt2}, which is one of the most common light-curve fitting techniques, is adopted in this work. The output of SALT2 are B-band apparent brightness (m$_B$), stretch factor (x$_1$) and colour (c) term. The distance modulus ($\mu_{SN}$) is then calculated as
\begin{equation}
\label{muEq}
\mu_{SN}=(m_B-M_B)+\alpha x_1-\beta c
\end{equation}
where M$_B$ is the ÒstandardisedÓ absolute SNe~Ia magnitude in the B-Band for x$_1$ = c = 0, $\alpha$ describes the overall stretch law for the sample, and $\beta$ is the colour law for the whole sample. These parameters are determined by minimising the scatter in the distance redshift relation or the Hubble residual (HR)
\begin{equation}
\label{HREq}
HR=\mu_{SN}(M_B,\alpha,\beta)-\mu_{cosmo}(z,H_0,\Omega_m,\Omega_{\Lambda})
\end{equation}
where H$_0$ is the Hubble constant, 
$\Omega_M$ is the density parameter for matter and $\Omega_{\Lambda}$ is the density parameter for the cosmological constant. 
We aim at detecting systematic trends in the derived Hubble residuals and adopt H$_0$=65 km s$^{-1}$ Mpc$^{-1}$, $\Omega_m$=0.3 and $\Omega_{\Lambda}$=0.7. The resulting light-curve parameters are M=-30.11, $\alpha$=0.12 and $\beta$=2.86.

For a negative Hubble residual, i.e. $\mu_{cosmo}>\mu_{SN}$, the cosmology places the SNe~Ia at a greater distance than the peak luminosity predicts. This also means that the SNe~Ia peak luminosity has not been fully corrected and it is too bright to match the cosmology.

\subsection{Host galaxy properties}
\label{hostprop}

The aim of this work is to analyse the host galaxy properties of SNe~Ia from a spectroscopic point of view, particularly using absorption line indices. For this purpose we need clean galaxy absorption spectra, free from contaminating emission lines. Absorption line indices are measured on the clean spectra and analysed using single stellar population models. We use the method detailed in \citet{thomas10} and \citet{johansson11} of which we provide a brief summary in the following section. 

\subsubsection{Kinematics and Lick indices}
\label{gandalf}

We utilise the fitting-code \texttt{GANDALF} \citep{sarzi06}, which is based on on the penalised pixel-fitting (\texttt{PPXF}) method of \citet{ppxf}, to obtain clean absorption spectra. A brief description of the code is given here, while the reader is referred to \citet{sarzi06} and \citet{ppxf} for details.
\texttt{GANDALF/PPXF }simultaneously
fits stellar templates and emission line Gaussians to galaxy spectra. The result 
is a separation of emission and absorption spectra. As stellar templates we adopt the single stellar population spectral energy distributions (SEDs) of \citet{mastro}. Besides the clean absorption spectra, other useful outputs of \texttt{GANDALF/PPXF} are stellar and gas kinematics, E(B-V) dust reddening and emission line fluxes/equivalent widths (EWs).

Fig.~\ref{SampleSpec} shows four examples of host galaxy spectra (black spectral energy distributions, SEDs) with varying stretch factor values (x$_1$) and signal-to-noise ratios (S/N) per pixel ($\sim$1 \AA~at 4500 \AA) in the g-band of the spectra, as given by the labels. The corresponding best-fit stellar templates from \texttt{GANDALF/PPXF} (red SEDs) are also shown. Prominent emission lines are visible in the galaxies hosting high stretch factor values.

The Lick system consists of definitions of absorption line indices for 25 prominent absorption features in the optical \citep{worthey94,trager98}. We measure the Lick indices on the clean absorption spectra, free from contaminating emission lines, using the latest definitions of \citet{trager98}. The resolution of the spectra are downgraded to the Lick/IDS resolution \citep[that of the models, $\sim$8-10 \AA,][]{worthey97} prior to measuring the indices. 
The index measurements are then corrected for velocity dispersion broadening, using the velocity dispersions found by \texttt{GANDALF /PPXF} together with the best fit stellar templates. The Lick indices are measured on the best fit stellar templates broadened to the Lick/IDS resolution, both before and after further broadening with respect to the velocity dispersion measurements. The difference between these two measurements gives the correction factor which we apply to the Lick indices measured on the galaxy spectra. Using the error vectors provided with each SDSS spectrum we estimate Lick index errors through Monte Carlo simulations.

It is important to note that these are all standard procedures. Hence, our derived Lick indices can be compared directly to other galaxy samples in the literature. This holds for samples with flux-calibrated spectra, as the Lick indices are sensitive to the instrumental configuration. When comparing Lick indices from different samples based on non-flux-calibrated spectra, the common procedure is to use standard-stars observed with both instrumental configurations to determine possible offsets/calibrations in the Lick indices \citep[e.g.][]{johansson10}. 
In this work we fit the TMJ models to the observed Lick indices. The TMJ models are based on 
the flux-calibrated stellar library MILES \citep{miles,johansson10}. Hence the models are no longer tied to the non-flux-calibrated Lick/IDS system, making standard star-derived offsets unnecessary when using flux-calibrated galaxy spectra. Lick index offsets are crucial when adopting stellar population models based upon non-flux calibrated stellar libraries, such as the Lick/IDS library \citep[e.g.][]{worthey94}, due to the sensitivity to the instrumental configuration of the Lick indices as mentioned above.

\subsubsection{Stellar population parameters}
\label{tmj}

From the derived Lick indices we determine the stellar population parameters age and total metallicity and the element abundance ratios accessible through integrated light spectroscopy of galaxies (O/Fe, Mg/Fe, C/Fe, N/Fe, Ca/Fe and Ti/Fe), using the iterative method described in detail in \citet{johansson11}. This method is based on the TMJ stellar population models of absorption line indices.
These are single stellar population models with variable element abundance ratios 
for the 25 Lick indices. 
Since several indices respond to variations of the same element abundances we have developed an iterative method. A $\chi^2$-minimisation routine is used at each step to find the best-fit model.

First we determine the traditional light-averaged stellar population parameters age, total metallicity, and $\alpha$/Fe ratio from indices sensitive to these three parameters only. In the subsequent steps we add in turn particular sets of indices that are sensitive to the element the abundance of which we want to determine. In each step we re-run the $\chi^2$-fitting code with a new set of models to derive the abundance of this element. This new set of models is a perturbation to the solution found for the base set and is constructed by keeping the stellar population parameters age, metallicity, and $\alpha$/Fe fixed while modifying the abundance of the element under consideration. The derivations of element abundances are iterated until the values do not further change. 
At the end of the sequence we re-determine the overall  $\chi^2$  using all indices together and re-derive the base parameters age, metallicity, and $\alpha$/Fe for the new set of element ratios. The whole procedure is iterated until the final  $\chi^2$  stops improving by more than 1\%. 

An enhanced $\alpha$/Fe ratio is characterised by a depression in Fe and reflects the ratio between total metallicity and iron abundance. Since O dominates the mass budget of Õtotal metallicityÕ,  the $\alpha$/Fe parameter derived through our procedure can be reasonably interpreted as O/Fe. In \citet{johansson11} we therefore re-named the parameter $\alpha$/Fe to O/Fe under the assumption that this ratio provides an indirect measurement of oxygen abundance, i.e.

\begin{equation}
[O/Fe] \equiv [\alpha/Fe] 
\end{equation}

\subsubsection{Stellar mass}
\label{data:masses}

We derive stellar masses (M$_*$) from photometry according to standard SED-fitting as in \citet{daddi05} and  \citet{M06} 
via normalization of the SED. The SED-fitting is performed using the \emph{HyperZ} code of \citet*{Bolzonella00} 
with SDSS spectroscopic redshifts and extinction corrected \emph{ugriz} magnitudes as input. 
Masses are derived using different sets of stellar population models, i.e. from \citet{M05} (M05) and \citet{bc03} (BC03). As is well known, the main difference between these models is the different treatment of the Thermally Pulsing Asymptotic Giant Branch (TP-AGB) contribution in the M05 models. The TP-AGB affects the luminosity at near-IR wavelengths for galaxies with stellar population ages between $\sim$ 0.2 and 2 Gyr \citep{M05}. 

\begin{figure}
\centering
\includegraphics[clip=true,trim=1cm 0cm 0cm 16cm,scale=0.51]{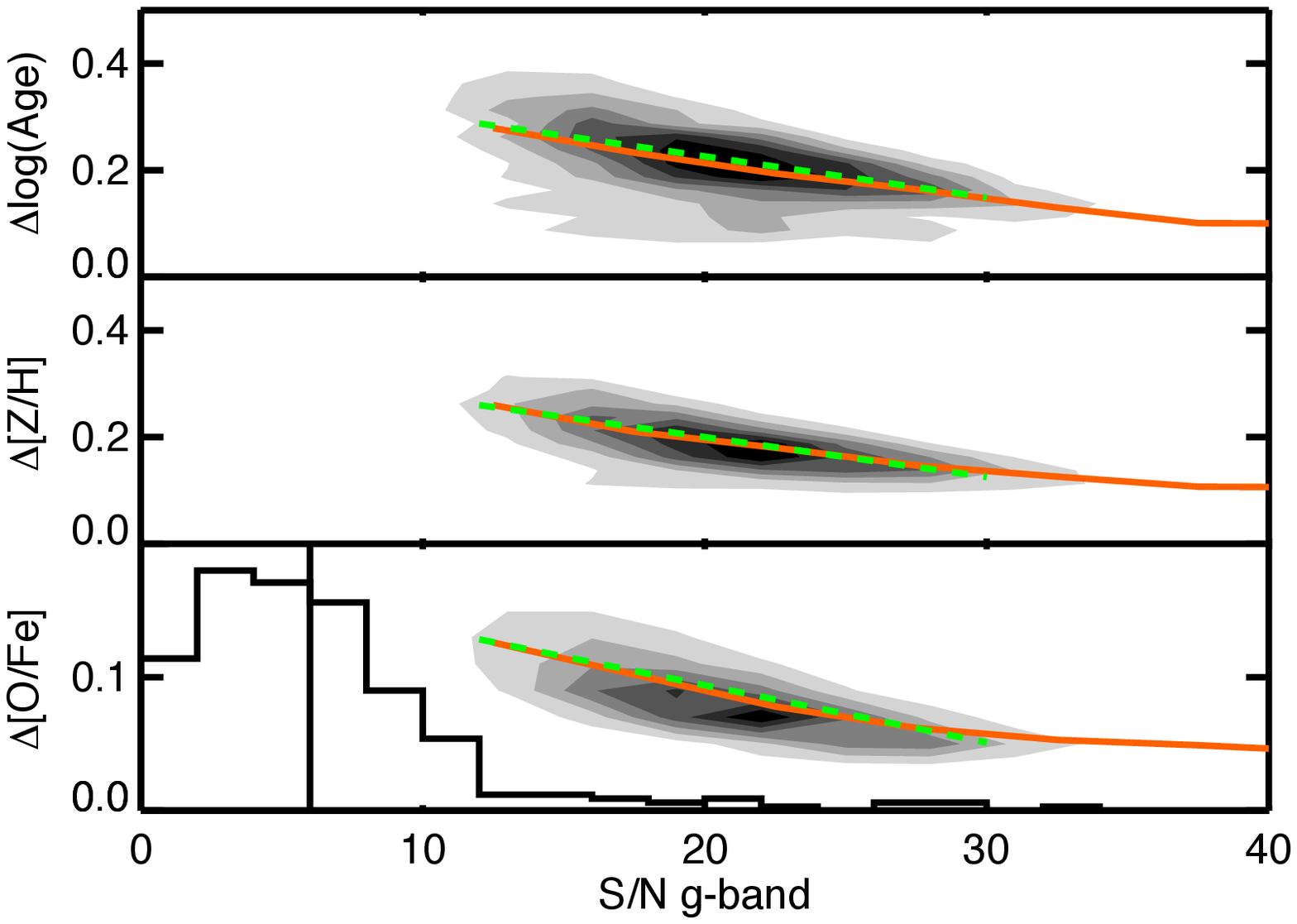}
\caption{The relationship between the error in stellar population parameters and quality of galaxy spectra, in terms of S/N per pixel ($\sim$1 \AA~at 4500 \AA) in the g-band \citep[remake of Fig.~4 from][]{johansson11}. Contours represent 3802 SDSS early-type galaxies from the MOSES sample \citep{thomas10} for error in age (upper panel), total metallicity (middle panel) and [O/Fe] (lower panel). Orange lines connect the mean errors in bins of g-band S/N, while green dashed lines show the linear behaviors below S/N=30. The distribution of the full sample of SNe~Ia host galaxies (292 objects) is presented by the histogram, scaled by multiplying the binned histogram-values with 0.18/60. The black vertical line represents the cut at S/N=6, which defines our final sample.}
\label{StoN}
\end{figure}

To obtain as robust total M$_*$ as possible, we exclude internal reddening in the fitting procedure, following \citet{pforr}. 
These authors find that the inclusion of galaxy internal reddening as a fit parameter 
generates masses that are often underestimated with respect to the true stellar mass.
This result occurs because dust reddening combined with young model ages produce young dusty solutions as best-fit, with low mass-to-light ratios \citep[the age-dust degeneracy, see][]{renzini06}. 
Furthermore, we exclude single burst templates which sometimes further increase the risk of getting underestimated stellar masses. 

To compare with literature values, which usually include reddening,  
we further calculate stellar masses for the following options: 
M05 models with reddening (M05$_{red}$), BC03 models without reddening and BC03 model with reddening (BC03$_{red}$).

\subsubsection{Gas-phase metallicity}
\label{data:gasZ}

Gas-phase metallicities are derived from the EWs of the emission lines measured by \texttt{GANDALF/PPXF} (see Section~\ref{gandalf}). Several calibrations for the relationship between metallicity and emission-line ratios are available in the literature. Relative metallicities generally agree between the various calibrations, while absolute metallicities frequently disagree. The calibration of \citet[][KD02]{kewley02} is commonly used in the literature, but it is based on the $[NII\lambda6583]/[OII\lambda3727]$ ratio. 
The large difference in wavelength between the [NII$\lambda$6583] and [OII$\lambda$3727] lines make this ratio sensitive to dust reddening. 
Instead we use the O3N2 index ($\log(([OIII\lambda5007]/H\beta)/([NII\lambda6583]/H\alpha))$ which is not sensitive to dust reddening due to the proximity of the emission lines in both the [OIII$\lambda$5007]/$H\beta$ and [NII$\lambda$6583]/H$\alpha$ ratios. The O3N2 index was calibrated with $12+\log[O/H]$ in \citet{pettini04} (PP04)

\begin{equation}
\label{PP04}
(12+\log[O/H])_{PP04}=
8.73-0.32\times O3N2
\end{equation}
\citet{kewley08} re-calibrated the metallicity from PP04 on to the KD02 metallicity and found the following relationship
\begin{multline}
\label{KD02}
(12+\log[O/H])_{KD02}=159.0567-\\
54.18511\times(12+\log[O/H])_{PP04}+\\
6.395364\times(12+\log[O/H])_{PP04}^2-\\
0.2471693\times(12+\log[O/H])_{PP04}^3
\end{multline}
Using Eq.~\ref{PP04}-\ref{KD02} we determine metallicities for objects with detected emission lines. The presence of an active galactic nucleus (AGN) may strongly affect the derived metallicities since a stellar ionising radiation field is assumed in the commonly used metallicity calibrations \citep{kewley08}. Hence we will therefore not consider gas metallicities for purely classified AGN hosts, while transition objects between star-formation and AGN are treated with caution.

\begin{figure}
\centering
\includegraphics[clip=true,trim=1.2cm 0.5cm 0cm 15cm,scale=0.53]{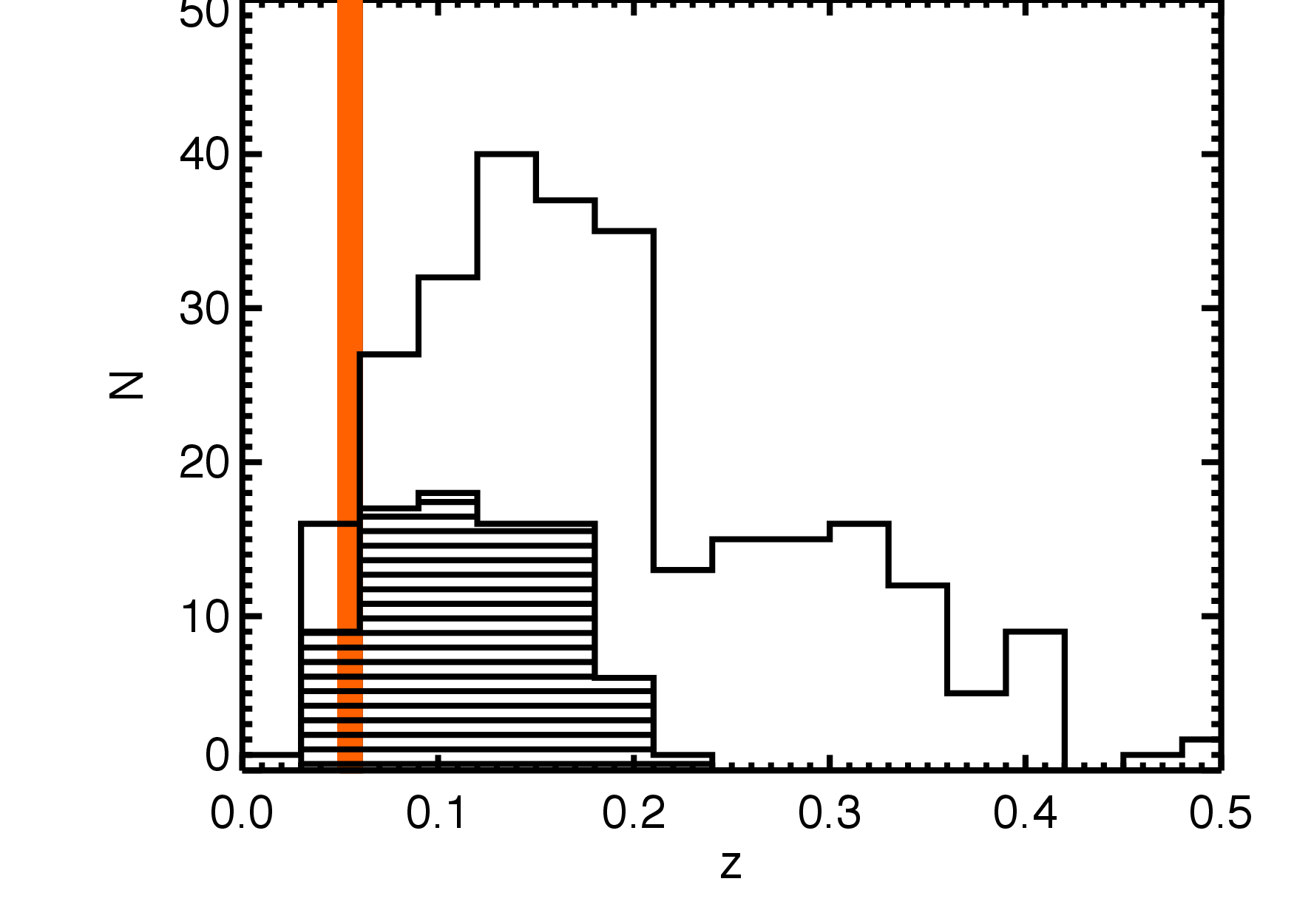}
\caption{Distribution in redshift for the full sample of 292 SNe~Ia (open histogram) and the final sample of 84 SNe~Ia (dash-filled histogram). The orange area represents the redshift range covered by the early-type galaxy sample used in Fig.~\ref{StoN} (represented by the contours) to show the relationship between stellar population parameter errors and quality of corresponding luminosity-weighted galaxy spectra.}
\label{zdistr}
\end{figure}

\subsection{Sample cuts}
\label{cuts}

The numbers of supernovae for the host spectroscopy sample are given in Table~\ref{Table_sel}, where ``Spec. Confirm''  gives just the spectroscopically confirmed SNe Ia, for the different selection cuts described in this section.

The individual objects of the host spectroscopy sample are presented in Table~\ref{ELtable_short} (the full version of this table is available in Appendix~A), where we provide SDSS-II Supernova Survey identification numbers, the SNe classification technique used (Sp=Spectroscopically, Ph=Photometrically) and host galaxy coordinates and redshifts. We also provide sample cut designation, where f=final host spectroscopy sample selected through the cuts presented below (m=Master sample described in Section~\ref{HR_ext}). 

\subsubsection{SN Ia cuts}
\label{SNcuts}

To avoid contamination by low quality data, we cut the sample to exclude the lowest quality SNe~Ia observations.  The accepted $\chi^2$ of the light-curve fit probability from SALT is constrained and we remove everything with a reduced $\chi^2>3$ \citep[][]{lampeitl10}. 
Following \citet{lampeitl10}, we retain SNe~Ia with SALT2 parameters inside the ranges -4.5$<$x$_1$$<$2.0 and -0.3$<$c$<$0.6, and remove anything with particularly large x$_1$/c uncertainties. Outside these ranges \citet{lampeitl10} found the derived parameters to be unreliable. 
These cuts of the fitted light-curve (LC) parameters reduce the sample to 196 objects (``LC fitter cuts''  in Table~\ref{Table_sel}).

It is possible that parameter gradients and local deviations from the integrated parameters will affect the results. 
Hence we discard SNe~Ia that are separated by more than 0.15' from the host galaxy, which is three times the SDSS fiber diameter (0.05'). This cut produces a sample of 183 objects. 

\begin{table}
\center
\caption{Numbers of the host spectroscopy sample before and after the different cuts. The second column gives the number of spectroscopically confirmed SNe~Ia and the third column both spectroscopically and photometrically identified SNe~Ia.}
\label{Table_sel}
\begin{tabular}{lcc}
\hline
\bf Selection & \bf Spec. Confirm & \bf Total$^a$ \\
\hline
     Full  &   138  &      292    \\ 
     LC fitter cuts   &   116  &      196   \\ 
     Separation cut   &  105  &  183   \\ 
     Host spec. quality cuts   &   63  &      84   \\ 
\hline
\end{tabular}\\
\flushleft
$^a$Total number used in the main analysis including both spectroscopically and photometrically identified SNe~Ia\\
\end{table}

\subsubsection{Host spectroscopy cuts}
\label{Hostcuts}

Fig.~\ref{StoN} shows the errors, represented by contours, in the stellar population parameters age (upper panel), total metallicity (middle panel) and [O/Fe] (lower panel) as a function of the median S/N ratios per pixel ($\sim$1 \AA~at 4500 \AA) in the g-band. This is the same format as Fig.~4 from \citet{johansson11}; the errors are for 3802 SDSS early-type galaxies of the sample of Morphologically Selected Ellipticals in SDSS \citep[MOSES,][]{schawinski07,thomas10}, derived using the same method based on absorption line indices adopted in this work (see Section~\ref{hostprop}). Orange lines connect median errors in bins of g-band S/N, while green dashed lines show the linear behaviors below S/N=30. The errors on stellar population parameters increase almost linearly with decreasing S/N over the S/N-range covered by the MOSES sub-sample. 
The quality, in terms of S/N in the g-band, of the 292 host galaxy spectra of our sample is illustrated by the histogram in the lower panel of Fig.~\ref{StoN}. 

The distribution in redshift-space for the full SNe~Ia sample is shown in Fig.~\ref{zdistr} (open histogram) together with the redshift-range covered by the MOSES sample (orange area). Only a small fraction of the SNe~Ia sample falls in and below the redshift-range of the MOSES sample. Thus the majority of the SNe~Ia host galaxies have a lower S/N than the MOSES sample (as seen in Fig.~\ref{StoN}). It is reasonable to assume that the linear behaviour of the error-S/N relationships can be extrapolated to lower S/N. However, at the lowest values this may not hold such that the errors increase significantly. To avoid including data with too large uncertainties in the stellar population parameters, we cut the sample at a specific S/N value to balance the quantity and the quality of the sample. 
By changing the S/N-limit by one unit we find that the sample size is consistently reduced by $\sim$15\% up to S/N=6, while above this value the sample reduction increases to $>$25\%. 
Hence, we only include host galaxies with S/N$>$6 in the g-band (vertical black line in the lower panel of Fig.~\ref{StoN}). 
We also discover one spectrum showing quasar features and two objects with telluric contaminated Mgb absorption features.
These quality cuts of the host galaxy spectra produce a final sample of 84 objects (``Host spec. quality cuts''  in Table~\ref{Table_sel}).
The redshift distribution of this sub-sample is shown by the dash-filled histogram in Fig.~\ref{zdistr}. The final host spectroscopy sample covers redshifts up to $\sim$0.2.

\begin{table}
\center
\caption{Specifications for the host spectroscopy sample.}
\label{ELtable_short}
\begin{tabular}{cccccc}
\hline
\bf ID & \bf Class.$^a$ & \bf Ra & \bf Dec &  \bf z & \bf Sample$^b$ \\
& & (deg)  & (deg) & &   \\
\hline
     691  &    Ph &      329.7300  &    -0.4990  &    0.1310  &  m   \\ 
     701  &    Ph &      334.6050  &     0.7976  &    0.2060  &  f  \\ 
     717  &    Ph &      353.6280  &     0.7659  &    0.1310  &     \\ 
     722  &    Sp &        0.7060  &     0.7519  &    0.0870  &  f  \\ 
     739  &    Sp &       14.5960  &     0.6794  &    0.1080  &  f  \\ 
     762  &    Sp &       15.5360  &    -0.8797  &    0.1920  &  m  \\ 
     774  &    Sp &       25.4640  &    -0.8767  &    0.0940  &  f  \\ 
\hline
\end{tabular}\\
\flushleft
$^a$Supernovae classification, Sp=Spectroscopic, Ph=Photometric\\ ~(see Section~\ref{data})\\
$^b$Sample selection, f=final host spectroscopy sample (see\\ ~Section~\ref{cuts}), m=Master sample (see Section~\ref{HR_ext})\\
(This table is available in its entirety in Appendix~A. A portion is shown here for guidance regarding its form and content.)
\end{table}

\begin{figure}
\centering
\includegraphics[clip=true,trim=0.2cm 6cm 0cm 3cm,scale=0.42]{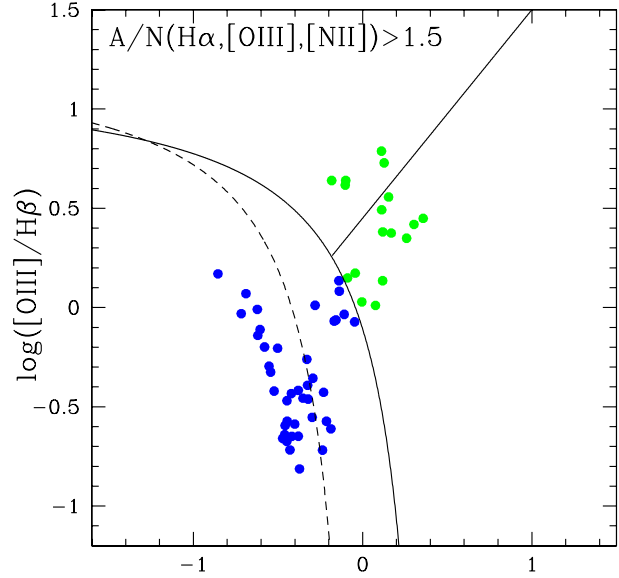}
\caption{BPT-diagram for the 59 galaxies with detected emission lines, i.e. AoN$>$1.5 for H$\alpha$, [OIII]($\lambda$5007) and [NII]($\lambda$6584). Blue points are star-forming galaxies and green points exhibit AGN activity only. The solid curved line is the theoretical star-formation limit from \citet[][Ke01]{kewley01} and the dashed curved line is the empirical separation of AGN and purely star-forming galaxies from \citet[][Ka03]{kauffmann03}. Objects that fall between these lines are transition objects, hosting both star-formation and AGN activity. The separation of LINER and Seyfert AGNs from \citet[][S07]{schawinski07} is indicated by the solid straight line. }
\label{bpt}
\end{figure}

\begin{figure*}
\includegraphics[clip=true,trim=0cm 0cm 0cm 0.6cm,angle=90,scale=0.75]{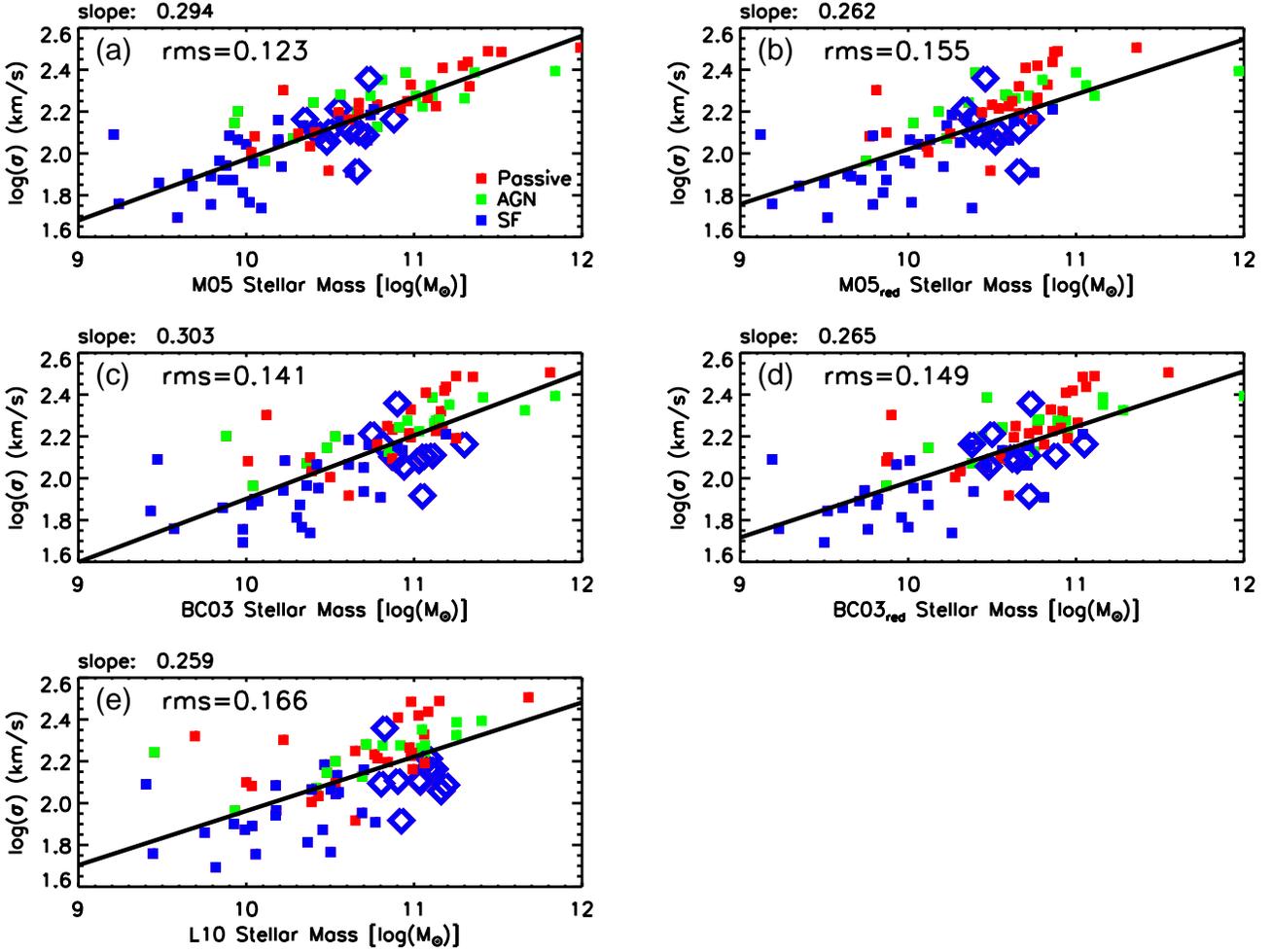}
\caption{Comparison between stellar velocity dispersion and the different stellar masses derived in this work, i.e. M05 models without reddening (M05, panel a), M05 models with reddening (M05$_{red}$, panel b), BC03 models without reddening (BC03, panel c), BC03 model with reddening (BC03$_{red}$, panel d) and the masses from \citet{lampeitl10} (L10, panel e). The data points are colour-coded according to their emission line classifications from Section~\ref{emission} and given by the labels in the upper right panel, where passive=passively evolving, AGN=AGN activity and SF=star-forming. 
Least-square fits to the $\sigma$-mass relationships are also included (solid lines) with the slopes given at the top of the panels. The standard deviations of the residuals to these fits are indicated by the labels. Star-forming  galaxies with high L10 masses (lower left panel) are high-lighted as large blue diamonds in all panels, i.e. these symbols represent the same objects in all panels.
}
\label{MvsS}
\end{figure*}

\section{Results}
\label{results}

Relationships between host galaxy parameters age, metallicity, element ratios, velocity dispersion and stellar mass with SNe~Ia properties stretch factor and Hubble residual are presented in this section. We do not find any significant trends between the host galaxy parameters and SALT2 colour, hence such relations are not further discussed.

For several of the studied relations we derive least-square fits. These are weighted using errors in the dependable parameter. 
Estimated errors of the fit-parameters, derived using standard routines for least-square fitting, are used for computing the significance of the derived fits. 
When stated so, the least-square fits are sigma-clipped at a 2$\sigma$ level, i.e., we first fit a line using all data points, remove the outliers deviating more than 2$\sigma$ from the derived fit and finally redo the least-square fitting using the new sample. 

\begin{figure*}
\centering
\includegraphics[clip=true,trim=2.3cm 1.5cm 3cm 0cm,angle=90,scale=0.67]{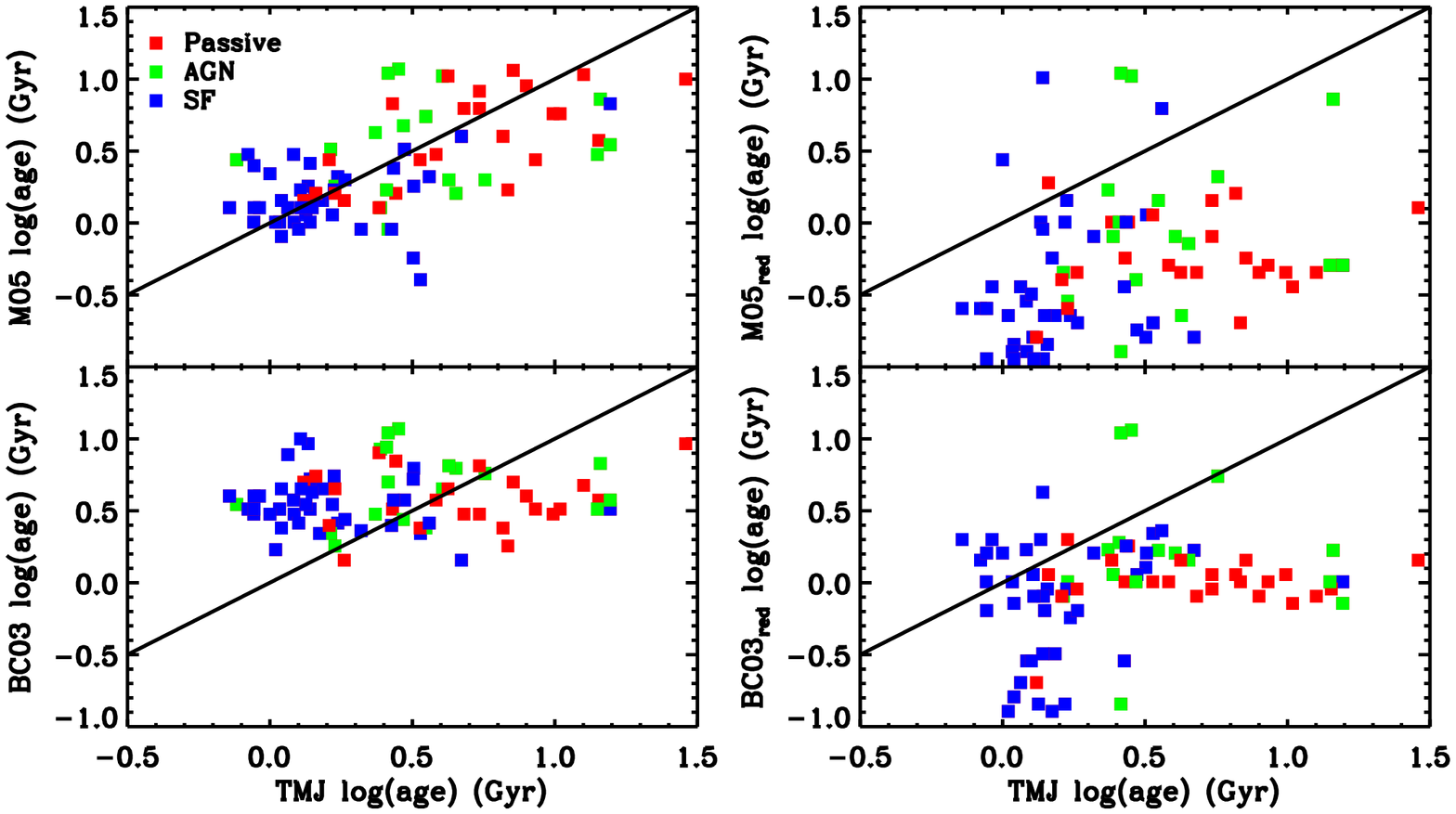}
\caption{Comparison between ages derived using absorption line indices (TMJ, see Section~\ref{tmj}) and the derived ages from each of the four SED-fitting cases, e.g. M05 models without reddening (M05, upper left panel), M05 models with reddening (M05$_{red}$, upper right panel), BC03 models without reddening (BC03, lower left panel) and BC03 models with reddening (BC03$_{red}$, lower right panel). The one-to-one relation is included in each of the four panels (solid lines). 
Data points are colour labelled according to their emission line classifications from Section~\ref{emission}.}
\label{agecomp}
\end{figure*}

\subsection{Emission line diagnostics}
\label{emission}

Emission line diagnostics are measured on the emission-line spectra separated from the galaxy absorption spectra (see Section~\ref{gandalf}). We use the emission line EWs and amplitude-to-noise (AoN) ratios to classify each galaxy as passively evolving, star-forming or hosting an AGN. Following \citet*{bpt} (BPT) the ratios [OIII]($\lambda$5007)/H$\beta$ and [NII]($\lambda$6584)/H$\alpha$ separate star-formation and AGN activity. The AoN for [OIII]($\lambda$5007), [NII]($\lambda$6584) and H$\alpha$ sets the limit for emission line detection. Due to the generally low S/N of the galaxy spectra (see Section~\ref{cuts}), we allow an AoN limit of 1.5, compared to 3.0 used in \citet{kauffmann03} and \citet{schawinski07}. Hence galaxies are classified as passively evolving if they do not have an AoN$>$1.5 for [OIII]($\lambda$5007), [NII]($\lambda$6584) and H$\alpha$. With this limit we find 25 out of the 84 galaxies in the final sample (see Section~\ref{cuts}) to be passively evolving, i.e. ~30 \% of the SNe~Ia in our sample occur in passively evolving galaxies.

Fig. \ref{bpt} shows the location of the emission-line detected galaxies in the BPT-diagram. The solid, curved line is the theoretical star-formation limit from \citet{kewley01}, i.e. galaxies that fall below this line form stars. The dashed, curved line is the empirical separation of AGN and purely star-forming galaxies from \citet{kauffmann03}. Objects that fall between these lines are transition objects, hosting both star-formation and AGN activity. For this work we are interested in detecting star-formation activity and consequently label everything below the \citet{kewley01} line as star-forming (blue points). The fraction of star-forming galaxies is 41 out of 84 ($\sim$50~\%) and 18 galaxies are AGN ($\sim$20~\%, green points). The separation of LINER and Seyfert AGNs from \citet{schawinski07} is also indicated by the solid straight line. Twelve of the 18 AGN labeled galaxies are LINERs and six are Seyfert AGNs. The colour coding (blue=SF, green=AGN) is kept throughout this paper, adding a red colour for passively evolving objects. 

\subsection{Velocity dispersion and stellar mass}
\label{veldisp}

In this section, we assess the stellar mass derivations 
and highlight ingredients in the SED-fitting that are responsible for systematic uncertainties in the derived masses. 
This exercise is important as we will later on use stellar velocity dispersion as a proxy for stellar mass. 
The evaluation of the stellar mass derivations is also important since stellar mass is extensively used in the literature when studying SNe~Ia host galaxies.

In Fig.~\ref{MvsS} we compare the velocity dispersion measurements from \texttt{GANDALF/PPXF} (see Section~\ref{gandalf}) to the various photometric stellar masses (Section~\ref{data:masses}), e.g. M05 models without reddening (M05, panel a), M05 models with reddening (M05$_{red}$, panel b), BC03 models without reddening (BC03, panel c) and BC03 model with reddening (BC03$_{red}$, panel e). A comparison between the velocity dispersion measurements and the masses from \citet[][L10, panel e]{lampeitl10} is also included. For this sample we find 81 out of 84 objects in common with our final sample. Data points are labeled according to the emission line classifications from Section~\ref{emission}. 
Objects from the L10 sample that are clearly offset from the general mass-sigma relation to higher masses are highlighted 
as large, open diamonds in each panel, i.e. these symbols represent the same objects in all panels.  
Least-square fits to the $\sigma$-mass relationships are also included in Fig.~\ref{MvsS} (solid lines), with the slopes given at the top of the panels. The standard deviations of the residuals to these fits are indicated by the labels. 

\citet{pforr} make a comprehensive study of the robustness of stellar population parameters that are derived from applying SED-fitting techniques to galaxy photometry. They use mock galaxies with known input stellar mass, age, etc. and study which SED-fit procedure is able to recover the input masses at best. 
They find that stellar masses are best reproduced when reddening is excluded in the SED-fitting procedure. In Fig.~\ref{MvsS} we do indeed find the tightest correlation with velocity dispersion for the M05 case without reddening, judging from the standard deviations of the fit residuals. It can further be seen that the M05 case without reddening make the most pronounced distinction between star-forming galaxies (blue points) having lower masses than galaxies without emission lines, which is expected. 

By comparing panel a and panel c in Fig.~\ref{MvsS}, we can evaluate the effect of the different models used in the SED-fitting. The BC03 models produce higher masses compared to the M05 models, in particular for star forming galaxies (blue data points). This result is well-known and due to the stronger TP-AGB phase in the M05 models that contribute significantly to the luminosity of young stellar populations (see Section~\ref{data:masses}). 
The lower amount of near-IR light from the TP-AGB phase at intermediate-ages in the BC03 models is compensated by finding a best fit at older ages, when the RGB takes over in producing near-IR light. The older age of the best fit implies then a higher M$_*$. In addition, the RGB takes over at older ages in the Padova tracks assumed in the BC03 models with respect to the M05 models, which contributes to the same effect (see Maraston et al. (2006) for details). 

The effect of including reddening in the fit can be appreciated by comparing panel a(c) to panel b(d) in Fig.~\ref{MvsS}. In this case the inclusion of reddening results in lower ages, hence, lower masses in particular for star-forming galaxies \citep[see][]{pforr}, a significantly larger scatter in the $\sigma$-mass relationship and consequently a flatter slope. 
The L10 masses are derived using the Pegase models and including reddening in the fit. These masses show a relationship with velocity dispersion similar to the BC03 with reddening case, also because of the similarity between the Pegase and 
BC03 theoretical SEDs \citep[see][]{M05}.

\begin{figure}
\includegraphics[clip=true,trim=2cm 0cm 2.8cm 3.7cm,angle=90,scale=0.39]{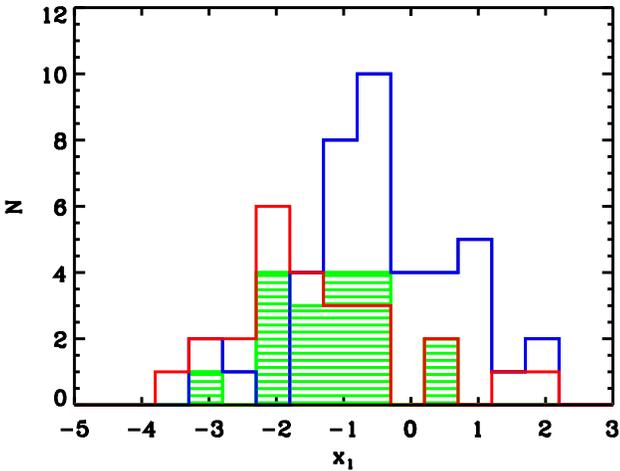}
\caption{Distribution of the stretch factor x$_1$ for the different emission classifications (see Section~\ref{emission}). The blue histogram represents star-forming, green AGN and red passively evolving galaxies. }
\label{histx1sng6}
\end{figure}

\begin{figure*}
\centering
\includegraphics[clip=true,trim=0.3cm 1cm 0.5cm 2.2cm,angle=90,scale=0.28]{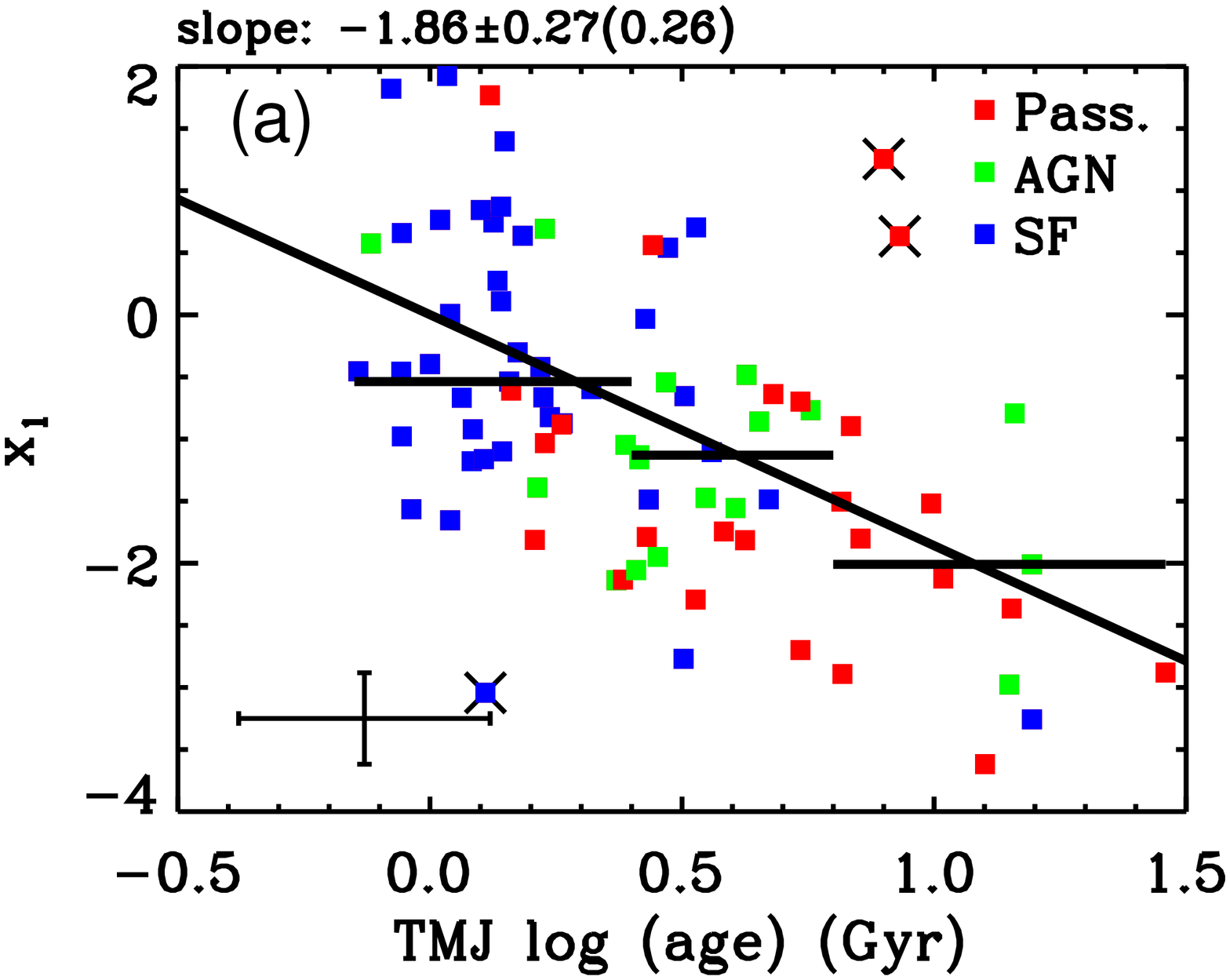}\includegraphics[clip=true,trim=0.3cm 1cm 0.5cm 2.2cm,angle=90,scale=0.28]{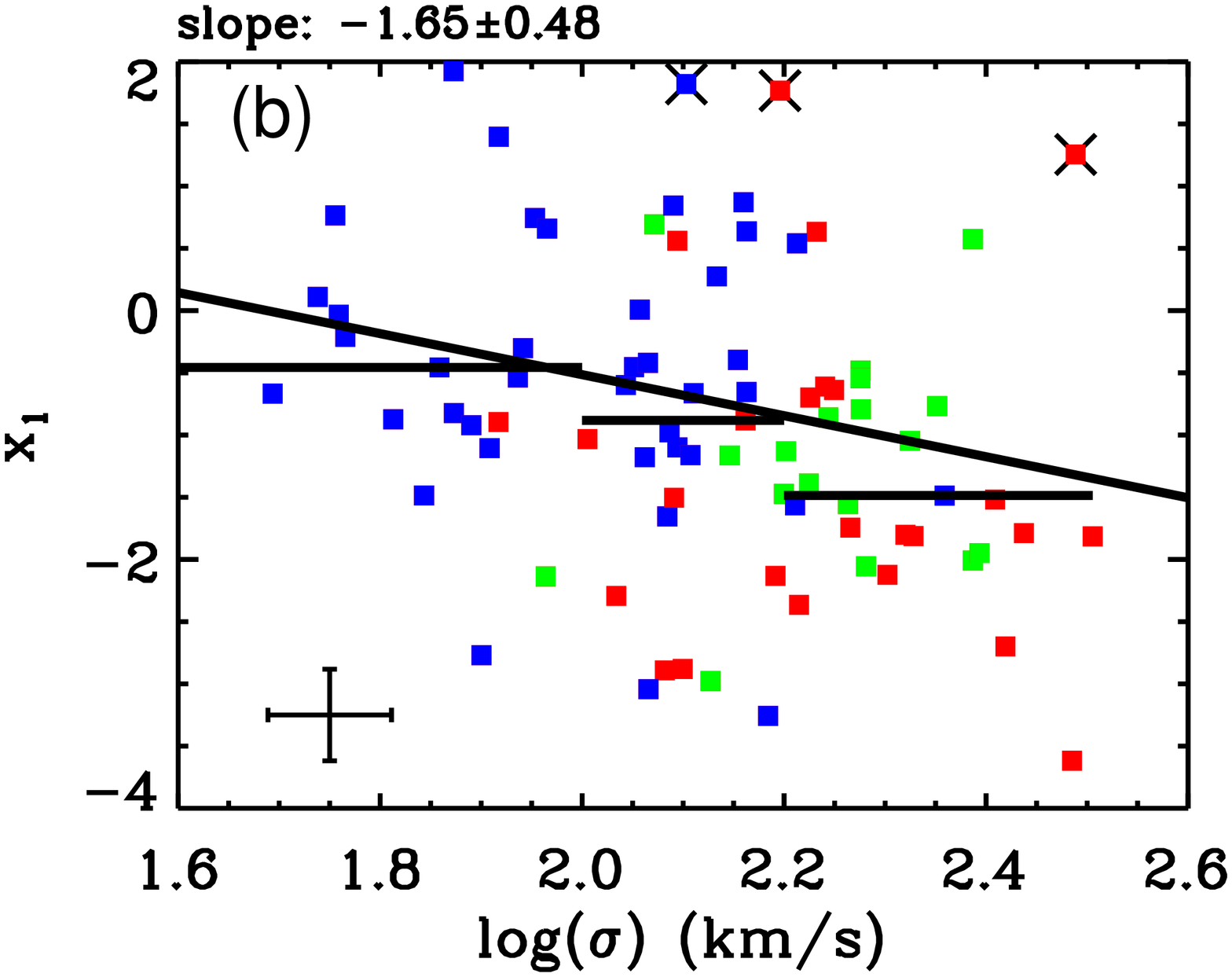}\includegraphics[clip=true,trim=0.3cm 1cm 0.5cm 2.2cm,angle=90,scale=0.28]{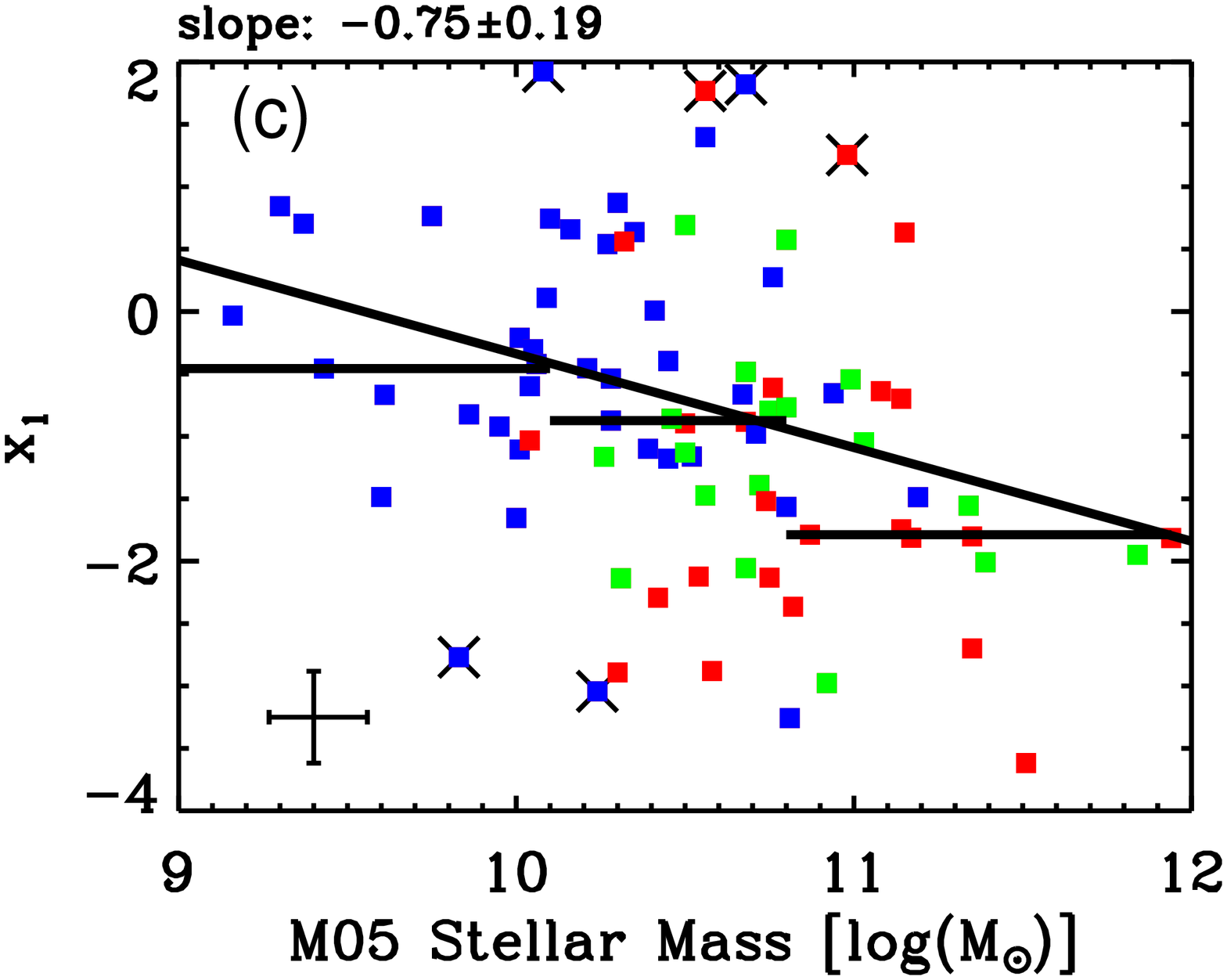}
\includegraphics[clip=true,trim=0.3cm 1cm 0.5cm 2.2cm,angle=90,scale=0.28]{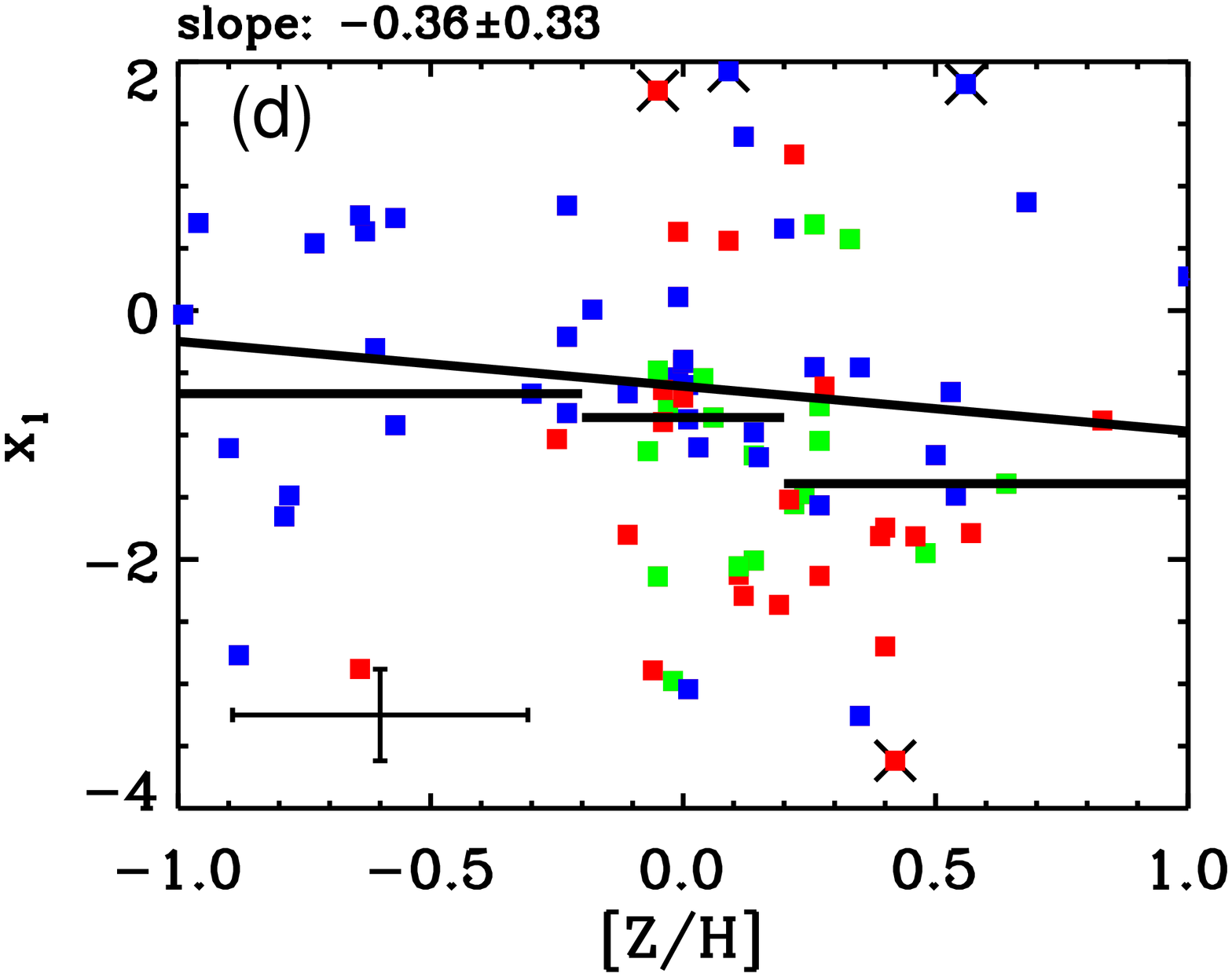}\includegraphics[clip=true,trim=0.3cm 1cm 0.5cm 2.2cm,angle=90,scale=0.28]{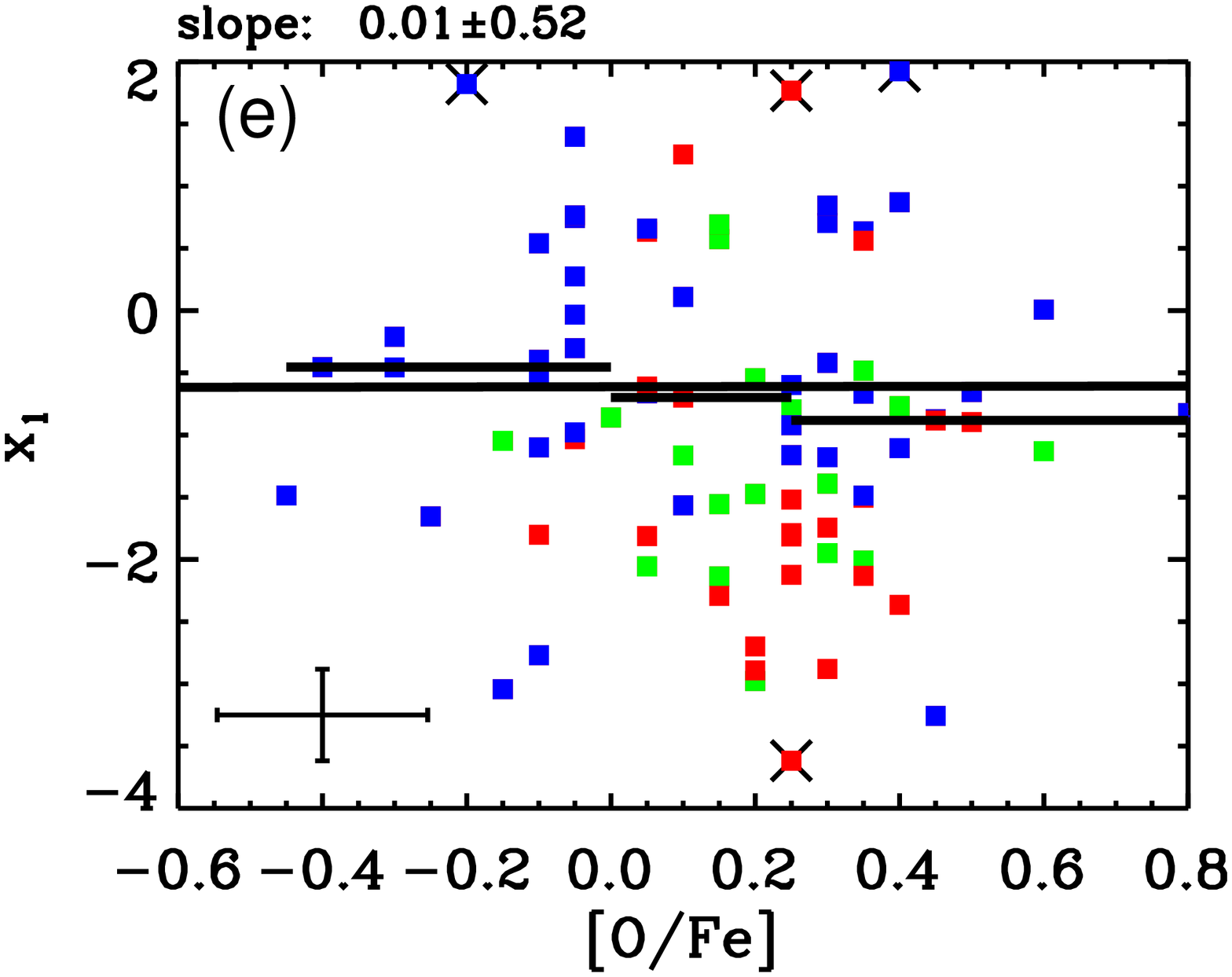}\includegraphics[clip=true,trim=0.3cm 1cm 0.5cm 2.2cm,angle=90,scale=0.28]{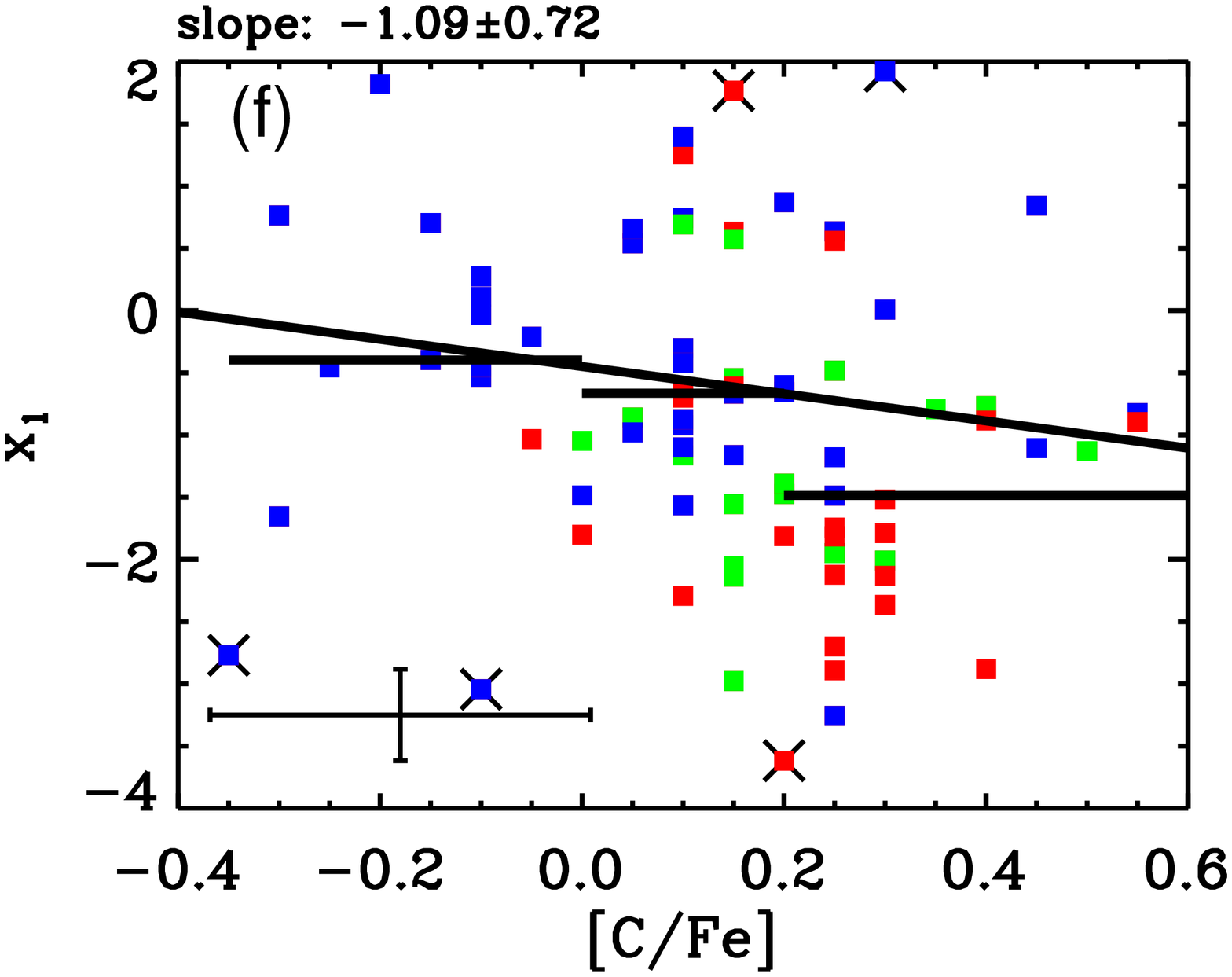}
\caption{Relationship with host galaxy age (TMJ, panel a), velocity dispersion (panel b), stellar mass (M05 without reddening case, panel c), Z/H (panel d), O/Fe (panel e) and C/Fe (panel f)  for the SALT2 stretch factor x$_1$. The data points are coloured according to the emission line classification of the host galaxies from Section~\ref{emission}, i.e. blue=SF, green=AGN and red=passively evolving. Solid lines are one time sigma-clipped (2$\sigma$ level) least-square fits and the over-plotted crosses are removed data points. The slopes of the fits are given at the top of the panels. The value in parentheses in panel a is an additional error estimate for the slope (see text for detail). Horizontal lines are median x$_1$-values in bins of the x-axis parameters with lengths indicating the width of the bins. 
Error bars in the lower left corners are average 1$\sigma$ errors.}
\label{agegrid_x1}
\end{figure*}

The objects with particularly high L10 masses compared to the velocity dispersion measurements, large open diamonds in each panel, shift significantly from panel to panel.  The masses of these objects are significantly lower for the M05 without reddening case compared to L10.

In Fig~\ref{agecomp} the ages derived using absorption line indices (TMJ, see Section~\ref{tmj}) are compared to those derived from each of the four SED-fitting cases, e.g. M05 models without reddening (M05, upper left panel), M05 models with reddening (M05$_{red}$, upper right panel), BC03 models without reddening (BC03, lower left panel) and BC03 models with reddening (BC03$_{red}$, lower right panel). The one-to-one relation is shown in each of the four panels (solid lines). The best agreement with the TMJ ages are found for the M05 case without reddening. This is also the case that shows the strongest correlation with velocity dispersion (see Fig.~\ref{MvsS}). The effect of model choice (compare the upper left to the upper right panel) is such that the BC03 models produce older ages for star-forming galaxies when compared to the TMJ ages. This result is due to the 
TP-AGB phase as already explained. 
The inclusion of reddening in the SED-fitting lead to an underestimation of the ages (compare the left to the right panels) with respect to those derived from absorption line indices.

Note that the agreement between TMJ ages and M05-SED-fit ages is remarkable and not simply due to the fact that the underlying population model is the same. Indeed, the results of SED-fit with reddening are in disagreement in spite of the use of the same model. This good result between photometrically-derived and optical-absorption derived ages is not found by \citet*{hansson12} in comparing ages from Lick indices from Gallazzi et al. (2005) based on BC03 models and SED-fit based on the same models. 

\begin{figure*}
\centering
\includegraphics[clip=true,trim=0cm 0cm 2cm 8cm,scale=0.39]{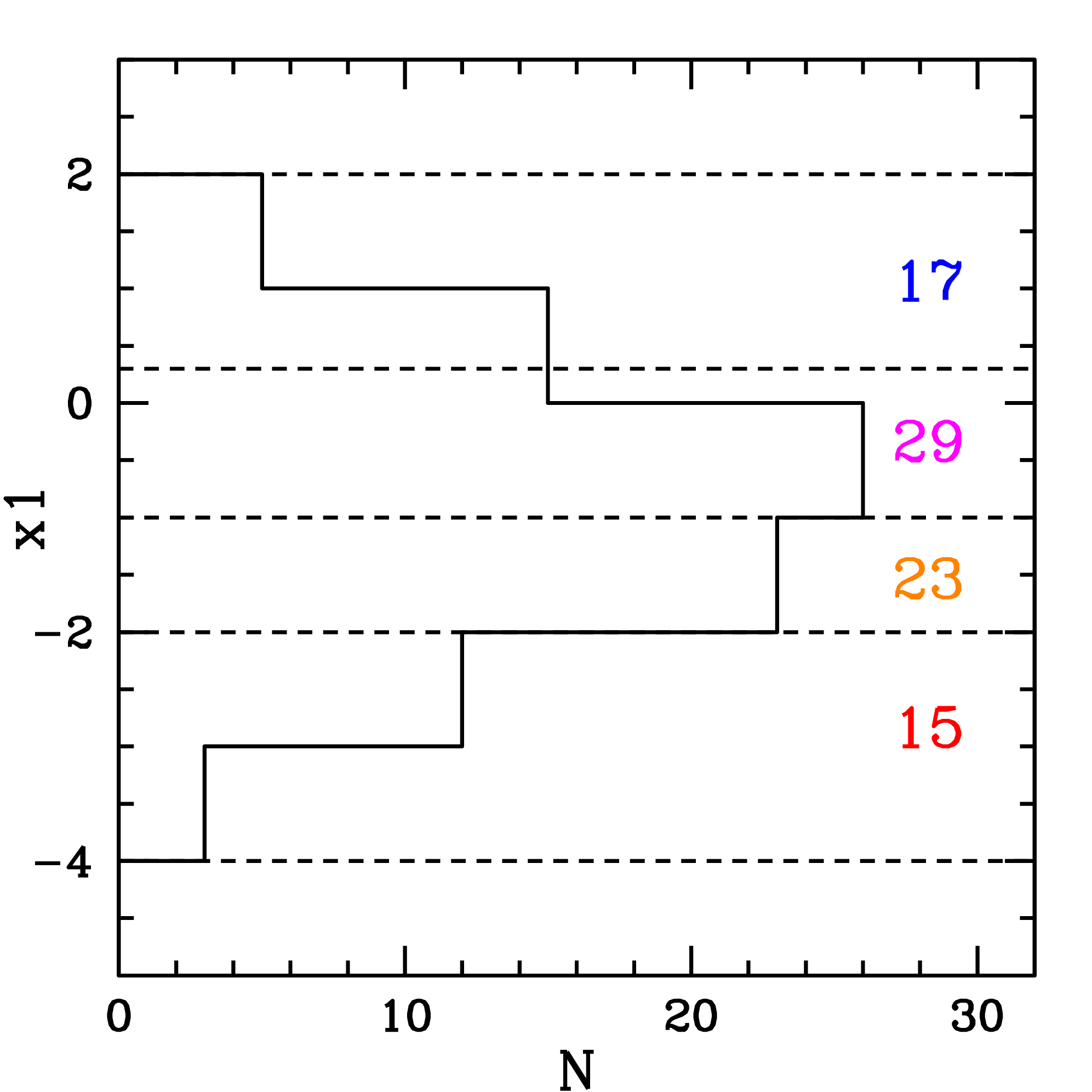}\includegraphics[clip=true,trim=4cm 2cm -10cm 8cm,scale=0.44]{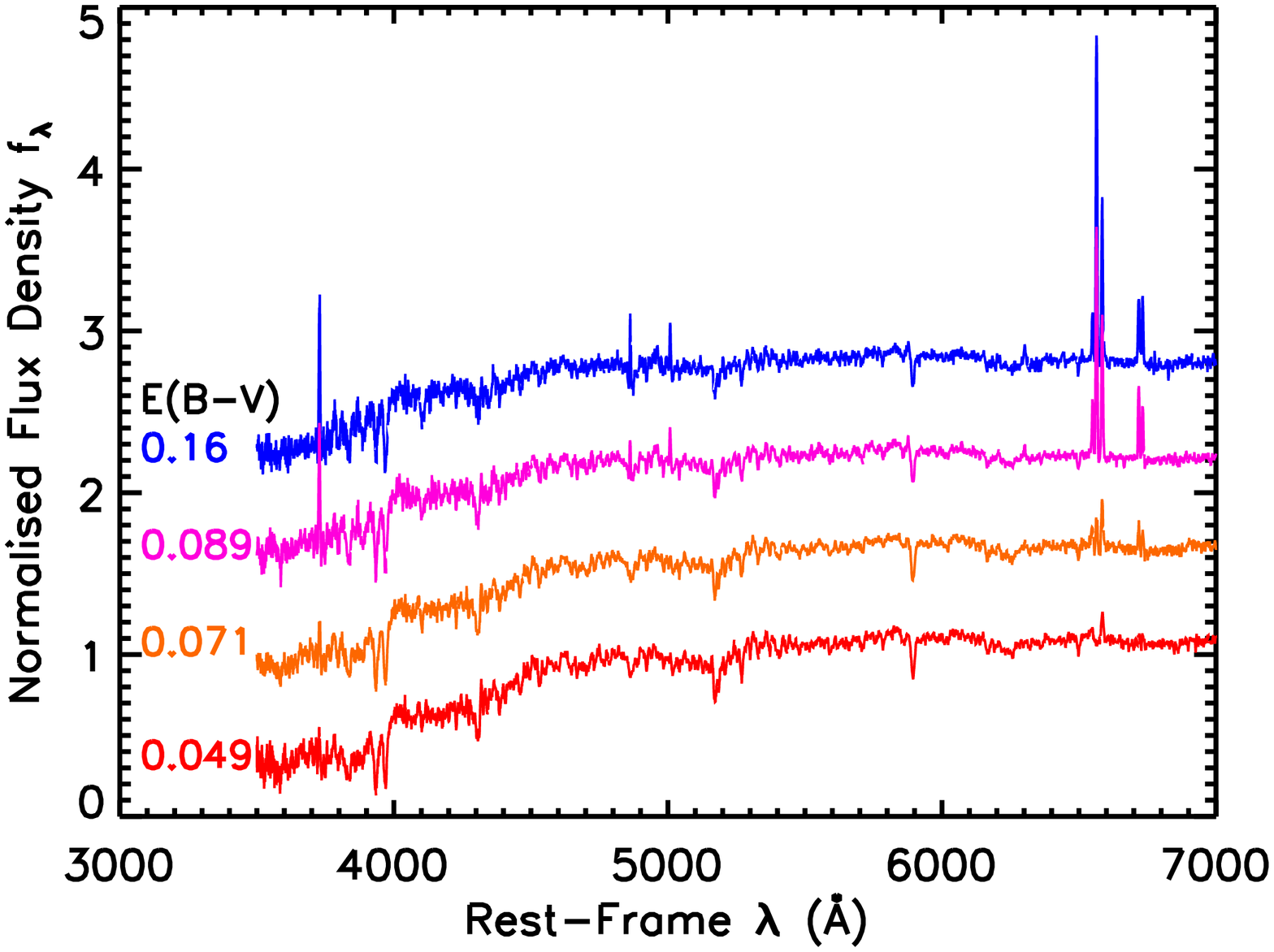}
\caption{Left hand panel: binning in x$_1$ for stacking of spectra, where the limits of the bins are indicated by the horizontal dashed lines. Coloured labels state the number of objects within each individual bin. The histogram is distribution of x$_1$. Right hand panel: resulting stacked spectra at rest-frame wavelength and for normalised flux density per unit $\lambda$~(f$_{\lambda}$), coloured according to the bin labels of the left hand panel. The labels, with corresponding colours, state the E(B-V) value for each stacked spectrum.}
\label{stackx1}
\end{figure*}

\subsection{Stretch factor and stellar populations}
\label{stretch}

\subsubsection{Final host spectroscopy sample}
\label{final}

Fig.~\ref{histx1sng6} shows the distribution in stretch factor x$_1$ for star-forming (blue histogram), AGN (green histogram) and passively evolving galaxies (red histogram), as classified according to their emission line strengths (see Section~\ref{emission}). It is clear that SNe~Ia observed in passive galaxies have the lowest x$_1$ values hence the shortest decline rates, i.e. star-forming galaxies contain the most luminous SNe~Ia.  
SNe~Ia observed in galaxies with AGN activity appear to have x$_1$ values falling in between those with star-forming and passively evolving host stellar populations.

Relationships between the stellar populations parameters (TMJ age, Z/H, O/Fe and C/Fe), velocity dispersion and stellar mass with SALT2 stretch factor x$_1$ are presented in Fig.~\ref{agegrid_x1}. The data points are colour coded according to the format in Fig.~\ref{histx1sng6}. 
The stellar masses are from the M05 without reddening case, since they showed the best agreement with stellar velocity dispersion (see Section~\ref{veldisp}). 
Solid lines are sigma-clipped least-square fits (see beginning of Section~\ref{results}). 
The slope and corresponding errors of the fits are given at the top of the panels and sigma-clipped data points are indicated by over-plotted crosses. Horizontal lines are median values in bins of the x-axis parameters, where the length of the lines indicate the width of the bins. Mean 1$\sigma$ errors are shown by the error bars in the lower left corners. 

Clear anti-correlations between stretch factor and stellar population age (panel a, Fig.~\ref{agegrid_x1}), galaxy velocity dispersion (panel b) and stellar mass (panel c) are found. Hence higher stretch factors, i.e. more luminous SNe~Ia, are found in younger stellar populations and lower stellar masses. 
We find the slopes of the least-square fits to be different from zero at a $>$6$\sigma$, 4$\sigma$ and $>$3$\sigma$ level for the x$_1$-log(age), x$_1$-mass and x$_1$-$\log(\sigma)$ relationships, respectively.
The fits are weighted with the x$_1$-errors. However, in panel a the errors in age seem to dominate the scatter. To evaluate the estimated error on the fitted slope in the x$_1$-log(age) relation we perturb the sample using errors in both x$_1$ and age through 1000 Monte-Carlo simulations. For each realization we perform a least-square fit and determine the standard deviation of the derived set of slopes. This value is given in parenthesis in panel a and is very close to the original error estimate.

The stretch-factor also anti-correlates with Z/H (panel d, Fig.~\ref{agegrid_x1}) and C/Fe (panel f), but these relations are significantly weaker than those for age, mass and velocity dispersion. The slopes of the least-square fits are shallower and differ from zero at less than a 1$\sigma$ level for Z/H and at a 1.5$\sigma$ level for C/Fe. The x$_1$-O/Fe relation (panel e, Fig.~\ref{agegrid_x1}) is flat with a large error on the fitted slope. 

To summarize, out of the parameters studied the stretch factor seems to primarily depend on the age of the host galaxy stellar populations. This age dependence is well in line with the correlation with star formation fraction discussed above.

Using the method described in Section~\ref{tmj} we also derive the stellar population parameters N/Fe, Mg/Fe, Ca/Fe and Ti/Fe. We find no strong correlations for x$_1$ with these parameters that would add valuable information to the analysis. Measurements of N/Fe, Ca/Fe and Ti/Fe require higher S/N than the other parameters, while the derived Mg/Fe and O/Fe ratios follow each other closely \citep{johansson11}. Hence we do not further discuss these abundance ratios.

\begin{figure*}
\centering
\begin{flushleft}
\includegraphics[clip=true,trim=0.3cm 0.8cm 0.5cm 1.2cm,angle=90,scale=0.27]{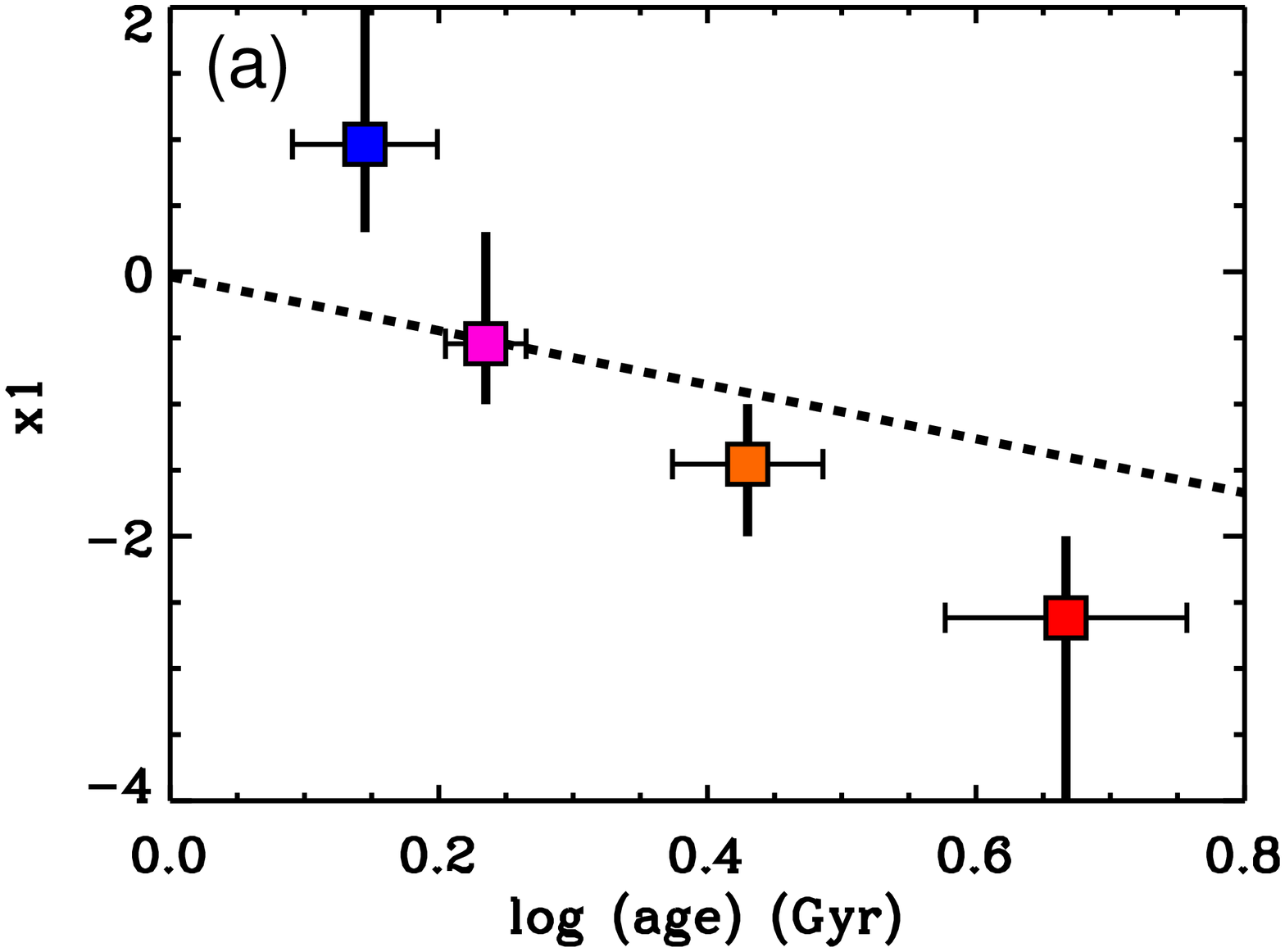}\includegraphics[clip=true,trim=0.3cm 0.8cm 0.5cm 2.2cm,angle=90,scale=0.27]{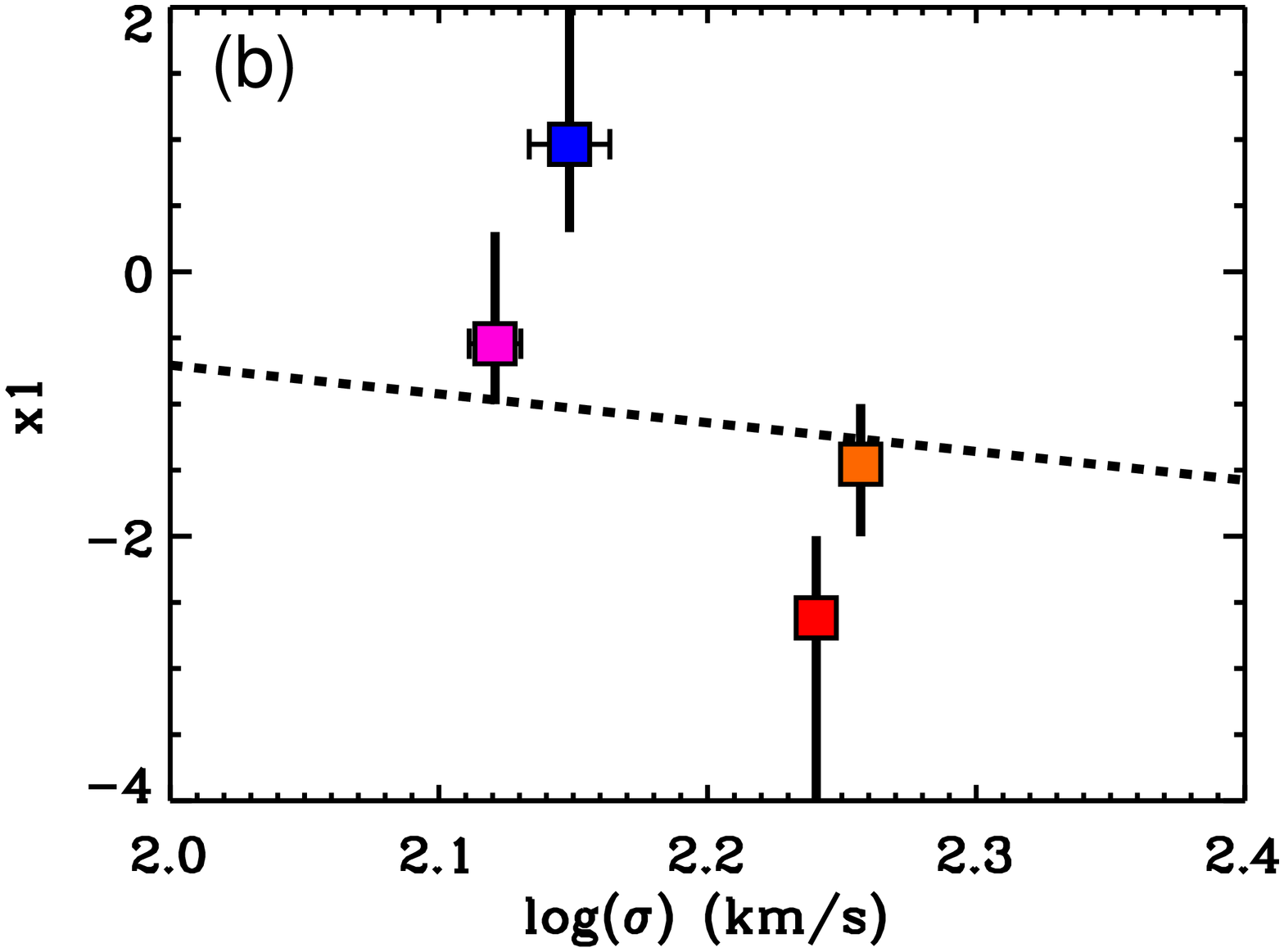}
\end{flushleft}
\includegraphics[clip=true,trim=0.3cm 0.8cm 0.5cm 2.2cm,angle=90,scale=0.27]{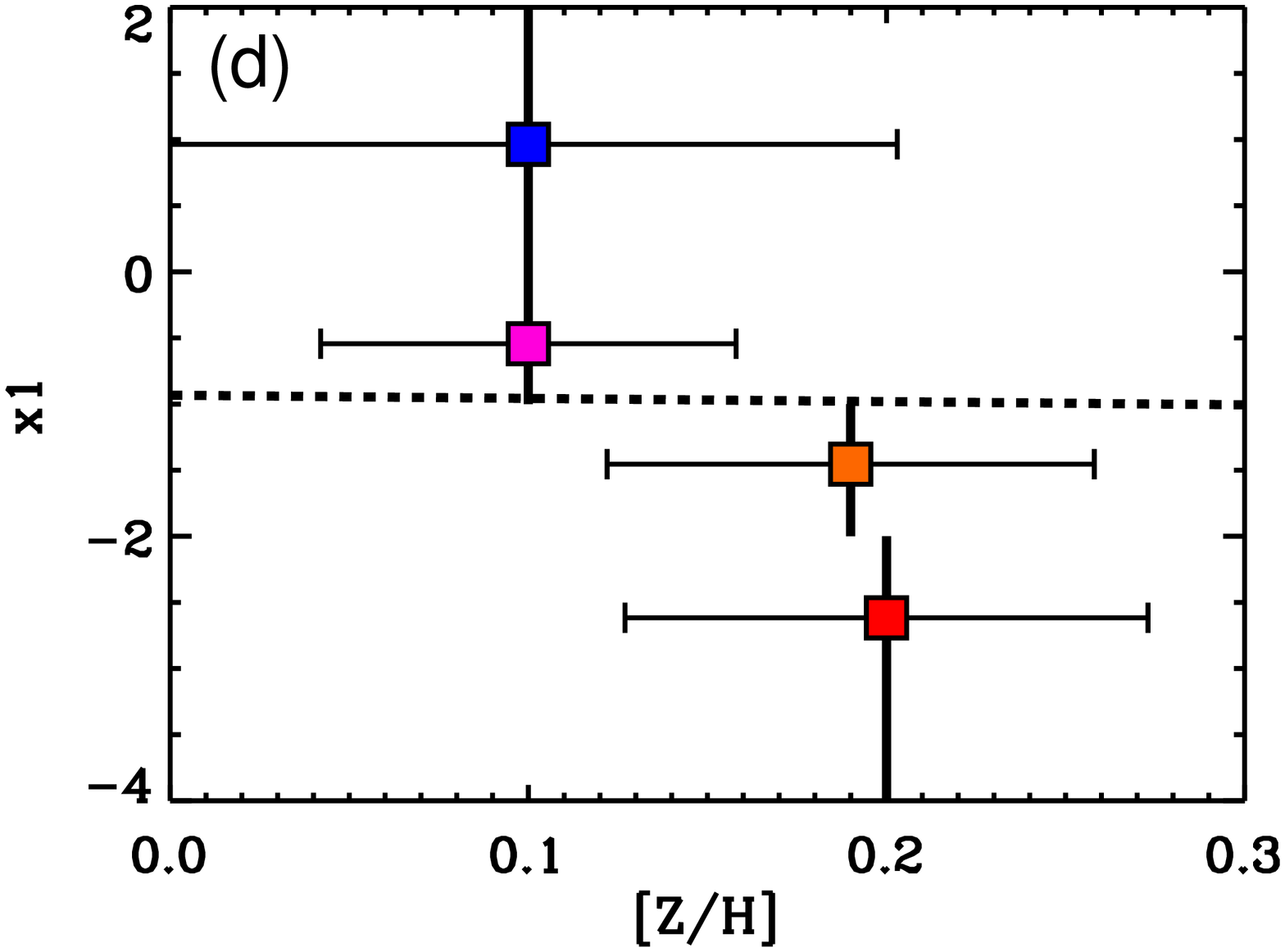}\includegraphics[clip=true,trim=0.3cm 0.8cm 0.5cm 2.2cm,angle=90,scale=0.27]{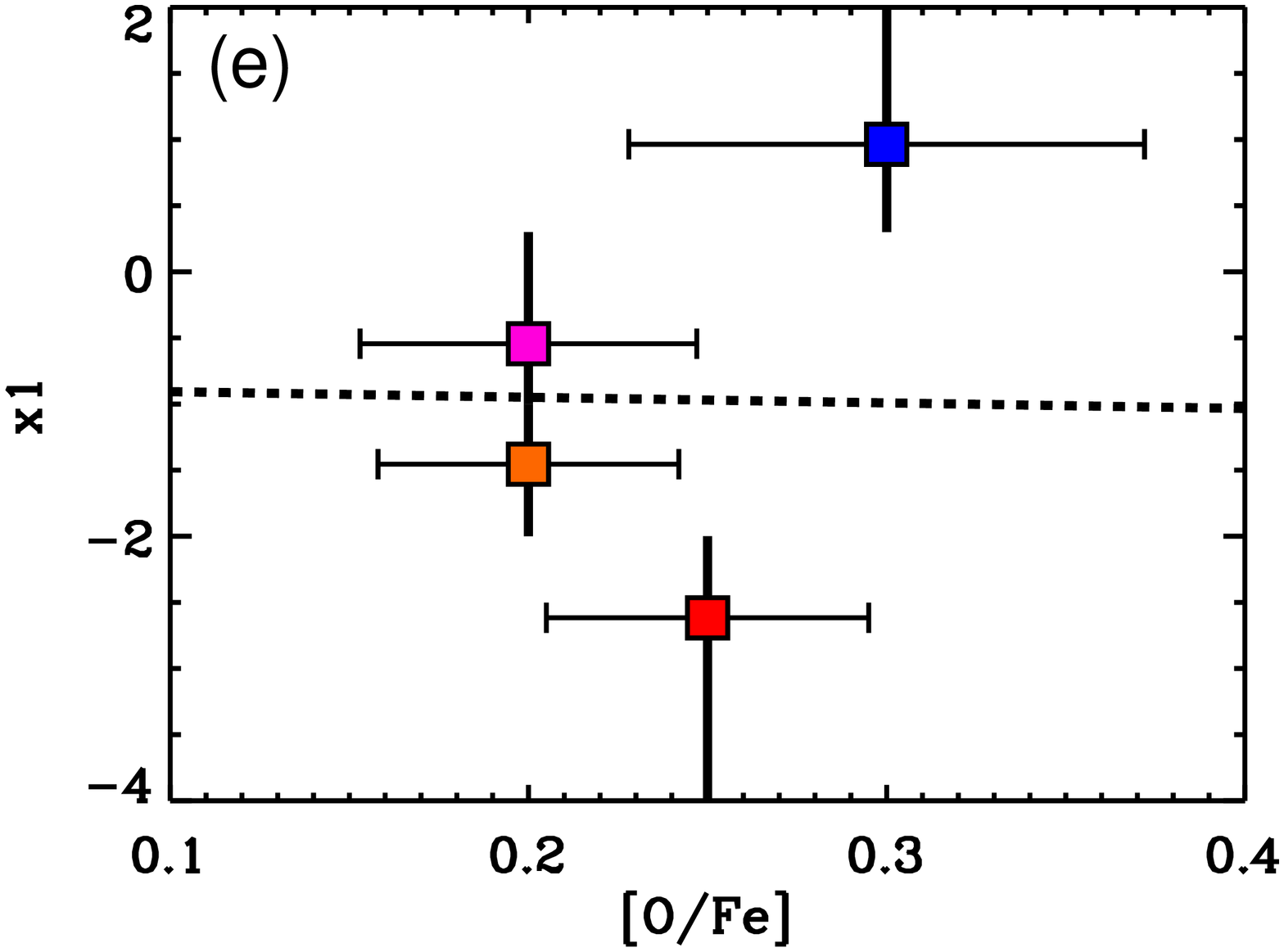}\includegraphics[clip=true,trim=0.3cm 0.8cm 0.5cm 2.2cm,angle=90,scale=0.27]{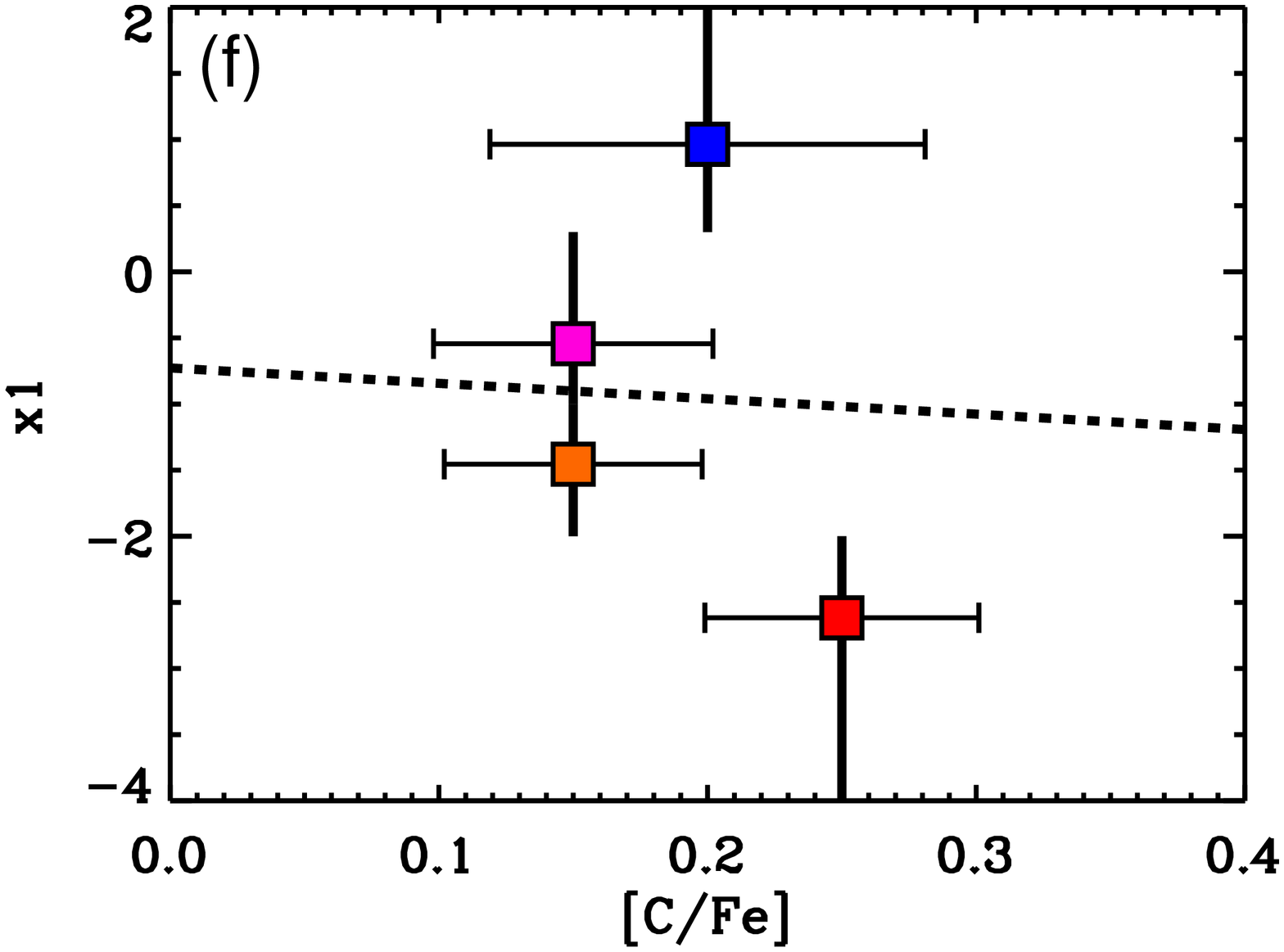}
\caption{Relationship with host galaxy age (panel a), velocity dispersion (panel b), Z/H (panel d), O/Fe (panel e) and C/Fe (panel f) for SALT2 x$_1$ for the stacked spectra. The data points are coloured according to the colours of the binned spectra in Fig.~\ref{stackx1}. The 1$\sigma$ errors for the x-axis parameters are indicated by the horizontal error bars and bin widths are shown by the vertical bars. The x$_1$ value for each data point is the average x$_1$ value in the corresponding bin. Dashed lines are the least-square fits for the corresponding relationships of the individual spectra from Fig.~\ref{agegrid_x1}.
}
\label{spp_stackx1}
\end{figure*}

\subsubsection{Stacked spectra}
\label{x1_stack}

Due to the low S/N of the individual spectra used in the previous section, in the following analysis we perform consistency checks by using stacked spectra with relatively high S/N. We stack the spectra of the final sample in bins of $x_1$. Hence, in practice we stack spectra of similar galaxy class because of the correlation between $x_1$ and emission line classification (see previous section). 
The spectra are first de-redshifted using the SDSS spectroscopic redshifts and linearly interpolated to the same wavelength binning.  We then normalise to the median flux density (f$_{\lambda}$) contained in the rest-frame wavelength range 5000-5500 \AA. The stacking is finally performed by taking the median flux density value in each wavelength pixel \citep{lee10}. Using the median value is a safeguard against contaminated data, telluric contamination, features of individual spectra, and does not bias against the very highest S/N spectra. Following \citet{lee10} we estimate the error in each pixel of the median stacked spectra with 
\begin{equation}
(S/N)_s=\sqrt{\sum(S/N)_i^2}
\end{equation}
where (S/N)$_s$ is the S/N of the stacked spectrum and (S/N)$_i$ is the S/N of the individual spectra.

The x$_1$ range covered by the final sample (see Section~\ref{cuts}) is divided into four bins, chosen to include over a dozen number of data points in each bin. The left hand panel of Fig.~\ref{stackx1} shows the widths of the bins separated by the dashed horizontal lines, together with the corresponding number of objects in each bin. The histogram is the x$_1$ distribution for the final sample. The procedures applied to the individual spectra to compute the stellar population parameters (see Section~\ref{gandalf} and~\ref{tmj}) are then applied to the stacked spectra.

The right hand panel of Fig.~\ref{stackx1} shows the stacked spectrum for each bin, following the colour coding of the left hand panel. The E(B-V) values produced by \texttt{GANDALF/PPXF} (see Section~\ref{gandalf}) for each stacked spectrum are given by the labels with corresponding colours. The stacked spectra show a number of distinct features and emission lines that clearly change as a function of stretch factor. The largest stretch factors are found in host galaxies with the most pronounced star formation activity and the highest dust extinction. The host galaxies of faint supernovae with small stretch factors, instead, are emission line-free and dust-free, and show strong absorption features in their spectra. 

An analogue to Fig.~\ref{agegrid_x1}, Fig.~\ref{spp_stackx1} shows the relationships between stretch factor and stellar population properties and galaxy velocity dispersion for the stacked spectra. 
The data points in each panel are coloured according to Fig.~\ref{stackx1} and the vertical bars indicate the size of the bins. The x$_1$ values are the mean within each bin. 1$\sigma$ error bars are shown for each data point in the horizontal direction. The 1$\sigma$ errors are similar for $\log$(age), [Z/H], [O/Fe] and [C/Fe], i.e. smaller than 0.1 dex. The least-square fits to the final sample from Section~\ref{final} are included for comparison (dashed lines).  The horizontal range covered in each panel has been truncated compared to Fig.~\ref{agegrid_x1} to better resemble the parameter range covered by the data of the stacked spectra.

Lower x$_1$ values show older ages (panel a, Fig.~\ref{spp_stackx1}), higher velocity dispersions (panel b) and higher total metallicities (panel d), in agreement with the case of the individual spectra (see Section~\ref{final}). However, only for age (panel a) we see this trend clearly for all data points adjacent in x$_1$-space and significantly above the 1$\sigma$ error level. For total metallicity the trend with x$_1$ is diminished by the error bar overlap, which is due to the short range covered by this parameter ($\sim$0.1 dex). A short parameter range is also found for [O/Fe] ($\sim$0.1 dex, panel e, Fig.~\ref{spp_stackx1}) and [C/Fe] ($\sim$0.1 dex, panel f), resulting in a significant error bar overlap. We see no clear trends for these parameters. Hence the result of the individual spectra from Section~\ref{final} is reproduced for the stacked spectra, i.e. x$_1$ show the strongest dependence on stellar population age. 

\begin{figure*}
\centering
\begin{flushleft}
\includegraphics[clip=true,trim=0.3cm 0.8cm 0.5cm 1.3cm,angle=90,scale=0.27]{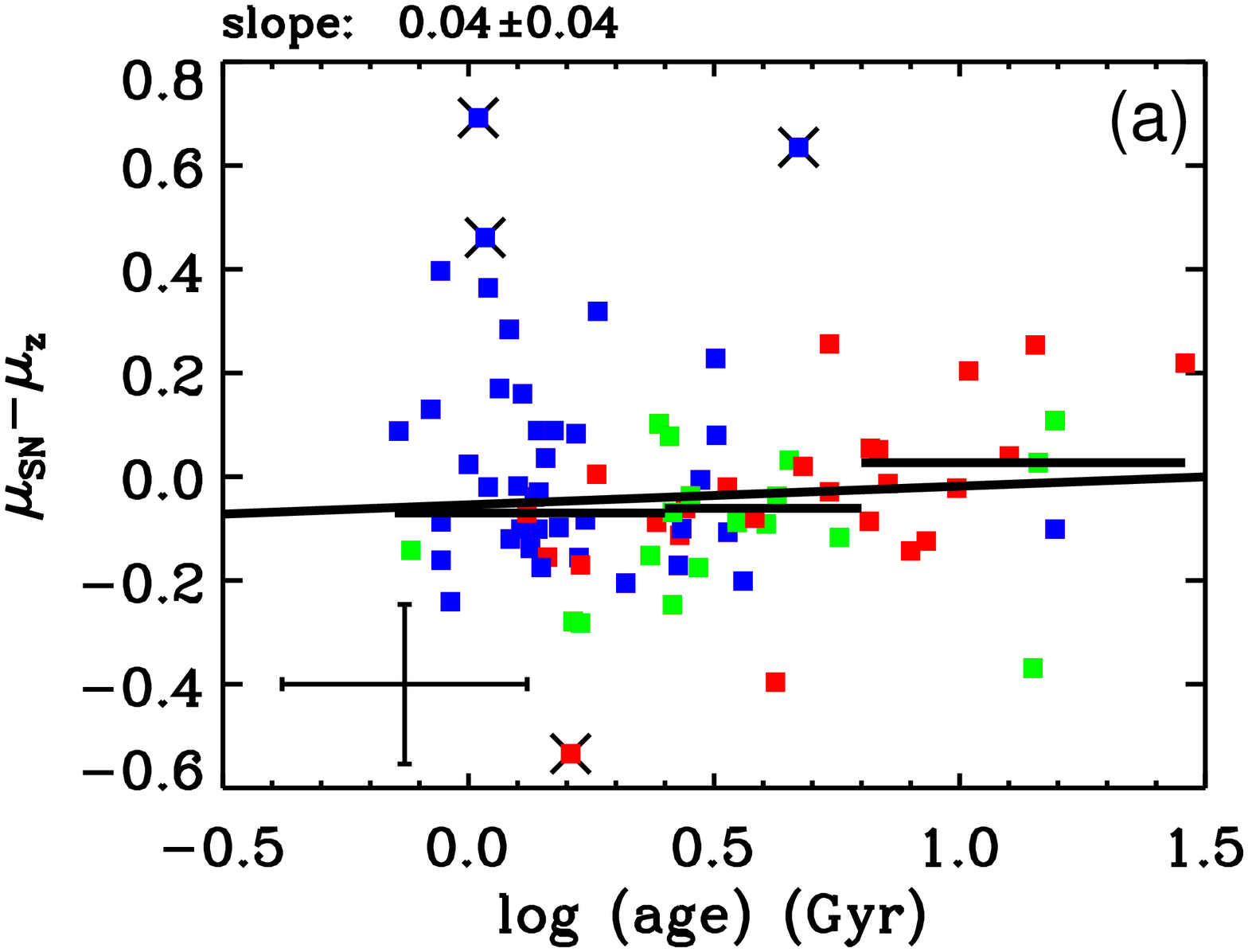}\includegraphics[clip=true,trim=0.3cm 0.8cm 0.5cm 1.8cm,angle=90,scale=0.27]{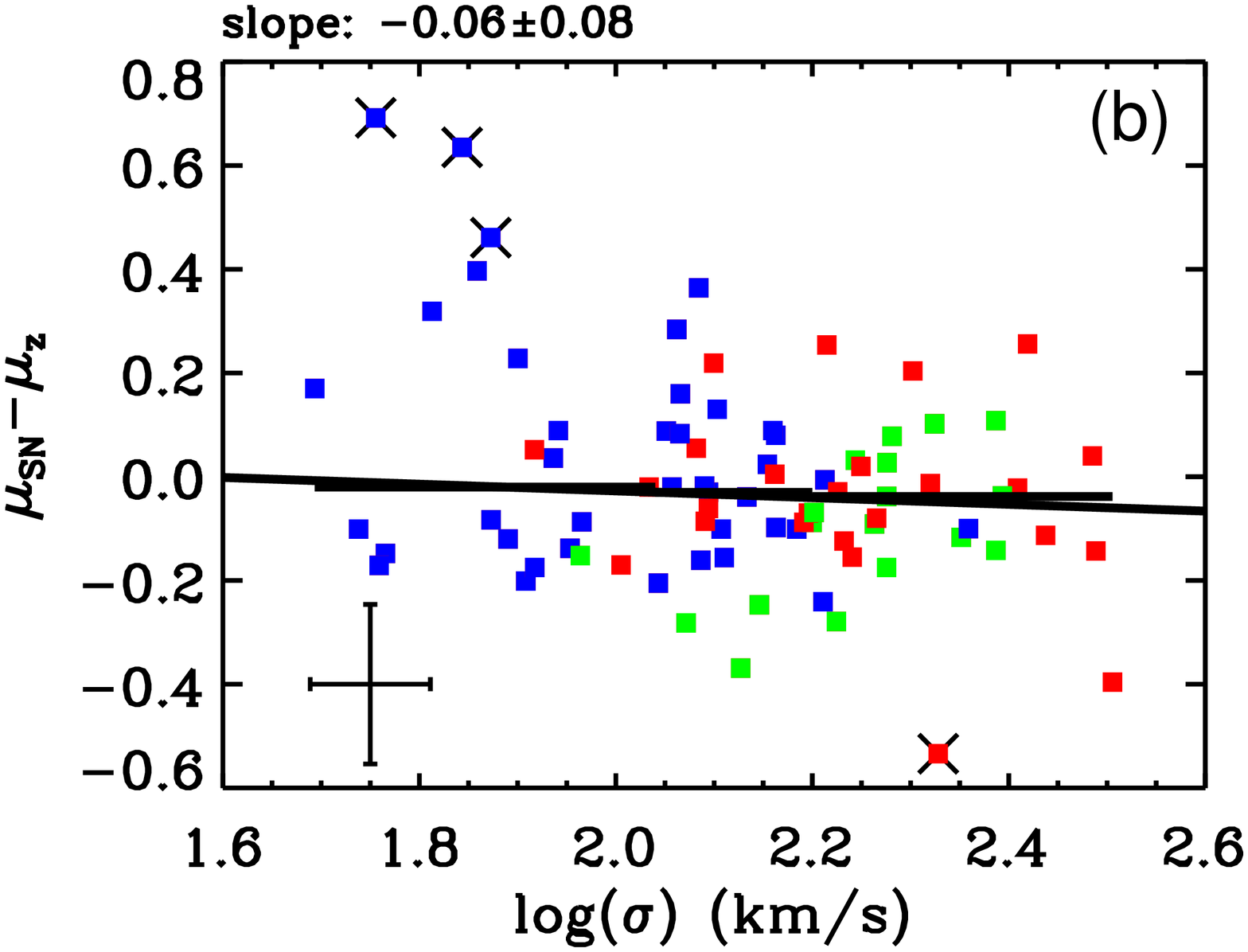}\includegraphics[clip=true,trim=7cm 0.8cm 0.5cm 2cm,angle=90,scale=0.43]{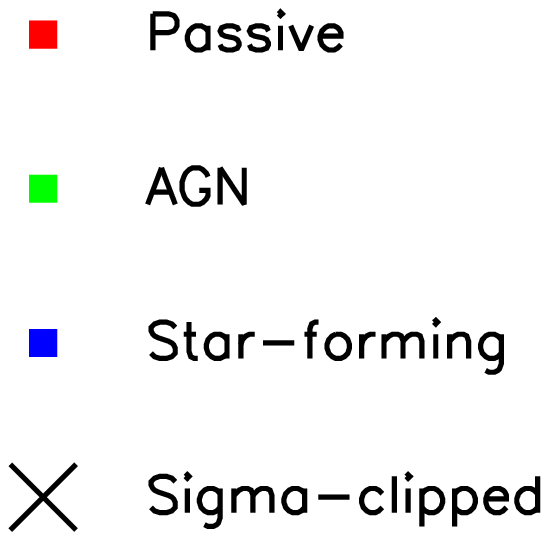}
\end{flushleft}
\includegraphics[clip=true,trim=0.3cm 0.8cm 0.5cm 1.8cm,angle=90,scale=0.27]{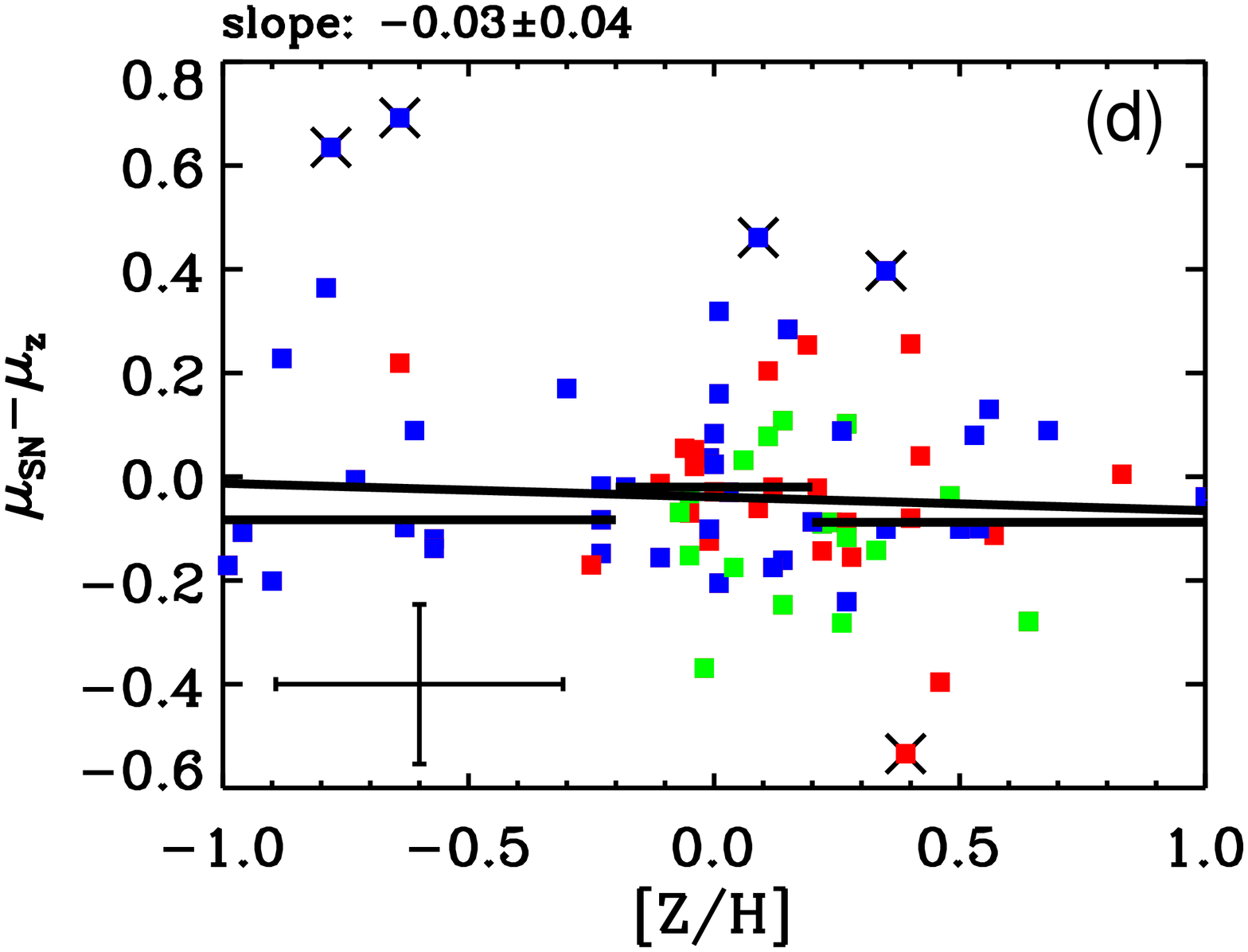}\includegraphics[clip=true,trim=0.3cm 0.8cm 0.5cm 1.8cm,angle=90,scale=0.27]{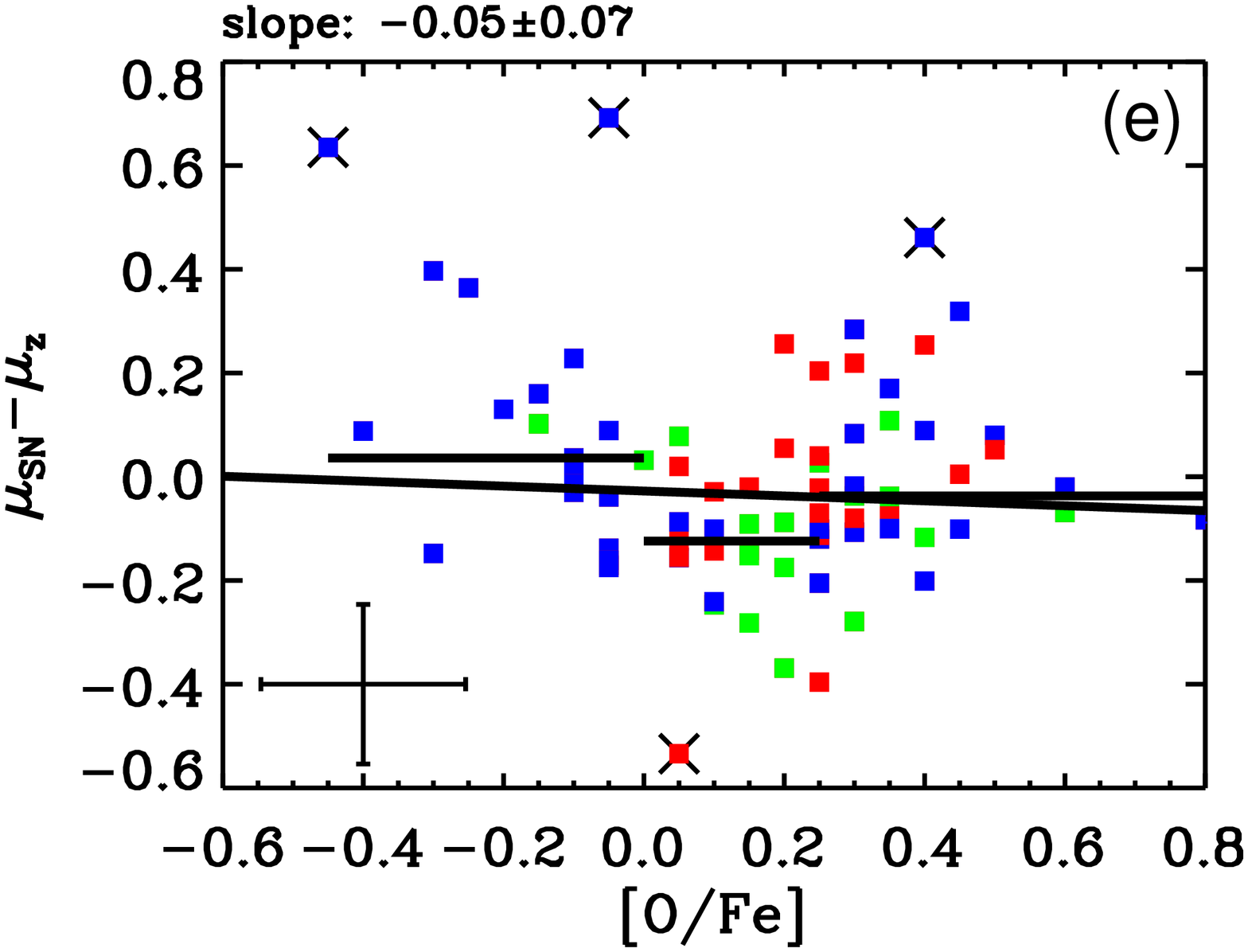}\includegraphics[clip=true,trim=0.3cm 0.8cm 0.5cm 1.8cm,angle=90,scale=0.27]{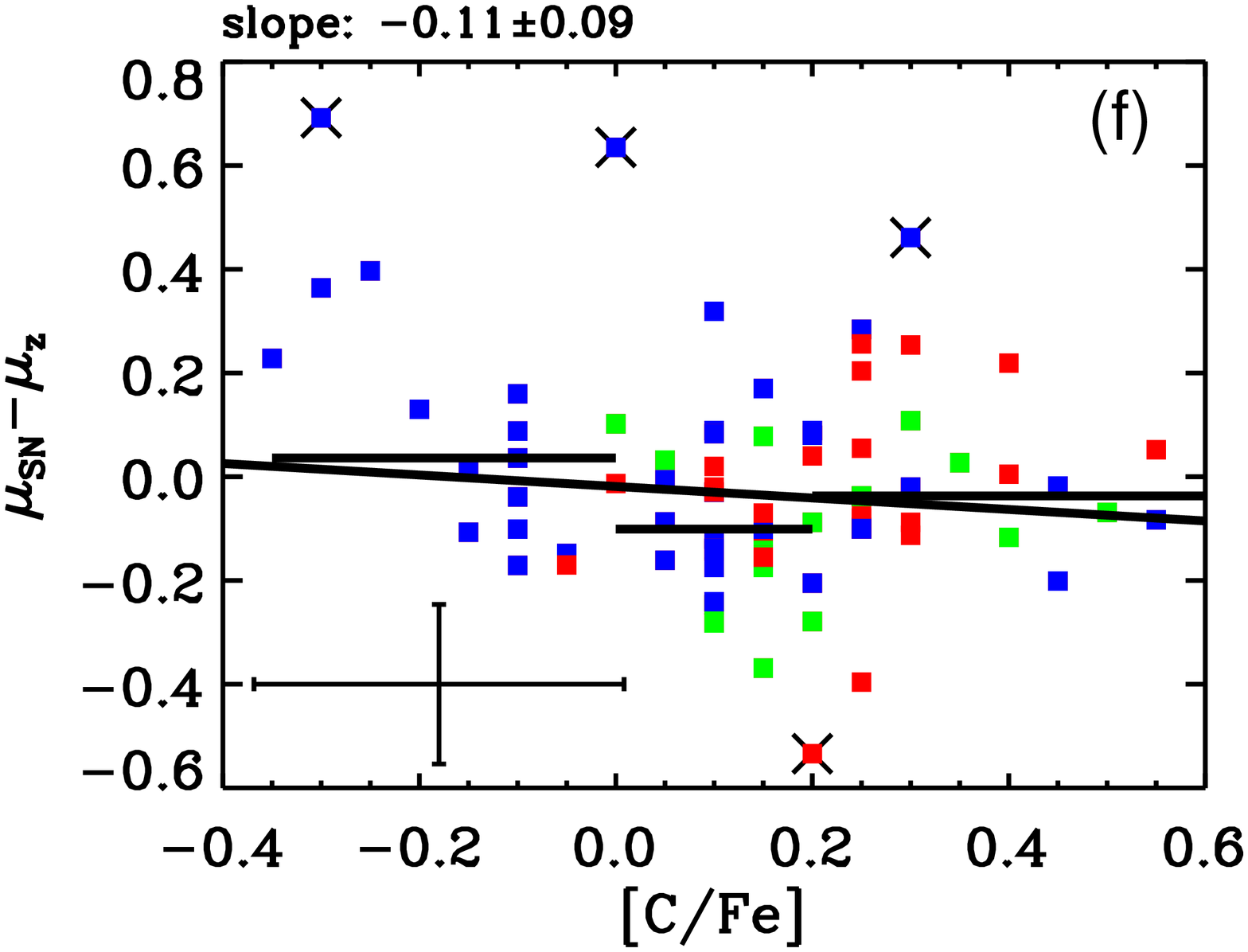}
\caption{Relationship with stellar population age (panel a), velocity dispersion (panel b), [Z/H] (panel d), [O/Fe] (panel e) and [C/Fe] (panel f) for Hubble residual. The data points are coloured according to the emission line classification of the host galaxies from Section~\ref{emission}, i.e. blue=SF, green=AGN and red=passively evolving. Solid lines are one time sigma-clipped (2$\sigma$ level) least-square fits and over-plotted crosses are removed data points. The slopes and corresponding errors are given at the top of the panels. Horizontal lines are median Hubble residuals in bins of the x-axis parameters with lengths according to the width of the bins.  
Error bars in the lower left corners are average 1$\sigma$ errors.}
\label{siggrid_res}
\end{figure*}

\subsubsection{Comparison with the literature}
\label{litcomp:x1}

In agreement with \citet{oemler79}, \citet{vdbergh90}, \citet{mannucci05}, \citet{sullivan06} and \citet{smith12}, we find a higher fraction of SNe~Ia events in star-forming compared to passively evolving galaxies. Star-forming galaxies also show slower decline rates, i.e more luminous SNe~Ia, compared to passively evolving galaxies, a pattern first noticed by \citet{sullivan06} and further confirmed by \citet{howell09}, \citet{neill09}, \citet{lampeitl10} and \citet{smith12}. 
For the first time we establish an anti-correlation between host galaxy velocity dispersion and stretch factor, suggesting faster decline rates in more massive galaxies. A similar anti-correlation is also found for stellar mass, in agreement with \citet{kelly10}, \citet{lampeitl10} and \citet{sullivan10}. 

For stellar metallicity and element abundance ratios we instead only find weak SALT2 x$_1$ dependencies.

Furthermore, we find a clear anti-correlation between stretch factor and luminosity-weighted stellar population age which is more prominent and well defined than found for velocity dispersion and stellar mass. This result is true for both individual objects and stacked host galaxy spectra and indicates that the x$_1$-mass relationship is a result of the correlation between galaxy mass and stellar population age. 

\begin{table*}
\center
\caption{Stellar populations parameters for stacked spectra in bins of Hubble residual. The stacking limits are given in column 1, the number of objects in each stack in column 2 and the mean Hubble residual in column 3. The stellar population parameters together with corresponding errors are given in columns 4-7 and velocity dispersion in column 8.}
\label{HRstack}
\begin{tabular}{cccccccc}
\hline
\bf Stack & N & $<$HR$>$ &  \bf $\mathbf{\log(age)}$ & \bf [Z/H] & \bf [O/Fe] & \bf  [C/Fe] & \bf $\sigma$  \\
& & &  \bf (Gyr) & & & & \bf (km s$^{-1}$) \\
\hline
HR$>$-0.05 & 43 & 0.13 & 0.44$\pm$0.04 & 0.08$\pm$0.06 & 0.30$\pm$0.05 & 0.25$\pm$0.05 & 148.2$\pm$2.5 \\
HR$<$-0.05 & 41 & -0.16 & 0.33$\pm$0.04 & 0.10$\pm$0.06 & 0.25$\pm$0.05 & 0.20$\pm$0.05 & 152.9$\pm$2.3\\
\hline
\end{tabular}
\end{table*}

\citet{hamuy00} and \citet{gallagher08} investigated absorption line indices allowing for distinction between age and metallicity effects. The former prefer metallicity over age as the main driver of SNe~Ia luminosity using only five objects. \citet{gallagher08} studied 29 SNe~Ia host galaxies and found age to be the dominant stellar population parameter affecting SNe~Ia luminosity. 
Two recent studies, \citet{howell09} and \citet{neill09}, derived host galaxy parameters from photometry. Both authors favour age over metallicity as the SNe~Ia luminosity-dependent factor, but determine the latter only indirectly by using derived masses and the mass-metallicity relationship from \citet{tremonti04}. 
In a recent study \citet{gupta11} also found faster decline rates for older stellar populations using photometry, but do not include metallicity or element abundance ratios in their study. 

\begin{figure}
\centering
\includegraphics[clip=true,trim=3cm 2.5cm -10cm 9.5cm,angle=0,scale=0.37]{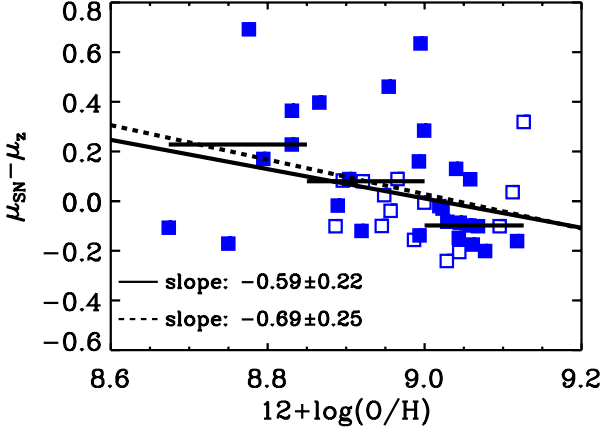}
\caption{Relationship between gas-phase metallicity and Hubble residual for purely star-forming galaxies (blue solid points) and transition galaxies (open blue points). The solid line and the dashed line are least-square fits to all data points and to solid data points, respectively. The slopes and corresponding errors are given by the labels. Horizontal lines are median Hubble residuals in bins of the x-axis parameters with lengths according to the width of the bins for all data points. }
\label{gaszgrid_res}
\end{figure}

To conclude, this analysis strengthens 
the emerging trend in the literature that host stellar population age is the main driver of 
SNe~Ia light-curve shape and luminosity, such that more luminous SNe~Ia events occur in galaxies with younger stellar populations.

\subsection{Hubble residual and stellar populations}
\label{HR}

\subsubsection{Final host spectroscopy sample}
\label{HR_finalsample}

Fig.~\ref{siggrid_res} shows Hubble residual as a function of stellar population age (panel a), [Z/H] (panel d), [O/Fe] (panel e), [C/Fe] (panel f) and velocity dispersion (panel b) for the final host spectroscopy sample. The same colour coding and symbols as in Fig.~\ref{agegrid_x1} are used. We do not find significant trends for any of the parameters studied ($<$2$\sigma$ level for all parameters). 
To be sure that the quality of the host spectra does not dilute possible trends, we have also stacked the spectra in two Hubble residual bins dividing the sample into sub-samples of roughly equal sizes. The stacking follows the procedure described in Section~\ref{x1_stack}. Table~\ref{HRstack}
presents the Hubble residual range, number of objects and mean Hubble residual for the two bins, together with the stellar population parameters  and velocity dispersion derived for the stacked spectra. We do not find significant differences for any of the stellar population parameters as well as for velocity dispersion between the two bins.

\begin{figure*}
\centering
\begin{flushleft}
\includegraphics[clip=true,trim=0.3cm 0.8cm 0.5cm 1.3cm,angle=90,scale=0.27]{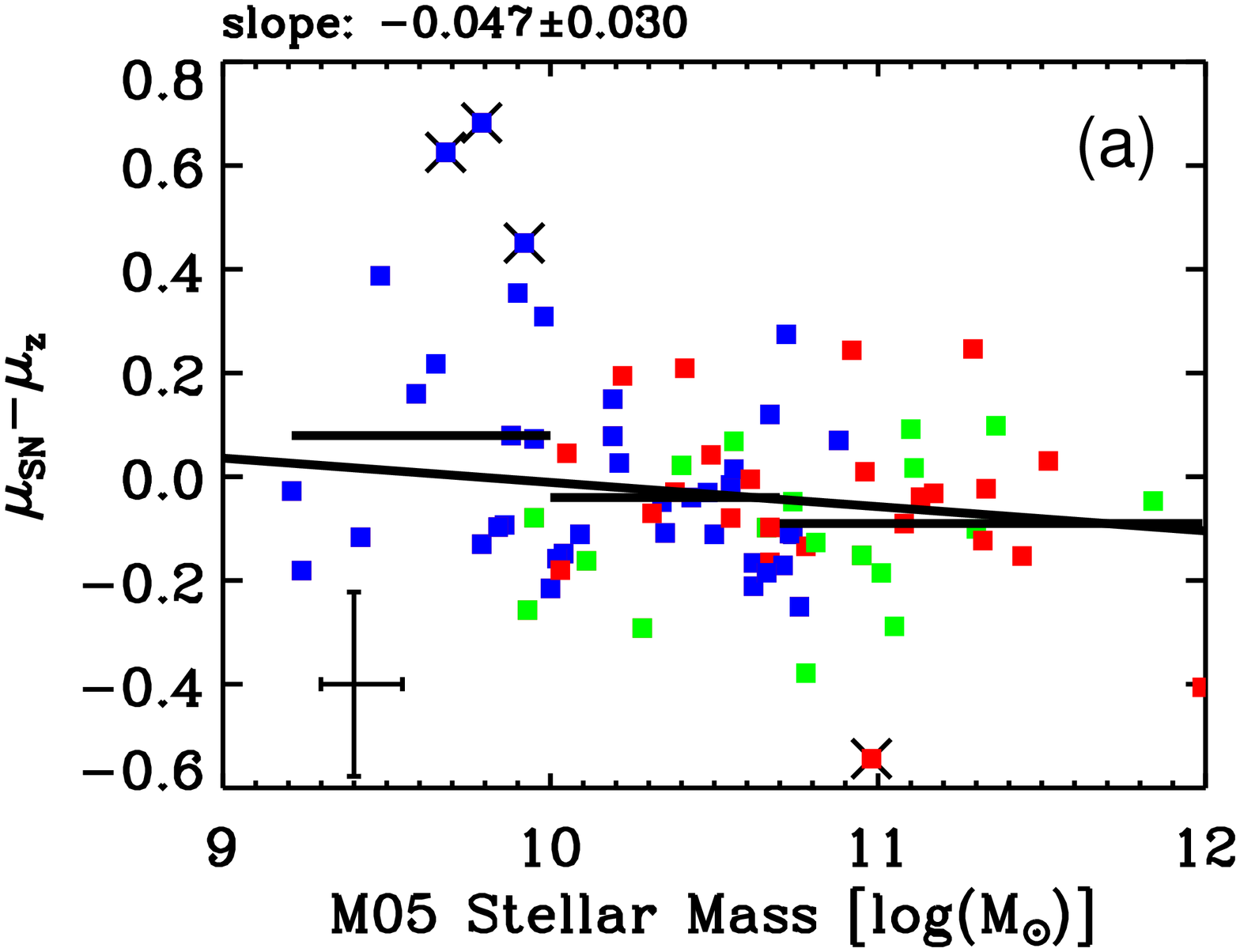}\includegraphics[clip=true,trim=0.3cm 0.8cm 0.5cm 1.8cm,angle=90,scale=0.27]{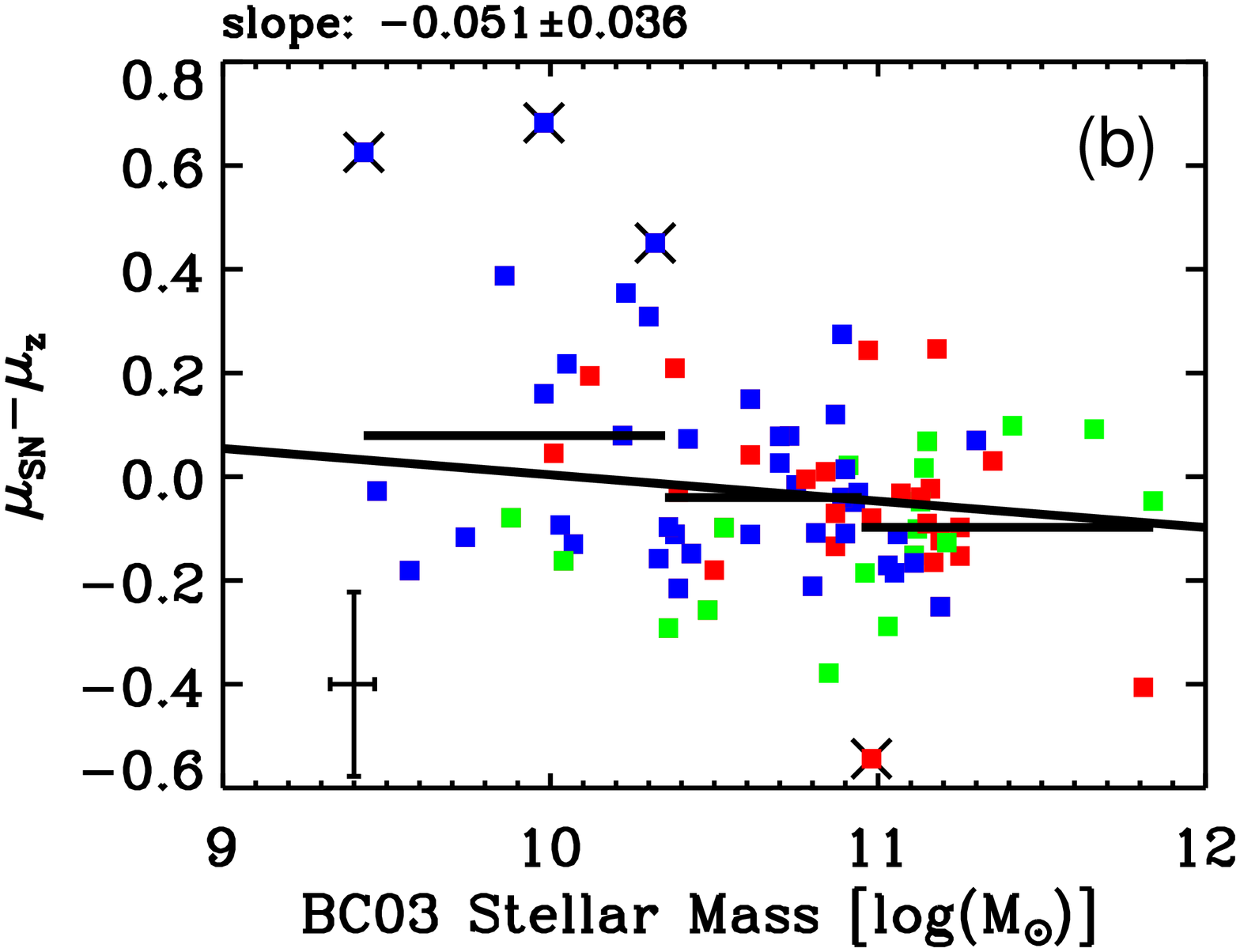}\includegraphics[clip=true,trim=7cm 0.8cm 0.5cm 2cm,angle=90,scale=0.43]{figs/symbols.ps}
\end{flushleft}
\includegraphics[clip=true,trim=0.3cm 0.8cm 0.5cm 1.8cm,angle=90,scale=0.27]{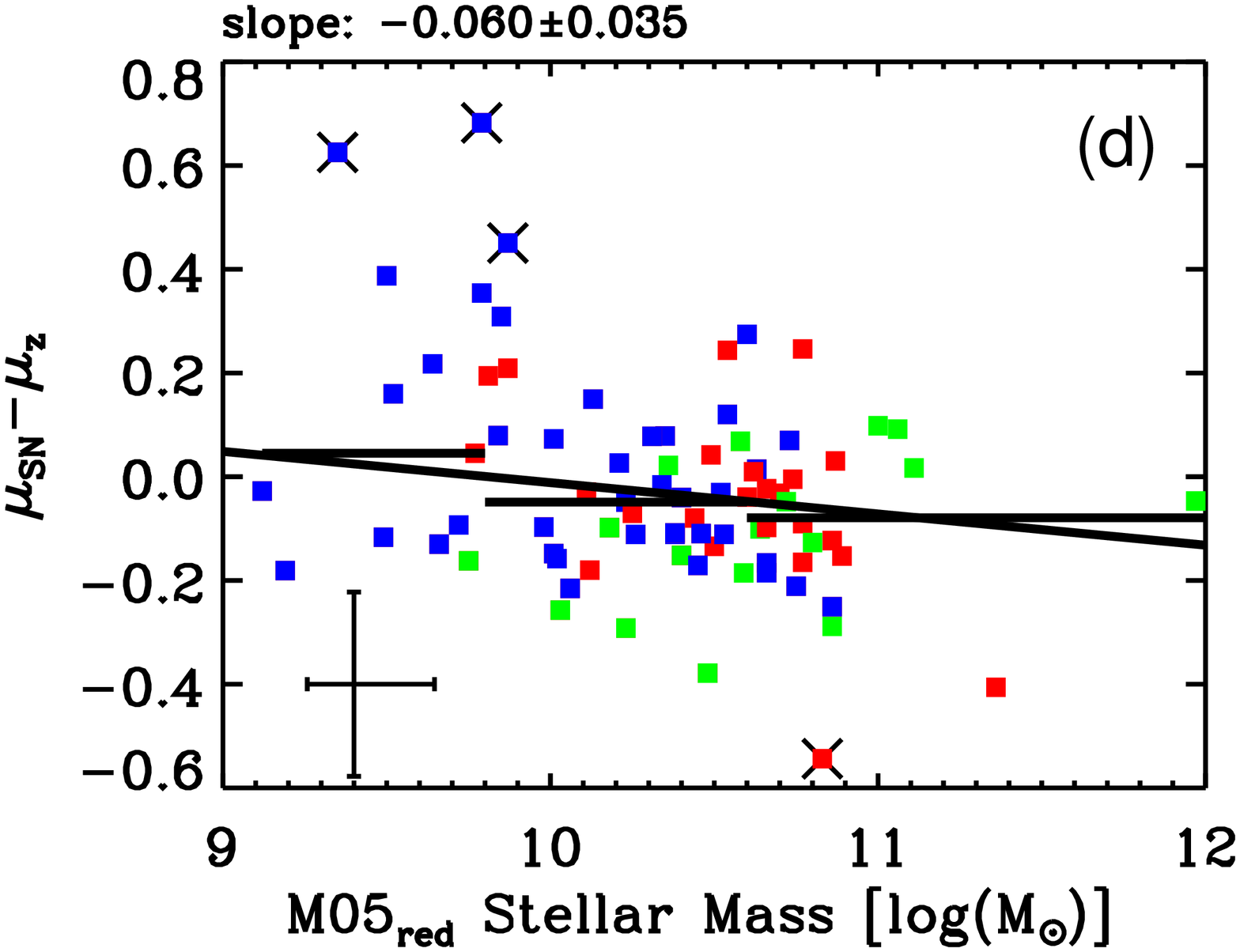}\includegraphics[clip=true,trim=0.3cm 0.8cm 0.5cm 1.8cm,angle=90,scale=0.27]{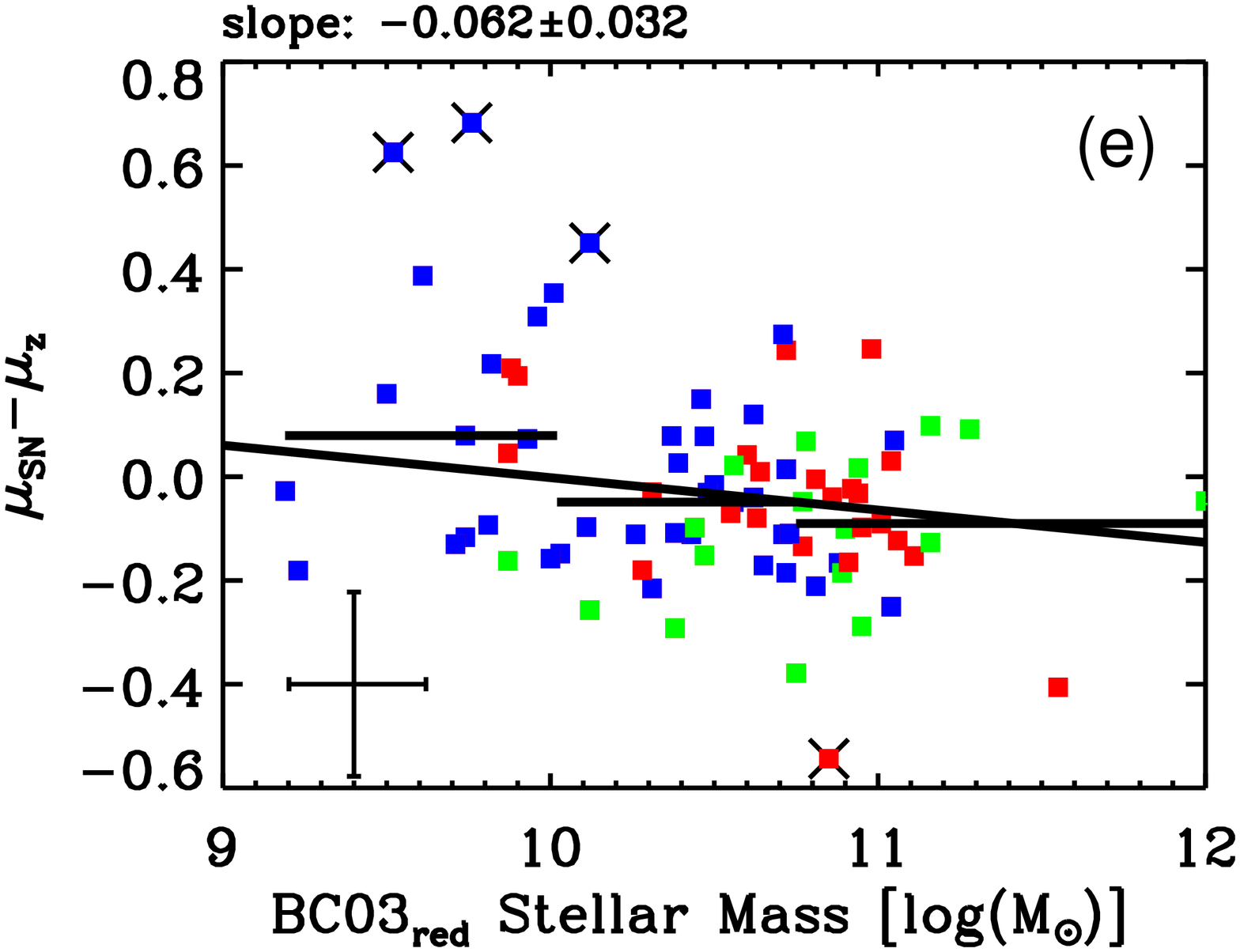}\includegraphics[clip=true,trim=0.3cm 0.8cm 0.5cm 1.8cm,angle=90,scale=0.27]{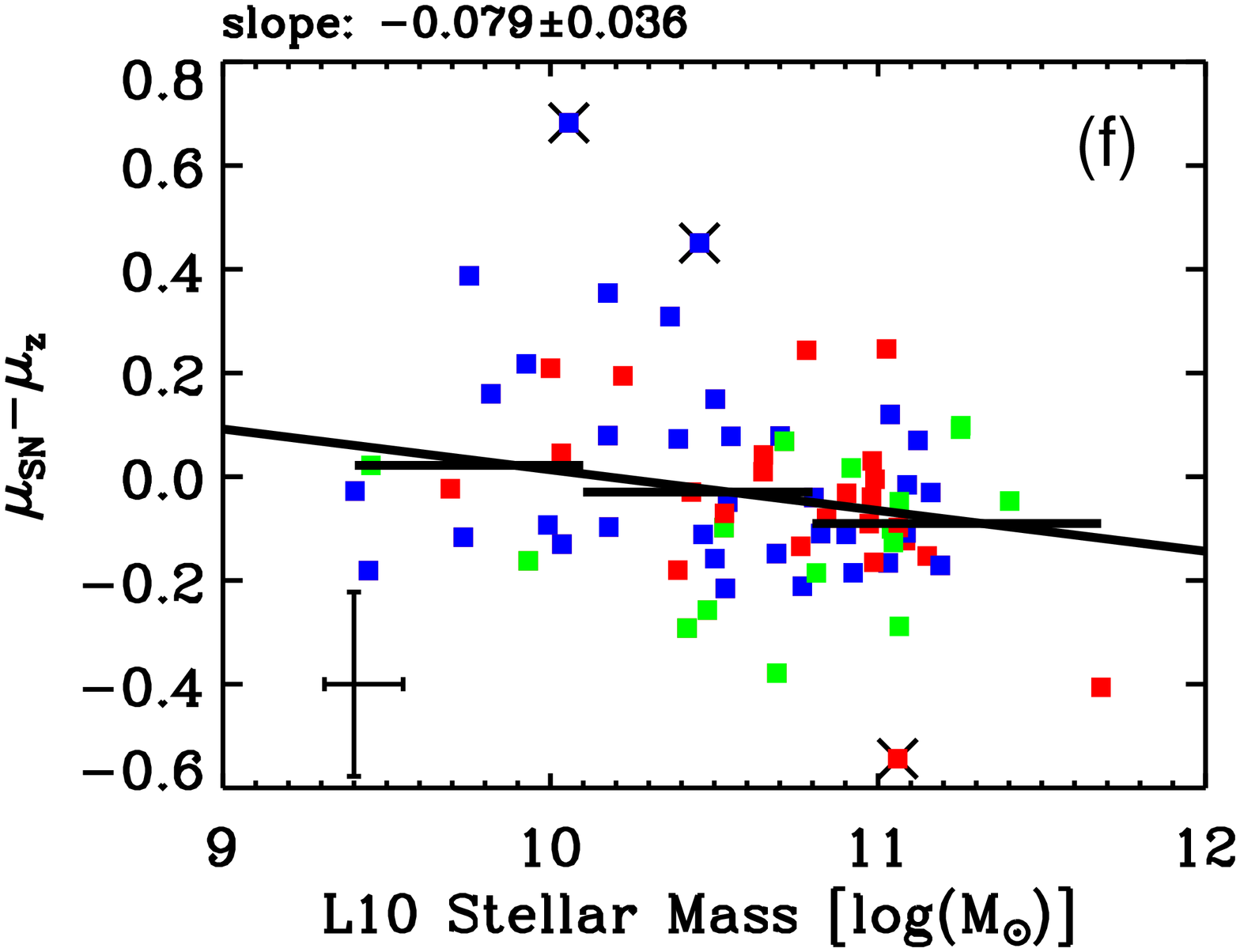}
\caption{Relationship between Hubble residual and the stellar masses derived for the different prescriptions described in Section~\ref{data:masses}, e.g. M05 models without reddening (M05, panel a), M05 models with reddening (M05$_{red}$, panel d), BC03 models without reddening (BC03, panel b), BC03 model with reddening (BC03$_{red}$, panel e) and masses from \citet{lampeitl10} (L10, panel f). The data points are coloured according to the emission line classification of the host galaxies from Section~\ref{emission}, i.e. blue=SF, green=AGN and red=passively evolving. Solid lines are one time sigma-clipped (2$\sigma$ level) least-square fits and over-plotted crosses are removed data points. The slopes and corresponding errors are given at the top of the panels. Horizontal lines are median Hubble residuals in bins of the x-axis parameters with lengths according to the width of the bins. Error bars in the lower left corner are average 1$\sigma$ errors.}
\label{massgrid_res}
\end{figure*}

The relationship between Hubble residual and gas-phase metallicity (see Section~\ref{data:gasZ}) is presented in Fig~\ref{gaszgrid_res}. As pointed out in Section~\ref{data:gasZ}, AGN activity may affect the derived metallicities. The gas-phase metallicities are thus shown for galaxies with detected emission lines and classified as star-forming. Galaxies in the transition region between star formation and AGN activity (see Section~\ref{emission}) are plotted as open symbols. Least-square fits are derived for all data points (solid line) and purely star-forming galaxies (solid points, dashed line) with slopes given by the labels. Since this is a small sample, we do not sigma-clip the least-square fits. Median values in bins of metallicity are shown for all data points as horizontal lines. These lines show a clear trend of decreasing Hubble residuals for higher gas metallicities. The least-square fits also indicate a decrease in Hubble residual with gas-phase metallicity with a significance of $\sim$2.5$\sigma$. 

This low statistical significance is caused by the two low-metallicity, low-Hubble residual data points. Considering the individual Hubble residual errors both points deviate more than 2$\sigma$ from the least-square fits. 
Ignoring these two outliers in the least-square fitting results in a significantly steeper slope of -1.58$\pm$0.32.
One possible explanation for the two outliers could be a failure in the emission-line template fitting (see Section~\ref{gandalf}), affecting the derived gas-phase metallicities. However, this is not the case for either of the two objects. We have further studied the location of these SNe events with respect to the SDSS fiber location of the host galaxy. For the lowest-metallicity object the SNe event is within the SDSS aperture and it is a spectroscopically confirmed Ia. The other outlier object, not spectroscopically confirmed Ia, is located in the outskirts of the galaxy, at a distance of about two times the SDSS aperture diameter (3") from the fiber center. This could possibly explain the deviation of this object if a significant positive gas-phase metallicity gradient or a local metallicity upturn is present.

Fig.~\ref{massgrid_res} shows the relationship between Hubble residual and the stellar masses derived for the different prescriptions described in Section~\ref{data:masses}, e.g. M05 models without reddening (M05, panel a), M05 models with reddening (M05$_{red}$, panel d), BC03 models without reddening (BC03, panel b) and BC03 model with reddening (BC03$_{red}$, panel e). The relationship between Hubble residual and the L10 stellar masses is also included (panel f). The data points are colour coded, as usual, with blue=star-forming, green=AGN and red=passively evolving galaxies. 
Solid lines are least-square fits that have been sigma-clipped one time at a 2$\sigma$ level (see beginning of Section~\ref{results}). The slope of the fits are given at the top of the panels and sigma-clipped data points are indicated by over-plotted crosses. Horizontal lines are median values in bins of the x-axis parameters, where the length of the lines indicate the width of the bins. 

We can see in Fig.~\ref{massgrid_res} that the method of stellar mass derivation affects the Hubble residual-mass relationship. The masses derived with the M05 models without reddening show the least significant trend with Hubble residual and hence the smallest difference in Hubble residual for high and low mass objects. These masses also showed the strongest correlation with velocity dispersion in Section~\ref{veldisp}. 
In comparison, the masses derived with reddening and the BC03 models display stronger trends with Hubble residual but are weaker than for the L10 masses.  The significance of the slopes of the least-square fits vary between 1.7$\sigma$ and 2.2$\sigma$.  

\begin{figure*}
\centering
\includegraphics[clip=true,trim=1.5cm 2.3cm -10cm 9.2cm,angle=0,scale=0.7]{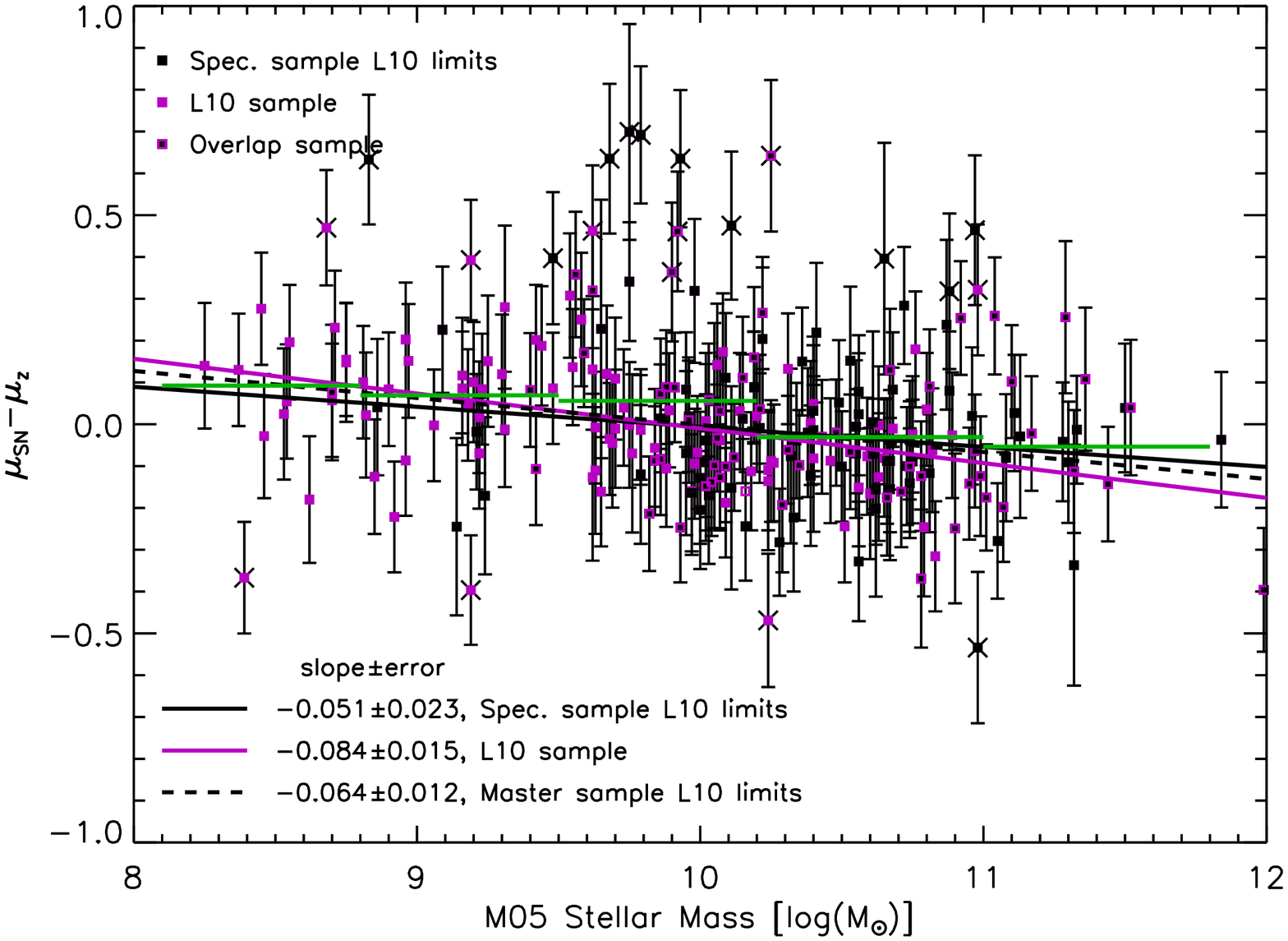}
\caption{Relationship between Hubble residual and stellar mass for the different samples studied (see text). The masses are derived using  the M05 models without reddening (see Section~\ref{data:masses}). The different samples are indicated by the upper labels with corresponding colours.
, i.e. black points=host spectroscopy sample, magenta points=L10 sample and black filled magenta points=overlap sample. 
The lines show least-square fits to each sample, i.e. solid black line=host spectroscopy sample, magenta line=L10 sample and dashed black line=Master sample. The slopes of the fits and corresponding errors are given by the lower labels. 
Over-plotted crosses are data points removed by twice sigma-clipping the Master sample. 
Green horizontal lines are median Hubble residual values for the Master sample, in bins of mass with lengths according to the width of the bins.
}
\label{fig_extsample}
\end{figure*}

\begin{figure*}
\centering
\includegraphics[clip=true,trim=2cm 2.3cm -10cm 10cm,angle=0,scale=0.8]{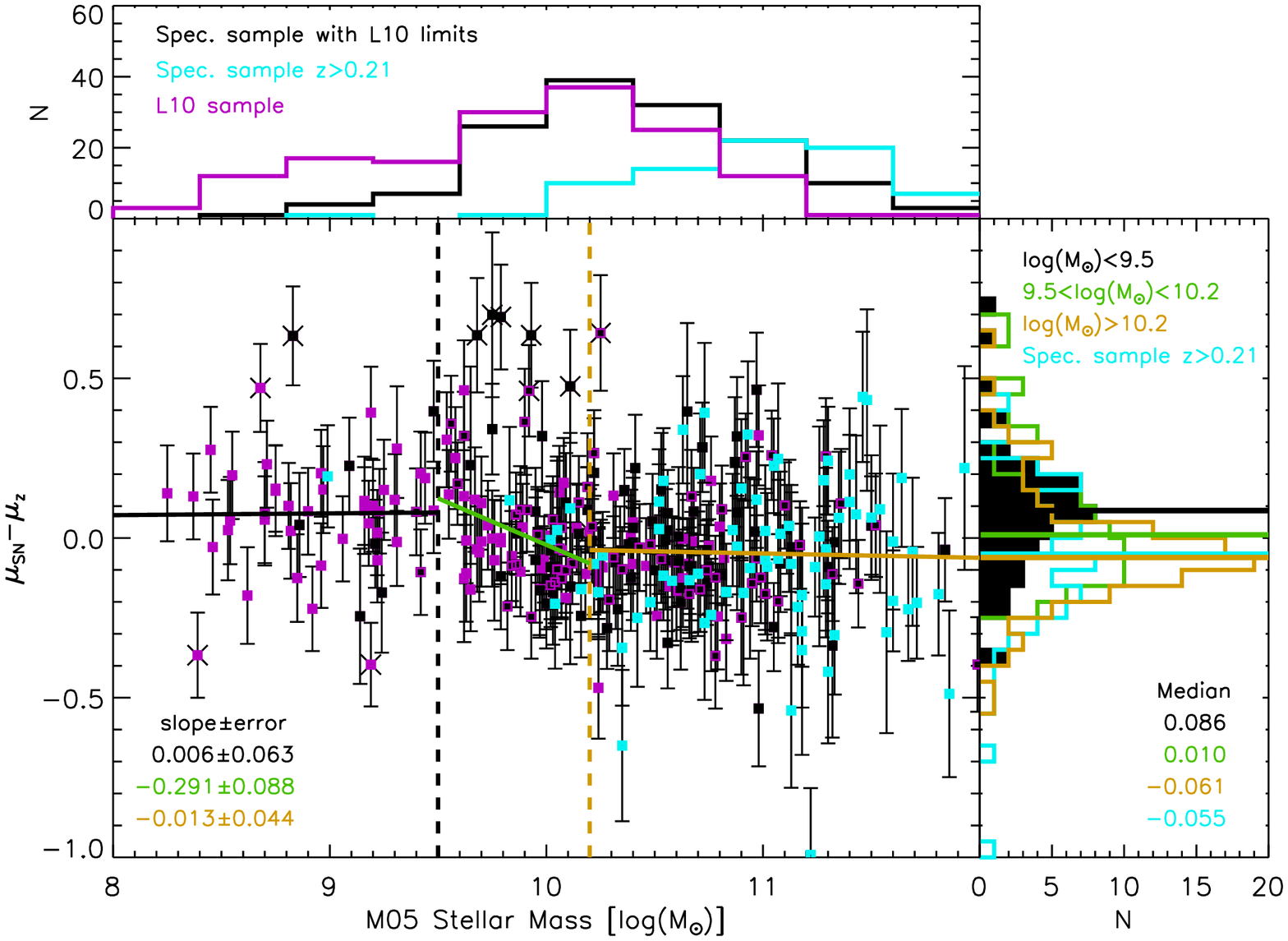}
\caption{\textit{Upper panel:} Histograms of the stellar mass distribution for the different samples studied (see text) as indicated by the labels. \textit{Main panel:} Re-make of Fig.~\ref{fig_extsample}, but also adding high redshift objects (z$>$0.21, cyan points, not used in the least-square fits). Vertical dashed lines indicate boundaries between different mass-ranges for which we derive least-square fits. The slope of each fit is given by the labels with corresponding colours. Over-plotted crosses are removed data points through one-time sigma-clipping. \textit{Right panel:} The Hubble residual distribution for each stellar mass bin of the main panel. The histograms are colour-coded according to the top labels. Lower labels give median values in each mass bin for the same colour coding.
}
\label{fig_segments}
\end{figure*}

\subsubsection{Extended photometric sample}
\label{HR_ext}

In addition to the SNe~Ia with both available host galaxy SDSS spectroscopy and photometry, we can add SNe~Ia with available SDSS photometry only to extend the sample for studying stellar masses. For this purpose we add the sample studied in L10. These authors did a number of quality, redshift and host galaxy type cuts that produced a sample of 162 objects. We followed the quality cuts of L10 for the host spectroscopy sample studied in this work, but we have also applied further quality cuts 
as described in Section~\ref{cuts}. These additional constraints are lifted here in order to have a homogeneous sample. Specifically, the cut in S/N of the host galaxy spectra  
can be removed when we focus on photometry only. For the purpose of having a homogeneous sample we now also apply the redshift cut from L10, i.e., selecting objects with z$<$0.21. In fact, the S/N cut of the final host spectroscopy sample removed objects with z$>$0.21 (see Fig.~\ref{zdistr}).

\begin{table*}
\center
\caption{Slopes and corresponding errors of least-square fits for the different samples studied and different stellar mass derivations.}
\label{MassModality}
\begin{tabular}{cccccc}
\hline
\bf Sample & \bf M05  & \bf M05$_{red}$ &  \bf BC03 & \bf BC03$_{red}$ & \bf Pegase (L10)\\
\hline
Spec. & -0.051$\pm$0.023 & -0.065$\pm$0.025 & -0.057$\pm$0.025 & -0.064$\pm$0.024 & -0.054$\pm$0.019 \\
L10 & -0.084$\pm$0.015 & -0.082$\pm$0.015 & -0.072$\pm$0.013 & -0.076$\pm$0.013 & -0.077$\pm$0.014 \\
Master & -0.064$\pm$0.012 & -0.067$\pm$0.013 & -0.062$\pm$0.013 & -0.071$\pm$0.012 & -0.068$\pm$0.011 \\
\hline
\end{tabular}
\end{table*}

The host galaxy type cut in L10 was made in order to be able to study two distinct types of host galaxies, passive and star-forming, by constraining the allowed error on derived star-formation rates. Since we are not interested in distinct types of galaxies in this work, we can ignore this cut. The host spectroscopy sample used in our study is drawn from the same parent sample as the L10 sample, we will therefore mainly just add galaxies rejected due to the galaxy-type cut in L10. This action results in a sub-sample of 145 objects drawn from the full host spectroscopy sample. Out of this sub-sample 60 are in common with the L10 sample, leaving a Master photometric sample consisting of 247 objects.

Fig.~\ref{fig_extsample} presents Hubble residual as a function of host stellar mass, derived with the M05 without reddening option, for the Master sample. The sub-samples (with M05 masses) are indicated by magenta points for the L10 sample and black points for the host spectroscopy sample with L10 limits. Open, black-filled magenta points are overlapping objects. The lines represent least-square fits to the different samples, i.e., black dashed line for the Master sample, black solid line for the host spectroscopy sample and magenta line for the L10 sample. 
The different samples were not individually sigma-clipped in order not to bias the results by removing different objects for the different samples.
Instead, the Master sample was sigma-clipped twice at a 2$\sigma$ level and the objects removed through this step were ignored when performing the least-squares fitting of the two sub-samples. Green horizontal lines are median Hubble residual values for the Master sample, in bins of mass with lengths according to the width of the bins.

It is clear that the slope of the fits, given by the labels, depends on the sample selection, where the L10 and host spectroscopy samples show the most and least prominent trends, respectively. The slope of the Master sample lies in between the values of the two sub-samples. The statistical significance of the slope is also sample dependent. The slope of the host spectroscopy sample is different from zero at a 2.2$\sigma$ level, while the corresponding numbers are 5.6$\sigma$ and 5.3$\sigma$ for the L10 and Master sample, respectively. For the L10 sample we recover the trend found in \citet{lampeitl10}, but with a slightly steeper slope due to the sigma-clipping procedure. We have repeated this analysis and confirm the above results for the different stellar mass derivations, including the procedure used in L10. Table~\ref{MassModality} presents slopes and errors of least-square fits for the different samples and different stellar mass derivation. The systematic variations apparent for the smaller sample (see Section~\ref{HR_finalsample}) are weaker when the number of galaxies is increased. It should also be noted that the Hubble residuals for the L10 sample have been re-derived using the $\alpha$ and $\beta$ values used in this work (see Section~\ref{SNIaprop}).

Two effects are responsible for the variation in slope of the different samples;
1. The median Hubble residual values (green horizontal lines in Fig.~\ref{fig_extsample}) show a drop around $\log M_{\odot}$=10.2 and are close to constant below and above this value. 2. The L10 and host spectroscopy samples have different stellar mass distributions, which is apparent in Fig.~\ref{fig_extsample}. This is better emphasized in Fig.~\ref{fig_segments}, where the top panel shows the stellar mass distributions for the different samples, i.e., magenta and black histograms represent the L10 and host spectroscopy samples, respectively. The host spectroscopy sample is lacking low-mass objects, while the L10 sample is deficient in high-mass galaxies. The high-mass bias of the host spectroscopy sample is not surprising as SDSS spectroscopic targeting required a magnitude of r$<$17.7 for the main galaxy sample \citep{strauss02}. 

The discrepancy in slope between the different samples is driven by the different mass ranges covered. This result is shown in the main panel of Fig.~\ref{fig_segments}, 
which is a remake of Fig.~\ref{fig_extsample}. Here we instead study the Hubble residual distributions in three different mass intervals: $\log M_{\odot}<$9.5, 9.5$<\log M_{\odot}<$10.2 and $\log M_{\odot}>$10.2. Least-square fits, derived without the cyan points, in these different regions are presented by the lines with slopes according to the labels with corresponding colours. For the high-mass bin the HR-mass relation is clearly flat. This appears to be the case also for the low-mass bin, but this regime is more sparsely populated. Higher Hubble residuals are found at low masses, while in the region around $\log M_{\odot}\sim$10.0 the Hubble residual varies significantly. Hence, the Hubble residual-mass relationship shows a step function where negative and positive values are found for high- and low-mass objects, respectively. This behaviour can also be found in the sample studied in \citet{sullivan10}. It is clear from their Fig.~4 that the Hubble residuals suddenly drop above $\log M_{\odot}\sim$10.0, while above and below this limit the average Hubble residuals stay roughly constant. 
Stellar masses around 10$^9$M$_\odot$ may not represent the total mass of a galaxy, but the mass of the star-forming fraction, which outshines the underlying older population \citep[the "tip of the iceberg" effect,][]{M10}. The simulations of \citet{pforr} indicate that when a 10$^{11}$M$_\odot$ galaxy has experienced a 1-10 \% by mass recent starburst, the SED-fit derived mass is closer to the mass of the burst rather than to the total mass. This is an interesting effect to keep in mind when a large mass range is considered.

We test the significance of the step function as compared to the linear fit over the full mass-range using the resulting $\chi^2$-values and an F-test for nested regression models. 
To have the same sample for both models we use the full Master sample (247 objects) and take into account that the step function has six parameters instead of two for the linear fit. 
We find that the F-test rejects the null hypothesis that the step function does not provide a significantly better fit than the linear fit
at a probability $>$90\% (93.5\%).

The Hubble residual distributions of the different mass bins are plotted in the right panel of Fig.~\ref{fig_segments} with colours according to the labels at the top of this panel. The median values of each mass bin is given by the lower right labels. The median Hubble residual of the low-mass bin is offset from the high-mass bin by 0.147 mag, while the median value of the intermediate mass bin is located between these values. A Kolmogorov-Smirnov test confirms that the distributions of the high- and low-mass bins are different at a significance level of over 99.9$\%$. 
Hence, the stellar mass distribution of the selected sample strongly affects the derived Hubble residual-mass trend. This would not be the case if there was a continuous  trend over the full stellar mass range. 

To further emphasize this point we added the sub-sample of objects with z$>$0.21 (cyan points) in Fig.~\ref{fig_segments}. In the top panel it is clear that this sample is distributed towards high stellar masses. This sub-sample exhibits low Hubble residuals similar to the high-mass bin of the Master sample, as can be seen in the main and right panels. Including this sub-sample in the Master sample significantly reduces the slope of the least-square fit from -0.065 to -0.048.

We confirm similar step functions for all stellar mass derivations, i.e. M05 with reddening, BC03 without reddening, BC03 with reddening and also for the stellar masses derived with the setup of L10.
We have also applied the cuts described in Section~\ref{cuts} to the Master sample and L10 sample and find that the results are not affected. However, since the cut in spectral resolution can not be applied to the L10 sample, it has not been considered for this consistency check.

\subsubsection{Comparison with the literature}
\label{litcomp:HR}

We find that Hubble residual depends on host galaxy mass in agreement with several recent reports in the literature \citep{kelly10,lampeitl10,sullivan10,gupta11}. 
However, we find that the Hubble residual-mass relationship behaves as a step function, which can also be seen in the work of \citet{sullivan10}, where the Hubble residuals mainly change around $\log M\sim$10.0.
For the first time, we study the relation between Hubble residual and stellar velocity dispersion, but find no significant trend. However, this may be due to the high-mass bias of the sample studied. 

It is believed that the Hubble residual-mass trend is the result of a more fundamental relationship between Hubble residual and SNe~Ia progenitor properties that could possibly be mimicked through host stellar populations. 
Several authors have therefore also studied relationships between Hubble residual and the stellar population parameters age and metallicity. \citet{gallagher08} presents the only study in the literature to date that is based on absorption line indices and find stellar metallicity to be the source of systematic Hubble residual variations for a small sample of $\sim$20 host galaxies. \citet{howell09} and \citet{sullivan10} instead infer gas-phase metallicities from stellar masses together with the mass-metallicity relationship from \citet{tremonti04} and find significant trends in these parameters with Hubble residual. Gas-phase metallicities are also inferred from stellar mass in a similar way by \citet{neill09}, but they instead find a stronger relation between stellar population age and Hubble residual. Similar to this work, \citet{dandrea} derive gas-phase metallicities directly from the emission line ratios and find a trend of higher Hubble residuals in galaxies of lower metallicity. Hayden et al. (in prep.) find that the scatter in the Hubble residuals are significantly reduced when gas-phase metallicity is added as further fit parameter besides stellar mass. \citet{gupta11} instead find stellar population age to correlate with Hubble residual. For the sample of 84 host galaxies studied in this work, we do not find significant trends of Hubble residuals with any of the absorption line derived stellar population parameters age, metallicity and element ratios. Again, this may be due to the high-mass bias of the sample studied. 

We have shown in the previous two sub-sections that the Hubble residual-mass trend is dependent on the mass-range spanned by the galaxies. The lack of a trend between Hubble residual and velocity dispersion as well as with stellar population age and metallicity could therefore be the result of the limited mass-range studied through the final host spectroscopy sample. The reason why we find a tentative relationship for gas-phase metallicity probably arises from the fact the star-forming galaxies, for which we can infer gas-phase metallicities, cover the stellar mass range where the Hubble residuals dramatically vary (around 10$^{10}$M$_\odot$; compare top left panel of Fig.~\ref{massgrid_res} and Fig.\ref{fig_extsample}) together with the fact that emission lines are more robustly derived in low-S/N spectra than absorption lines.

\section{Discussion}
\label{disc}

We derive host galaxy stellar population parameters for SNe~Ia from the SDSS-II Supernova survey. These parameters are derived from absorption line indices through comparison with stellar population model predictions (see Section \ref{hostprop}). Due to the low S/N of the host galaxy spectra we focus on the highest quality objects, considering both the SNe~Ia observations and host galaxies, which results in a final sample of 84 objects (see Section \ref{cuts}). 

We find a strong relationship between SALT2 stretch factor and stellar population age (see Section \ref{stretch}). 
For SALT2 colour on the other hand, we do not find dependencies on any of the host galaxy parameters, in agreement with previous studies.
 We also derive photometric stellar masses for the host galaxies using a variety of methods and models. 
 The Hubble residual-stellar mass relation is found to behave as a step function, where the trend is flat at the high-mass end (see Section \ref{HR_ext}). This is also the mass-regime covered by our host spectroscopy sample, which may explain a lack of trends between Hubble residual and the stellar population parameters studied.

\subsection{Progenitor systems}
\label{dtimes}

Considering several different progenitor systems, theoretical models of SNe~Ia explosions find SNe~Ia rates (SNRs) peaking at delay times below or close to 1 Gyr \citep{yungelson00,greggio05,ruiter10}. The SNRs then smoothly decline and become 10-100 times lower at delay times of $\sim$10 Gyr. Thus theory implies that the delay times of SNe~Ia span the range from relatively quick events to explosions delayed by a Hubble time. 

Using the host galaxy stellar population ages as delay-time proxies, observations support the theory as we find a wide range of luminosity-weighted ages from below 1 Gyr to $>$10 Gyr. Similarly \citet{gupta11} find SNe~Ia events in stellar populations with mass-weighted ages $>$10 Gyr and down to $\sim$2 Gyr, while \citet{howell09} and \citet{neill09} present luminosity-weighted ages from $\sim$100 Myr up to $>$10 Gyr. The work of \citet{brandt10} suggests delay-times of $<$400 Myr and $>$2.4 Gyr, indicative of two different progenitor channels. All these studies agree with theory such that a higher fraction of SNe~Ia events are found in young star-forming galaxies compared to old passively evolving galaxies. If SN~Ia peak luminosity increases with progenitor mass, the wider range of stellar masses in star-forming galaxies together with short life-times for massive stars result in the higher SNe~Ia fractions. 

Indeed, considering the strong anti-correlation between stretch factor and stellar population age (see Section \ref{stretch}), there is a clear connection such that those SNe~Ia with the shortest delay times, hence most massive progenitors, are the most luminous and vice versa. Contrary to \citet{brandt10}, the relation between stretch factor and stellar population age we found shows a smooth behaviour which may be indicative of a single progenitor system.

\citet{yungelson00} show that  the SNR for a DD system peaks at delay times of $\sim$100 Myr, while the analogous for a SD system is $\sim$1 Gyr. Thus the lower age limit of the stellar populations hosting SNe~Ia could be used to constrain possible progenitor systems. However, it may be unreasonable to attribute the integrated light of a galaxy to be a good proxy for SNe~Ia delay times, considering the range of stellar populations, age and metallicity, possibly present within an instrumental aperture. An alternative approach could be to use observations confined to the vicinity of the SNe~Ia. However, it is likely that a mixing of stellar populations occur over a long time span. The probability of identifying the "true" parent stellar population for a SNe~Ia progenitor should increase for  younger stellar populations. It may therefore be possible to constrain the lower delay time limit using the stellar populations of the vicinity of the SNe~Ia. This information can then in turn constrain possible progenitor systems when comparing to theoretical SNe~Ia models. 

\subsection{Hubble residual minimisation}
\label{disc:HR}

The reports of Hubble residual dependencies on host galaxy mass \citep{kelly10,sullivan10,gupta11,conley11} have led to the inclusion of this parameter to further reduce the scatter in the redshift-distance relation \citep[e.g.][]{lampeitl10}. In this work we find systematic uncertainties in the derived masses, due to the adopted SED-fitting prescription, to affect the Hubble residual-mass relationship (see Section \ref{HR_finalsample}). However, these systematic uncertainties become weaker when the number of galaxies is increased (see Section \ref{HR_ext}). 

We also find that the stellar mass range covered affects the derived Hubble residual-mass trend (see Section \ref{HR_ext}). This trend behaves as a step function, where the low- and, in particular, the high-mass end show flat trends. Due to this step function, the stellar mass range sampled will result in a systematic variation of the derived Hubble residual-mass slope.
Hence, the non-linear behaviour of the Hubble residual-mass relationship should be accounted for when using stellar mass as a further parameter for minimising the Hubble residuals.

The dependence of Hubble residual on stellar mass is believed to be a consequence of a dependency on a more fundamental parameter. 
The sampling of the high-mass regime of our host spectroscopy sample, where the trend between Hubble residual and stellar mass is flat, hampers a conclusion on which parameter is driving Hubble residual variations.

Interestingly, the step in the Hubble residual-mass plane appears at around $\sim$2$\times10^{10}$M$_{\odot}$ which is close to the evolutionary transition mass of low-redshift galaxies, discovered by \citet{kauffmann03a}. They find that below $\sim$3$\times10^{10}$M$_{\odot}$, galaxies are typically disc-like objects with young stellar populations and low surface mass densities. Above this limit, they instead find an increasing fraction of bulge-dominated galaxies with old stellar populations. Hence, the transition mass marks a change in the stellar populations of galaxies. The similar mass-range shared between the sudden shift in the Hubble residual and transition in galaxy properties is therefore most likely not coincidental. The stellar populations below and above the transition mass bare witness of the properties of the SNe~Ia progenitors with high and low Hubble residuals, respectively. 
If the Hubble residual-mass relation is indeed flat at the low-mass end (only tentative due to low number statistics) as it is at the high-mass end, the step function may indicate two samples of SNe~Ia with high and low Hubble residuals, i.e. showing a weak difference in peak luminosity not accounted for by the light-curve fitting techniques. 

The overlap in mass for the galaxy evolutionary transition and change in Hubble residual tells us that the latter does not change significantly in old, passively evolving stellar populations. Instead there is a strong variation in lower mass galaxies with younger stellar populations, higher SFRs and lower gas-phase metallicities \citep{tremonti04}. Indeed several authors suggest gas-phase metallicity to be the fundamental parameter driving Hubble residual variations \citep[][Hayden et al. in prep.]{howell09,sullivan10,dandrea}. Also in this work we find tentative results for such a relation between Hubble residual and gas-phase metallicity (see Section \ref{HR_finalsample}). 
If this is indeed the fundamental parameter, it represents progenitor metallicity. This is in agreement with theoretical predictions of prompt SNe~Ia in star-forming galaxies \citep{yungelson00,greggio05,ruiter10}.
The flat slope at masses above $\sim$2$\times$10$^{10}M_\odot$ is then expected as the gas-phase metallicities show a flat behaviour in this mass-regime \citep[][Hayden et al. in prep.]{tremonti04}. However, prompt SNe~Ia are not expected in old stellar populations present in passively evolving, massive galaxies. 
The derivation of the full range of stellar population parameters and gas-phase properties for a high quality data set, covering the full mass range, is required to determine the fundamental parameter driving Hubble residual variations.

Future data from IFU spectroscopy, sampling the local environment of SNe~Ia will be quite useful in identifying the fundamental parameter responsible for the scatter in the Hubble residuals. For such data sets it will be important to sample the full stellar mass range, in particular the mass range around 10$^{10}$M$_{\odot}$. 

\section{Conclusions}
\label{conc}

We present an analysis of the stellar populations of SNe~Ia host galaxies using SDSS-II spectroscopy. Using the stellar population models of absorption line indices from \citet{TMJ11} we derive the stellar population parameters age, metallicity and element abundance ratios to study relationships with SNe~Ia properties. We also measure stellar velocity dispersion from stellar template fitting and gas-phase metallicity from emission lines when detected. Furthermore, we derive stellar masses and revisit the correlation between Hubble residual and stellar mass that has recently received much attention in the literature. The stellar masses are derived from SDSS-II photometry from repeat Stripe 82 observations and several SED-fit methods. 
From a sample of 292 SNe~Ia we select 84 objects depending on the quality of the host galaxy spectroscopy and accuracy of the SNe~Ia properties  (see Section \ref{cuts}).

We find a larger fraction of SNe~Ia together with typically higher SALT2 stretch factor values (i.e. more luminous SNe~Ia with slower declining light-curves) in star-forming compared to passively evolving galaxies (see Section \ref{stretch}), in agreement with the literature. 
Hence, previous studies suggest that the decline rate and peak-luminosity of SNe~Ia depend on stellar population age. 
With the large parameter space covered in this work, we indeed find that SALT2 stretch factor values show a strong dependence on stellar population age in the sense of a clear anti-correlation (see Section \ref{stretch}). 
As a result of this trend we also find anti-correlations with velocity dispersion and galaxy mass,  
while only weak anti-correlations are also found for stellar metallicity and the element abundance ratios studied. 

To ensure that the quality of the host galaxies are not affecting the selected sample, we stacked the spectra in bins of stretch factor. This exercise confirms the above results that stellar age is a strong candidate for the main driver of SNe~Ia luminosity (see Section \ref{x1_stack}). 
Hence, SNe~Ia peak-luminosity is closely related to the age of the stellar progenitor systems, where more luminous SNe~Ia appear in young stellar populations. 

The peak-luminosity variation of SNe~Ia is corrected for through light-curve fitting. However, scatter in the redshift-distance relation even after these corrections may introduce uncertainties in derived cosmological parameters. It has been debated which parameter that could be used to further minimize this scatter. While stellar mass has already been introduced for this purpose, stellar population age and metallicity as well as gas-phase metallicity have been proposed as fundamental parameters for the Hubble residual dependency.

However, we identify no statistically significant trends of Hubble residuals with any of the stellar population parameters studied or with stellar velocity dispersion (see Section \ref{HR_finalsample}). 
Instead we find tentative results confirming a trend between Hubble residual and gas-phase metallicity as previously reported. 
For the Hubble residual-stellar mass relationship, our selected sample shows a weak trend that is affected by the method of stellar mass derivation. Stellar masses that show the tightest correlation with stellar velocity dispersion produce the weakest trend with Hubble residual. 
To study this in more detail we extend the sample to include 102 SNe~Ia with available SDSS host galaxy photometry and lift the spectroscopic quality constraints, resulting in a sample of 247 objects for deriving stellar masses from SDSS photometry. With the better statistics of this larger sample the systematic variations arising from the stellar mass derivation are weakened (see Section \ref{HR_ext}).

For the extended photometric sample it is clear that the reported Hubble residual-mass relation is strongly dependent on the stellar mass range studied and behaves as a step function. In the high-mass regime, the relation between Hubble residual and stellar mass is flat. Since our sample with available host galaxy spectroscopy mainly probes this high-mass regime, it is not surprising that we do not find any significant Hubble residual trend with the stellar population parameters studied.  Below a stellar mass of $\sim2\times10^{10} M_{\odot}$, i.e., close to the evolutionary transition mass of low-redshift galaxies reported in the literature, the trend changes dramatically such that lower-mass galaxies exhibit fainter SNe~Ia after light-curve corrections. 
We conclude that the non-linear behaviour of the Hubble residual-mass relationship should be accounted for if stellar mass is to be used as a further parameter for minimising the Hubble residuals. 
However, it is crucial to find the fundamental parameter driving systematic variations in the Hubble residual. 

\section*{ACKNOWLEDGMENTS}

We thank 
Rita Tojeiro for helpful discussions.

Funding for the SDSS and SDSS-II has been provided by the Alfred P. Sloan Foundation, the Participating Institutions, the National Science Foundation, the U.S. Department of Energy, the National Aeronautics and Space Administration, the Japanese Monbukagakusho, the Max Planck Society, and the Higher Education Funding Council for England. The SDSS Web Site is http://www.sdss.org/.

The SDSS is managed by the Astrophysical Research Consortium for the Participating Institutions. The Participating Institutions are the American Museum of Natural History, Astrophysical Institute Potsdam, University of Basel, University of Cambridge, Case Western Reserve University, University of Chicago, Drexel University, Fermilab, the Institute for Advanced Study, the Japan Participation Group, Johns Hopkins University, the Joint Institute for Nuclear Astrophysics, the Kavli Institute for Particle Astrophysics and Cosmology, the Korean Scientist Group, the Chinese Academy of Sciences (LAMOST), Los Alamos National Laboratory, the Max-Planck-Institute for Astronomy (MPIA), the Max-Planck-Institute for Astrophysics (MPA), New Mexico State University, Ohio State University, University of Pittsburgh, University of Portsmouth, Princeton University, the United States Naval Observatory, and the University of Washington.


{}

\newpage
\appendix
\begin{table}
\section{}
\center
\caption{Properties of the host galaxies}
\label{ELtable}
\begin{tabular}{cccccc}
\hline
\bf SN ID & \bf Class.$^a$ & \bf Ra & \bf Dec &  \bf Redshift & \bf Sample$^b$ \\
& & (deg)  & (deg) & &   \\
\hline
     691  &    Ph &      329.7300  &    -0.4990  &    0.1310  &  m   \\ 
     701  &    Ph &      334.6050  &     0.7976  &    0.2060  &  f  \\ 
     717  &    Ph &      353.6280  &     0.7659  &    0.1310  &     \\ 
     722  &    Sp &        0.7060  &     0.7519  &    0.0870  &  f  \\ 
     739  &    Sp &       14.5960  &     0.6794  &    0.1080  &  f  \\ 
     762  &    Sp &       15.5360  &    -0.8797  &    0.1920  &  m   \\ 
     774  &    Sp &       25.4640  &    -0.8767  &    0.0940  &  f  \\ 
    1031  &    Ph &       45.8490  &     1.0435  &    0.1160  &     \\ 
    1032  &    Sp &       46.7960  &     1.1200  &    0.1300  &  f  \\ 
    1041  &    Ph &       49.8790  &     1.1956  &    0.0660  &  f  \\ 
    1112  &    Sp &      339.0170  &    -0.3749  &    0.2580  &     \\ 
    1126  &    Ph &      323.9900  &     1.0194  &    0.1150  & m    \\ 
    1351  &    Ph &        2.2660  &    -1.1065  &    0.1760  &  f  \\ 
    1371  &    Sp &      349.3740  &     0.4297  &    0.1190  &  f  \\ 
    1415  &    Ph &        6.1060  &     0.5990  &    0.2120  &     \\ 
    1495  &    Ph &       36.8480  &     0.1053  &    0.2050  &     \\ 
    1511  &    Ph &       42.2580  &     0.0757  &    0.1870  &     \\ 
    1580  &    Sp &       45.3250  &    -0.6451  &    0.1830  &  f  \\ 
    1588  &    Ph &       52.1480  &    -0.8348  &    0.1510  &  f  \\ 
    1748  &    Ph &      353.1120  &    -0.4822  &    0.3400  &     \\ 
    1825  &    Ph &      316.8200  &     0.3590  &    0.1560  &  m   \\ 
    1906  &    Ph &      309.6490  &     0.9133  &    0.2000  &  f  \\ 
    2056  &    Ph &      320.0370  &    -0.3899  &    0.1900  &     \\ 
    2092  &    Ph &      336.4560  &    -0.2390  &    0.1430  &     \\ 
    2331  &    Ph &       16.6430  &     1.0873  &    0.2630  &     \\ 
    2361  &    Ph &        7.6140  &    -0.4533  &    0.1540  &     \\ 
    2561  &    Sp &       46.3440  &     0.8597  &    0.1190  &  m  \\ 
    2689  &    Sp &       24.9000  &    -0.7579  &    0.1620  &  f  \\ 
    2766  &    Ph &       52.7120  &    -0.6186  &    0.1500  &  f  \\ 
    2992  &    Sp &       55.4970  &    -0.7829  &    0.1270  &   m  \\ 
    3241  &    Sp &      312.6530  &    -0.3542  &    0.2590  &     \\ 
    3293  &    Ph &       39.6390  &     0.2862  &    0.1320  &     \\ 
    3565  &    Ph &        2.5900  &     0.7100  &    0.2900  &     \\ 
    3592  &    Sp &       19.0530  &     0.7905  &    0.0870  &  f  \\ 
    3888  &    Ph &        7.4840  &    -0.4557  &    0.1180  &  f  \\ 
    3892  &    Ph &       18.6860  &    -0.4518  &    0.3500  &     \\ 
    3901  &    Sp &       14.8500  &     0.0026  &    0.0630  &  f  \\ 
    4019  &    Ph &        1.2620  &     1.1464  &    0.1810  &  m   \\ 
    4065  &    Ph &       47.9980  &     1.0513  &    0.1310  &  f  \\ 
    4181  &    Ph &       37.8160  &    -1.1343  &    0.3420  &     \\ 
    4651  &    Ph &       37.3760  &    -0.7475  &    0.1520  &     \\ 
    4690  &    Ph &       32.9300  &     0.6877  &    0.1990  & m    \\ 
    4966  &    Ph &      313.4490  &    -1.0414  &    0.3130  &     \\ 
    5230  &    Ph &      334.7930  &     0.8388  &    0.3220  &     \\ 
    5736  &    Sp &       22.8660  &    -0.6288  &    0.1440  &   m  \\ 
    5785  &    Ph &      328.5980  &     0.0844  &    0.1480  &  f  \\ 
    5944  &    Sp &       29.2020  &    -0.2126  &    0.0460  &  m   \\ 
    5966  &    Sp &       16.1910  &     0.5133  &    0.3100  &     \\ 
    6057  &    Sp &       52.5540  &    -0.9745  &    0.0670  &  f  \\ 
    6213  &    Ph &      344.1060  &    -0.4500  &    0.1090  &     \\ 
    6295  &    Sp &       23.6740  &    -0.6042  &    0.0800  &     \\ 
    6332  &    Ph &      325.9660  &    -0.7183  &    0.1520  &     \\ 
    6406  &    Sp &       46.0890  &    -1.0631  &    0.1250  &  m   \\ 
    6558  &    Sp &       21.7020  &    -1.2382  &    0.0570  &  f  \\ 
    6638  &    Ph &       45.0390  &    -1.2376  &    0.3260  &     \\ 
    6683  &    Ph &      327.2140  &     0.6126  &    0.5090  &     \\ 
\hline
\end{tabular}
\flushleft
$^a$Supernovae classification, Sp=Spectroscopic, Ph=Photometric\\ ~(see Section~\ref{data})\\
$^b$Sample selection, f=final host spectroscopy sample (see\\ ~Section~\ref{cuts}), m=Master sample (see Section~\ref{HR_ext})\\
\end{table}

\begin{table}
\center
\contcaption{}
\begin{tabular}{cccccc}
\hline
\bf SN ID & \bf Class.$^a$ & \bf Ra & \bf Dec &  \bf Redshift & \bf Sample$^b$ \\
& & (deg)  & (deg) & &   \\
\hline
    6813  &    Ph &       27.2700  &     0.0566  &    0.4110  &     \\ 
    6851  &    Ph &       52.1050  &    -0.0488  &    0.3050  &     \\ 
    6962  &    Sp &       38.8610  &     1.0745  &    0.0940  &  f  \\ 
    7335  &    Sp &      318.8830  &    -0.3546  &    0.1970  & m    \\ 
    7350  &    Ph &        7.5630  &    -0.7872  &    0.1550  &     \\ 
    7431  &    Ph &      340.9550  &    -0.2746  &    0.3500  &     \\ 
    7876  &    Sp &       19.1830  &     0.7936  &    0.0760  &  f  \\ 
    8004  &    Ph &      347.5290  &    -0.5580  &    0.3510  &     \\ 
    8195  &    Ph &      331.0060  &    -0.8957  &    0.2690  &     \\ 
    8280  &    Ph &        8.5730  &     0.7955  &    0.1850  &  m   \\ 
    8555  &    Ph &        2.9160  &    -0.4150  &    0.1980  &   m  \\ 
    8888  &    Ph &        5.1660  &     0.3214  &    0.3990  &     \\ 
    9117  &    Ph &       46.9000  &     0.9884  &    0.2720  &     \\ 
    9133  &    Ph &       16.6430  &     0.4606  &    0.2670  &     \\ 
    9558  &    Ph &       15.9840  &     0.3810  &    0.3910  &     \\ 
    9633  &    Ph &       33.9190  &     1.0952  &    0.1960  &  m   \\ 
    9739  &    Ph &      323.6950  &    -0.8790  &    0.1200  &     \\ 
    9817  &    Ph &        5.0380  &     0.5950  &    0.2250  &     \\ 
   10028  &    Sp &       17.7420  &     0.2762  &    0.0650  &  f  \\ 
   10096  &    Sp &       29.4300  &    -0.1794  &    0.0780  &  f  \\ 
   10434  &    Sp &      329.9570  &    -1.1924  &    0.1040  &  f  \\ 
   10690  &    Ph &      347.5000  &     1.0822  &    0.3120  &     \\ 
   10805  &    Sp &      344.9280  &    -0.0136  &    0.0450  &  f  \\ 
   11102  &    Ph &      315.1590  &    -0.9996  &    0.1540  &  f  \\ 
   11120  &    Ph &      325.4590  &     0.3693  &    0.1070  &  f  \\ 
   11172  &    Ph &      322.4130  &    -0.2022  &    0.1360  &  f  \\ 
   11294  &    Ph &       20.2970  &     0.3737  &    0.1170  &  m   \\ 
   11296  &    Ph &       20.5190  &     0.3106  &    0.1760  & m    \\ 
   11306  &    Ph &       56.7380  &    -0.5176  &    0.2740  &     \\ 
   11311  &    Ph &       47.0150  &     0.4335  &    0.2050  &  m   \\ 
   11404  &    Ph &       26.9260  &     0.8671  &    0.1950  &     \\ 
   11697  &    Ph &        0.1350  &    -0.5553  &    0.2480  &     \\ 
   12110  &    Ph &      338.6770  &     0.0114  &    0.2820  &     \\ 
   12310  &    Ph &      351.5230  &     1.0208  &    0.1510  &     \\ 
   12326  &    Ph &      326.2440  &     1.0698  &    0.2930  &     \\ 
   12779  &    Sp &      309.4730  &     1.2194  &    0.0800  &  f  \\ 
   12780  &    Sp &      322.1570  &     1.2302  &    0.0490  &  m   \\ 
   12781  &    Sp &        5.4080  &    -1.0106  &    0.0840  &  f  \\ 
   12825  &    Ph &      359.5440  &    -1.1823  &    0.1500  &  f  \\ 
   12843  &    Sp &      323.8790  &    -0.9796  &    0.1670  &  f  \\ 
   12856  &    Sp &      332.8650  &     0.7556  &    0.1720  &  m   \\ 
   12897  &    Sp &       18.4240  &    -0.1027  &    0.0170  &     \\ 
   12903  &    Ph &      327.6490  &     0.3851  &    0.0530  & m    \\ 
   12909  &    Ph &       50.7700  &     0.2366  &    0.1310  &     \\ 
   12917  &    Ph &       17.0230  &     1.2028  &    0.2030  &  m   \\ 
   12923  &    Ph &        9.9910  &     0.6400  &    0.1130  &  f  \\ 
   12930  &    Sp &      309.6830  &    -0.4764  &    0.1470  &  f  \\ 
   12950  &    Sp &      351.6670  &    -0.8406  &    0.0830  &  f  \\ 
   12964  &    Ph &      309.9510  &    -0.0693  &   -0.0000  &  f  \\ 
   12971  &    Sp &        6.6480  &    -0.3033  &    0.2350  &     \\ 
   12979  &    Sp &       11.6010  &     0.0024  &    0.1160  &     \\ 
   12983  &    Sp &       16.4580  &     0.1454  &    0.2650  &     \\ 
   13007  &    Ph &        4.8410  &     0.6031  &    0.1050  &  f  \\ 
   13013  &    Ph &       14.3390  &     0.8482  &    0.1070  &     \\ 
   13035  &    Ph &      337.6120  &     0.9692  &    0.1310  &     \\ 
   13063  &    Ph &      338.9400  &    -1.1845  &    0.1340  & m    \\ 
   13068  &    Ph &      345.3330  &    -0.6541  &    0.0750  &  f  \\ 
   13070  &    Sp &      357.7850  &    -0.7466  &    0.1990  &  m   \\ 
\hline
\end{tabular}
\flushleft
$^a$Supernovae classification, Sp=Spectroscopic, Ph=Photometric\\ ~(see Section~\ref{data})\\
$^b$Sample selection, f=final host spectroscopy sample (see\\ ~Section~\ref{cuts}), m=Master sample (see Section~\ref{HR_ext})\\
\end{table}

\begin{table}
\center
\contcaption{}
\begin{tabular}{cccccc}
\hline
\bf SN ID & \bf Class.$^a$ & \bf Ra & \bf Dec &  \bf Redshift & \bf Sample$^b$ \\
& & (deg)  & (deg) & &   \\
\hline
   13071  &    Ph &      358.6180  &    -0.7188  &    0.1790  &     \\ 
   13099  &    Sp &      359.8190  &    -1.2507  &    0.2660  &     \\ 
   13113  &    Ph &       19.7740  &    -1.2338  &    0.1220  &  m   \\ 
   13135  &    Sp &        4.1740  &    -0.4252  &    0.1050  &  f  \\ 
   13254  &    Sp &       42.0590  &    -0.3468  &    0.1810  &  m   \\ 
   13354  &    Sp &       27.5650  &    -0.8867  &    0.1580  &  f  \\ 
   13458  &    Ph &       16.4620  &    -0.2496  &    0.3190  &     \\ 
   13511  &    Sp &       40.6110  &    -0.7942  &    0.2380  &     \\ 
   13522  &    Ph &       21.6000  &    -0.1613  &    0.1670  &     \\ 
   13545  &    Ph &       52.3430  &     0.5963  &    0.2140  &     \\ 
   13601  &    Ph &       16.4310  &    -0.5327  &    0.2440  &     \\ 
   13610  &    Sp &      326.0140  &     0.7255  &    0.2980  &     \\ 
   13633  &    Ph &        4.6650  &     0.0055  &    0.3880  &     \\ 
   13998  &    Ph &      324.0760  &    -0.0256  &    0.1230  &     \\ 
   14137  &    Ph &       56.5190  &    -0.4009  &    0.3120  &     \\ 
   14153  &    Ph &       49.3390  &     0.0785  &    0.1820  &  f  \\ 
   14193  &    Ph &      359.0010  &     1.0131  &    0.1510  &     \\ 
   14269  &    Ph &       54.7590  &     0.2792  &    0.2810  &     \\ 
   14279  &    Sp &       18.4880  &     0.3714  &    0.0450  &  f  \\ 
   14284  &    Sp &       49.0490  &    -0.6010  &    0.1810  & m    \\ 
   14318  &    Sp &      340.4250  &    -0.1369  &    0.0580  &  f  \\ 
   14340  &    Ph &      345.8270  &    -0.8553  &    0.2780  &     \\ 
   14377  &    Sp &       48.2640  &    -0.4718  &    0.1390  & m    \\ 
   14398  &    Ph &       50.7100  &     0.2729  &    0.1180  &     \\ 
   14421  &    Sp &       31.8300  &     1.2520  &    0.1750  &  f  \\ 
   14753  &    Ph &       44.3410  &     0.6866  &    0.1350  &     \\ 
   14816  &    Sp &      336.7160  &     0.5064  &    0.1070  &  f  \\ 
   14844  &    Ph &      353.4150  &     0.1766  &    0.4750  &     \\ 
   14961  &    Ph &       15.9190  &     0.9313  &    0.3710  &     \\ 
   15129  &    Sp &      318.9020  &    -0.3217  &    0.1980  &  f  \\ 
   15136  &    Sp &      351.1620  &    -0.7179  &    0.1490  &  f  \\ 
   15161  &    Sp &       35.8430  &     0.8190  &    0.2500  &     \\ 
   15222  &    Sp &        2.8520  &     0.7020  &    0.1990  &    m \\ 
   15234  &    Sp &       16.9580  &     0.8286  &    0.1360  &  m   \\ 
   15262  &    Ph &      342.4330  &    -0.4042  &    0.2470  &     \\ 
   15421  &    Sp &       33.7410  &     0.6027  &    0.1850  &  m   \\ 
   15425  &    Sp &       55.5610  &     0.4784  &    0.1600  &  f  \\ 
   15443  &    Sp &       49.8670  &    -0.3180  &    0.1820  &   m  \\ 
   15454  &    Ph &      327.8590  &    -0.8492  &    0.3830  &     \\ 
   15456  &    Sp &      331.8630  &    -0.9029  &    0.0890  &  m   \\ 
   15467  &    Sp &      320.0200  &    -0.1774  &    0.2100  &     \\ 
   15533  &    Ph &      353.8630  &    -0.3733  &    0.3440  &     \\ 
   15568  &    Ph &      356.8460  &     1.0501  &    0.2070  &     \\ 
   15587  &    Ph &       54.4180  &     0.9984  &    0.2190  &     \\ 
   15648  &    Sp &      313.7190  &    -0.1958  &    0.1750  &  f  \\ 
   15727  &    Ph &      351.4000  &    -0.0145  &    0.1050  &     \\ 
   15753  &    Ph &       57.4170  &    -0.0277  &    0.1590  &     \\ 
   15765  &    Ph &       32.8480  &     0.2462  &    0.3050  &     \\ 
   15784  &    Ph &      356.6760  &    -0.6167  &    0.2770  &     \\ 
   15950  &    Ph &        6.1890  &     0.9841  &    0.2200  &  f  \\ 
   15971  &    Ph &       40.1130  &     0.5263  &    0.3160  &     \\ 
   16038  &    Ph &       55.5710  &    -0.2378  &    0.4200  &     \\ 
   16069  &    Sp &      341.2450  &    -1.0064  &    0.1290  &  f  \\ 
   16099  &    Sp &       26.4210  &    -1.0546  &    0.1970  &  m   \\ 
   16163  &    Ph &       31.4990  &    -0.8558  &    0.1550  &  f  \\ 
   16172  &    Ph &       57.7050  &    -0.2182  &    0.2190  &     \\ 
   16211  &    Sp &      348.1630  &     0.2660  &    0.3110  &     \\ 
   16215  &    Sp &       18.4070  &     0.4237  &    0.0470  &     \\ 
   \hline
\end{tabular}
\flushleft
$^a$Supernovae classification, Sp=Spectroscopic, Ph=Photometric\\ ~(see Section~\ref{data})\\
$^b$Sample selection, f=final host spectroscopy sample (see\\ ~Section~\ref{cuts}), m=Master sample (see Section~\ref{HR_ext})\\
\end{table}

\begin{table}
\center
\contcaption{}
\begin{tabular}{cccccc}
\hline
\bf SN ID & \bf Class.$^a$ & \bf Ra & \bf Dec &  \bf Redshift & \bf Sample$^b$ \\
& & (deg)  & (deg) & &   \\
\hline
   16287  &    Sp &       46.6650  &     0.0620  &    0.1070  &  m   \\ 
   16259  &    Sp &      352.0330  &     0.8582  &    0.1190  &  m   \\ 
   16276  &    Sp &       20.5760  &     1.0082  &    0.0560  &   m  \\ 
   16280  &    Sp &       14.1220  &    -1.2268  &    0.0380  &  f  \\ 
   16314  &    Sp &      320.9290  &    -0.8430  &    0.0630  &  m   \\ 
   16333  &    Sp &      328.9940  &    -1.0703  &    0.0720  & m    \\ 
   16392  &    Sp &       27.9520  &     0.2638  &    0.0590  &  f  \\ 
   16462  &    Ph &       17.0410  &    -0.3859  &    0.2450  &     \\ 
   16482  &    Sp &      328.7080  &     0.9307  &    0.2110  &     \\ 
   16543  &    Ph &       34.3940  &     0.6481  &    0.2890  &     \\ 
   16666  &    Ph &       52.0300  &     0.1031  &    0.1330  &  m   \\ 
   16692  &    Sp &      320.3830  &     0.9949  &    0.0340  &  f  \\ 
   16789  &    Sp &       43.7840  &     0.2335  &    0.3250  &     \\ 
   17117  &    Sp &       40.6020  &    -0.7952  &    0.1400  &  f  \\ 
   17134  &    Sp &       54.5680  &    -0.1101  &    0.0870  &     \\ 
   17135  &    Sp &       56.5290  &     0.3903  &    0.0310  &     \\ 
   17171  &    Sp &      326.5030  &    -1.2193  &    0.1600  &  f  \\ 
   17176  &    Sp &      334.4030  &     0.6133  &    0.0930  &  m   \\ 
   17186  &    Sp &       31.6160  &    -0.8981  &    0.0800  &  m   \\ 
   17206  &    Ph &       45.9860  &     0.7282  &    0.1560  &  m   \\ 
   17215  &    Sp &       54.9270  &     1.0927  &    0.1810  &  m   \\ 
   17219  &    Ph &        3.1940  &    -0.1031  &    0.1940  &   m  \\ 
   17240  &    Sp &        8.6420  &    -1.2160  &    0.0730  &  f  \\ 
   17280  &    Sp &       55.7920  &     0.1040  &    0.1310  &  f  \\ 
   17316  &    Ph &        3.7660  &    -0.6243  &    0.1040  &     \\ 
   17332  &    Sp &       43.7720  &    -0.1477  &    0.1830  &  m   \\ 
   17340  &    Sp &       41.2130  &     0.3653  &    0.2570  &     \\ 
   17366  &    Sp &      315.7850  &    -1.0312  &    0.1390  &  m   \\ 
   17376  &    Ph &      330.6440  &     0.6728  &    0.0720  &     \\ 
   17434  &    Ph &       18.4380  &    -0.0730  &    0.1790  &  m   \\ 
   17497  &    Sp &       37.1360  &    -1.0428  &    0.1450  & m    \\ 
   17458  &    Ph &      358.1590  &    -0.3380  &    0.0810  &     \\ 
   17473  &    Ph &       12.2590  &     0.5474  &    0.1700  & m    \\ 
   17547  &    Ph &      339.1610  &     1.0913  &    0.0610  &     \\ 
   17605  &    Sp &      309.2030  &     0.0985  &    0.1460  &  f  \\ 
   17629  &    Sp &       30.6360  &    -1.0899  &    0.1370  &  f  \\ 
   17784  &    Sp &       52.4620  &     0.0545  &    0.0370  &  m   \\ 
   17880  &    Sp &       44.9740  &     1.1601  &    0.0730  &  m   \\ 
   17886  &    Sp &       54.0070  &     1.1048  &    0.0410  &  f  \\ 
   17907  &    Ph &       56.2950  &     0.4094  &    0.1940  &  m   \\ 
   17958  &    Ph &       34.4350  &    -0.7129  &    0.2760  &     \\ 
   18030  &    Sp &        4.9330  &    -0.4001  &    0.1560  &  m   \\ 
   18047  &    Ph &       22.0740  &    -0.6586  &    0.3590  &     \\ 
   18100  &    Ph &       29.4320  &     0.4243  &    0.4980  &     \\ 
   18201  &    Ph &       47.3120  &    -0.6450  &    0.2930  &     \\ 
   18224  &    Ph &      348.1750  &    -0.3129  &    0.3390  &     \\ 
   18273  &    Ph &      334.7390  &    -0.6303  &    0.3160  &     \\ 
   18298  &    Sp &       18.2670  &    -0.5400  &    0.1200  &  f  \\ 
   18454  &    Ph &       57.7460  &    -0.2836  &    0.3480  &     \\ 
   18612  &    Sp &       12.2880  &     0.5966  &    0.1150  &  f  \\ 
   18630  &    Ph &      347.9800  &    -0.2641  &    0.3590  &     \\ 
   18697  &    Sp &       11.2240  &    -0.9969  &    0.1070  &  f  \\ 
   18721  &    Sp &        3.0780  &    -0.0777  &    0.4030  &     \\ 
   18751  &    Sp &        5.7220  &     0.7759  &    0.0710  &  f  \\ 
   18764  &    Ph &        4.0140  &     1.1744  &    0.2520  &     \\ 
   18801  &    Ph &       49.7600  &    -0.4582  &    0.2450  &     \\ 
   18809  &    Sp &       50.8810  &     0.6673  &    0.1320  &  f  \\ 
   18835  &    Sp &       53.6850  &     0.3555  &    0.1230  &  f  \\ 
\hline
\end{tabular}
\flushleft
$^a$Supernovae classification, Sp=Spectroscopic, Ph=Photometric\\ 
~(see Section~\ref{data})\\
$^b$Sample selection, f=final host spectroscopy sample (see\\ 
~Section~\ref{cuts}), m=Master sample (see Section~\ref{HR_ext})\\
\end{table}

\begin{table}
\center
\contcaption{}
\begin{tabular}{cccccc}
\hline
\bf SN ID & \bf Class.$^a$ & \bf Ra & \bf Dec &  \bf Redshift & \bf Sample$^b$ \\
& & (deg)  & (deg) & &   \\
\hline
   18855  &    Sp &       48.6340  &     0.2689  &    0.1280  &  m   \\ 
   18890  &    Sp &       16.4430  &    -0.7595  &    0.0660  &  f  \\ 
   18903  &    Sp &       12.2510  &    -0.3233  &    0.1560  &  f  \\ 
   18927  &    Sp &       46.6840  &    -0.7573  &    0.3610  &     \\ 
   18959  &    Sp &       36.4080  &     0.7089  &    0.4010  &     \\ 
   19090  &    Ph &      357.1620  &    -0.4065  &    0.3120  &     \\ 
   19155  &    Sp &       31.2650  &     0.1751  &    0.0770  &  f  \\ 
   19317  &    Ph &      310.4550  &     1.0649  &    0.1790  &     \\ 
   19353  &    Sp &       43.1130  &     0.2517  &    0.1540  &  m   \\ 
   19616  &    Sp &       37.1000  &     0.1860  &    0.1650  &  f  \\ 
   19626  &    Sp &       35.9280  &    -0.8265  &    0.1130  &  f  \\ 
   19681  &    Ph &       20.4120  &    -0.4702  &    0.3510  &     \\ 
   19778  &    Ph &      349.6880  &    -0.4976  &    0.3960  &     \\ 
   19787  &    Ph &        0.2810  &    -0.0979  &    0.1970  &  f  \\ 
   19794  &    Sp &      359.3190  &     0.2485  &    0.2970  &     \\ 
   19968  &    Sp &       24.3490  &    -0.3117  &    0.0560  &  f  \\ 
   19969  &    Sp &       31.9100  &    -0.3240  &    0.1750  &  f  \\ 
   20047  &    Ph &      326.5910  &     0.6311  &    0.3740  &     \\ 
   20064  &    Sp &      358.5860  &    -0.9172  &    0.1050  &  m   \\ 
   20084  &    Sp &      347.9770  &    -0.5791  &    0.0910  &  f  \\ 
   20141  &    Ph &      357.5410  &    -0.5245  &    0.3410  &     \\ 
   20314  &    Ph &        4.1790  &     0.7225  &    0.2100  &     \\ 
   20331  &    Ph &        7.5400  &     1.2459  &    0.1840  & m    \\ 
   20350  &    Sp &      312.8070  &    -0.9578  &    0.1290  &  f  \\ 
   20386  &    Ph &       55.7980  &    -0.4896  &    0.3080  &     \\ 
   20420  &    Sp &      338.8710  &     0.4822  &    0.1510  &     \\ 
   20480  &    Ph &      357.2740  &     0.9182  &    0.1680  &     \\ 
   20528  &    Sp &       43.1210  &    -1.1394  &    0.1360  & m    \\ 
   20625  &    Sp &        5.6830  &    -0.4794  &    0.1080  &  f  \\ 
   20626  &    Ph &        8.4760  &    -0.5930  &    0.2760  &     \\ 
   20678  &    Ph &       17.5050  &     1.2353  &    0.2060  &     \\ 
   20721  &    Ph &      323.1850  &    -0.6227  &    0.2120  &     \\ 
   20726  &    Ph &       42.2800  &    -0.1595  &    0.3200  &     \\ 
   20787  &    Ph &       50.5110  &    -0.4425  &    0.2700  &     \\ 
   20788  &    Ph &       51.6730  &    -0.4778  &    0.3930  &     \\ 
   20889  &    Sp &       52.3810  &     0.5168  &    0.2090  &  m   \\ 
   20979  &    Ph &       44.7730  &    -0.2587  &    0.1270  &     \\ 
   21034  &    Sp &       28.1420  &     1.2441  &    0.1090  &  f  \\ 
   21062  &    Sp &      333.4290  &     0.3966  &    0.1390  & m    \\ 
   21442  &    Ph &        6.9600  &     0.6003  &    0.2130  &     \\ 
   21502  &    Sp &      353.6000  &    -0.8902  &    0.0890  &  f  \\ 
   21510  &    Sp &        7.3520  &     0.8311  &    0.1500  &     \\ 
   21669  &    Sp &       11.6140  &    -1.0609  &    0.1240  &  f  \\ 
   21709  &    Ph &      326.5280  &    -1.0493  &    0.1590  & m    \\ 
   21872  &    Ph &        7.6130  &    -0.5915  &    0.2290  &     \\ 
   22006  &    Ph &       48.7830  &    -0.9626  &    0.3960  &     \\ 
   22075  &    Sp &       29.9640  &     1.2166  &    0.1300  &  f  \\ 
\hline
\end{tabular}
\flushleft
$^a$Supernovae classification, Sp=Spectroscopic, Ph=Photometric\\ ~(see Section~\ref{data})\\
$^b$Sample selection, f=final host spectroscopy sample (see\\ ~Section~\ref{cuts}), m=Master sample (see Section~\ref{HR_ext})\\
\end{table}

\label{lastpage}

\end{document}